\documentstyle[12pt,epsf,epsfig,rotating]{article}

\input axodraw.sty

\catcode`@=11
\def\citer{\@ifnextchar [{\@tempswatrue\@citexr}{\@tempswafalse\@citexr[]}}

\def\@citexr[#1]#2{\if@filesw\immediate\write\@auxout{\string\citation{#2}}\fi
  \def\@citea{}\@cite{\@for\@citeb:=#2\do
    {\@citea\def\@citea{--\penalty\@m}\@ifundefined
       {b@\@citeb}{{\bf ?}\@warning
       {Citation `\@citeb' on page \thepage \space undefined}}%
\hbox{\csname b@\@citeb\endcsname}}}{#1}}
\catcode`@=12

\newcommand{\lsim}{\raisebox{-0.13cm}{~\shortstack{$<$ \\[-0.07cm] $\sim$}}~}
\newcommand{\gsim}{\raisebox{-0.13cm}{~\shortstack{$>$ \\[-0.07cm] $\sim$}}~}
\newcommand{\tgb}{\mbox{tg}\beta}
\newcommand{\ctgb}{\mbox{ctg}\beta}
\newcommand{\tg}{\mbox{tg}}
\newcommand{\non}{\nonumber}

\newcommand{\lessim}{\lsim}
\newcommand{\STS}{\vspace{1mm}}
\newcommand{\GS}{\vspace{5mm}}

\newcommand{\ee}{$e^+e^-$\ }

\newcommand{\demi}{1\! /\! 2}

\newcommand{\W}{{\it W} }

\newcommand{\SM}{Standard Model }

\newcommand{\ra}{\to }
\newcommand{\epem}{e^+e^- }
\newcommand{\cs}{cross section }

\newcommand{\Hs}{Higgs-strahlung }
\newcommand{\p}{particle }
\newcommand{\ps}{particles }
\newcommand{\cp}{coupling }
\newcommand{\cps}{couplings }
\newcommand{\sce}{scenario }
\newcommand{\sces}{scenarios }
\newcommand{\ssy}{supersymmetric }

\begin{document}

\setcounter{page}{0}

\begin{titlepage}

\begin{flushright}
CERN-TH/97-379 \\
DESY 97-261 \\
hep-ph/9803257 \\
December 1997
\end{flushright}

\vspace*{1cm}

\renewcommand{\thefootnote}{\fnsymbol{footnote}}

\begin{center}
{\large \sc Electroweak Symmetry Breaking and Higgs Physics\footnote{
Lectures  at 36. Internationale Universit\"atswochen f\"ur Kern- und
Teilchenphysik, Schladming 1997.}
}

\vspace*{2cm}

{\large Michael Spira$^1$ and Peter M.~Zerwas$^2$}

\vspace*{2cm}

$^1$ {\it CERN, Theory Division, CH-1211 Geneva 23, Switzerland}

$^2$ {\it Deutsches Elektronen-Synchrotron DESY, D-22603 Hamburg, Germany}

\vspace*{2cm}


\end{center}
\begin{abstract}
An introduction to electroweak symmetry breaking and Higgs physics is
presented for  the Standard Model and its supersymmetric extensions. A brief overview
 will also be given on strong interactions of the electroweak gauge bosons
 in alternative scenarios. In addition to the theoretical basis, the 
present experimental status of Higgs physics and implications for future
 experiments at the LHC and $e^+e^-$ linear colliders are discussed.
\end{abstract}

\vspace*{\fill}

\begin{flushleft}
CERN-TH/97-379 \\
DESY 97-261 \\
hep-ph/9803257 \\
December 1997
\end{flushleft}

\end{titlepage}

\section{Introduction}
\paragraph{1.}
Revealing  the physical mechanism that is
responsible for the breaking of 
electroweak symmetries is one of the key
problems in particle physics. If the fundamental
particles -- leptons, quarks and gauge
bosons -- remain weakly interacting up to
very high energies, the sector in which
the electroweak symmetry is broken
must contain one or more fundamental
scalar Higgs bosons with light masses
of the order of the symmetry-breaking
scale $v\sim 
246$ GeV. The masses of the fundamental
particles are generated through the
interaction with the scalar background
Higgs field, which is  non-zero in the ground
state \cite{1}. Alternatively, the symmetry
breaking could be generated dynamically
by new strong forces characterized by
an interaction scale $\Lambda \sim
1$ TeV \cite{2}. If global symmetries of
the strong interactions are broken
spontaneously, the associated Goldstone
bosons can be absorbed by the gauge
fields, generating the masses of the
gauge particles. The masses of leptons
and quarks can be generated through
interactions with the fermion condensate.

\paragraph{2.}
A simple mechanism for the breaking of
the electroweak symmetry is incorporated in 
the Standard Model (SM) \cite{3}. To
accommodate all observed phenomena, a
complex isodoublet scalar field is introduced
 through self-interactions; this acquires
a non-vanishing vacuum expectation
value, breaking spontaneously the electroweak
symmetry SU(2)$_I\times$ U(1)$_Y$
down to the electromagnetic U(1)$_{EM}$
symmetry. The interactions of the gauge
bosons and fermions with the background
field generate the masses of these
particles. One scalar field component
is not absorbed in this process,
manifesting itself as the physical
Higgs particle $H$.

The mass of the Higgs boson is the
only unknown parameter in the symmetry-breaking 
sector of the Standard
Model, while all couplings are fixed
by the masses of the particles, a
consequence of the Higgs mechanism
{\it per se}. However, the mass of the Higgs
boson is constrained in two ways. Since
the quartic self-coupling of the Higgs
field grows indefinitely with rising
energy, an upper limit on the Higgs
mass can be derived from demanding
the SM particles to remain weakly
interacting up to a scale $\Lambda$
\cite{4}. On the other hand, stringent
lower bounds on the Higgs mass follow
from requiring the electroweak vacuum
to be stable \cite{5}. If the Standard
Model is valid up to scales near the 
Planck scale, the SM Higgs mass is
restricted to a narrow window between
130 and 190 GeV. For Higgs masses
either above or below this window, new
physical phenomena are expected to
occur at a scale $\Lambda$ between
$\sim 1$ TeV and the Planck scale. For Higgs
masses near 700 GeV, the scale of
new strong interactions would be as
low as $\sim 1$ TeV \cite{4,6}. 

The electroweak observables are affected
by the Higgs mass through radiative
corrections \cite{7}. Despite  the weak 
logarithmic dependence, the high-precision
electroweak data indicate a preference
for light Higgs masses close to
$\sim 100$ GeV \cite{8}. At the 95\% CL, the data
require a value of the Higgs mass within
the canonical range of the Standard Model.
By searching directly for the SM Higgs particle,
the LEP experiments have set a lower
limit of $M_H\gsim  84$ to $88$ GeV
 on the Higgs mass \cite{9}. If the
Higgs boson will not be found at LEP2
with a mass of less than about 100 GeV
\cite{10}, the search will continue at the
Tevatron, which may reach masses up to $\sim 120$ GeV
\cite{11}. The proton collider LHC can sweep
the entire canonical Higgs mass range
of the Standard Model \cite{12}. The properties
of the Higgs particle can be analysed very
accurately at $e^+e^-$ linear colliders \cite{13}, thus establishing
the Higgs mechanism experimentally.

\paragraph{3.}
If the Standard Model is embedded in a 
Grand Unified Theory (GUT) at high energies,
the natural scale of electroweak symmetry
breaking would be expected close to the
unification scale $M_{GUT}$.
Supersymmetry \cite{14} provides a solution of
this hierarchy problem. The quadratically
divergent contributions to the radiative
corrections of the scalar Higgs boson
mass are cancelled by the destructive
interference between bosonic
and fermionic loops in sypersymmetric theories \cite{15}. The Minimal
Supersymmetric extension of the Standard
Model (MSSM) can be derived as an effective
theory from supersymmetric grand unified
theories. A strong indication for the realization
of this physical picture in nature is the
excellent agreement between the value of
the electroweak mixing angle $\sin^2 \theta_W$
predicted by the unification of the gauge
couplings, and the experimentally measured value. If the
gauge couplings are unified in the
minimal supersymmetric theory at a scale
$M_{GUT} = {\cal O}(10^{16}~\mbox{GeV})$, 
the electroweak mixing angle is predicted
to be $\sin^2\theta_W = 0.2336 \pm 0.0017$
\cite{16} for a mass spectrum of the supersymmetric
particles of order $M_Z$ to 1 TeV.
This theoretical prediction is matched very well by 
 the experimental result
$\sin^2\theta_W^{exp} = 0.2316 \pm 0.0003$
\cite{8}; the difference between the two numbers
is less than 2 per mille.

In the MSSM, the Higgs sector is built up 
by two Higgs doublets \cite{17}. The doubling is
necessary to generate masses for up- and
down-type fermions in a supersymmetric
theory and to render the theory anomaly-free. 
The Higgs particle spectrum consists
of a quintet of states: two CP-even
scalar neutral ($h,H$), one CP-odd pseudoscalar
neutral ($A$), and a pair of charged ($H^\pm$)
Higgs bosons \cite{19}. The masses of the heavy
Higgs bosons, $H,A,H^\pm$, are expected to be of order $v$, 
but they may extend up to the TeV range. By contrast,
since the quartic Higgs self-couplings are 
determined by the gauge couplings, the mass
of the lightest Higgs boson $h$ is constrained
very stringently. At tree level, the mass
has been predicted to be smaller than the
$Z$ mass \cite{19}. Radiative corrections,
increasing as the fourth power of the
top mass, shift the upper limit to a value
between $\sim 100$ GeV and $\sim 130$
GeV, depending on the parameter $\tgb$,
the ratio of the vacuum expectation values
of the two neutral scalar Higgs fields.

A general lower bound of 73 GeV has been 
experimentally established
for the Higgs particle $h$ at
LEP \cite{9}. Continuing this search, the
entire $h$ mass range can be covered for
$\tgb \lsim 2$,
a value compatible with the unification
of the $b$ and $\tau$
masses at high energies. The search for $h$
masses in excess of $\sim 100$ GeV
and the search for the heavy Higgs bosons
will continue at the Tevatron, the LHC and $e^+e^-$
linear colliders. In these machines the
mass range can be covered up to $\sim 1$ TeV
 \citer{11,13}.

\paragraph{4.}
Elastic-scattering amplitudes of massive
vector bosons grow indefinitely with
energy if they are calculated as a
perturbative expansion in the coupling
of a non-Abelian gauge theory. As a
result, they violate the unitarity beyond
a critical energy scale of 
$\sim 1.2$ TeV. This problem can be solved
by introducing a light Higgs boson. In
alternative scenarios, the $W$ bosons may
become strongly interacting at TeV energies,
thus damping the rise of the elastic-scattering 
amplitudes. Naturally, the strong
forces between the $W$ bosons may be traced
back to new fundamental interactions
characterized by a scale of order 1 TeV \cite{2}.
If the underlying theory is globally
chiral-invariant, this  symmetry may be broken
spontaneously. The Goldstone bosons 
associated with the spontaneous 
breaking of the symmetry can be absorbed by  gauge
bosons to generate their masses and to build
up the longitudinal degrees of freedom of the wave functions.

Since the longitudinally polarized $W$ bosons
are associated with the Goldstone modes
of chiral symmetry breaking, the scattering
amplitudes of the $W_L$
bosons can be predicted for high energies
by a systematic expansion in the energy.
The leading term is parameter-free, a
consequence of the chiral symmetry-breaking
mechanism {\it per se}, which is independent of
the particular structure of the dynamical theory. The 
higher-order terms in the chiral expansion however 
are defined by the detailed structure 
of the underlying theory. With rising
energy the chiral expansion is expected to diverge
and new resonances may be generated in
$WW$ scattering at mass scales between 1
and 3 TeV. This picture is analogous to
pion dynamics in QCD, where the threshold
amplitudes can be predicted in a chiral
expansion, while at higher energies vector
and scalar resonances are formed in $\pi \pi$
scattering.

Such a scenario can be studied in $WW$
scattering experiments, where the $W$ bosons
are radiated, as quasi-real particles \cite{22},
off high-energy quarks in the proton
beams of the LHC \cite{12}, \citer{23,23B} or off electrons
and positrons in TeV linear colliders \cite{13,24,24a}.

\paragraph{5.}
This report is divided into three parts.
A basic introduction and a summary of the
main theoretical and experimental results
will be presented in the next section on
the Higgs sector of the Standard Model.
Also the search for the Higgs particle
at future $e^+e^-$
and hadron colliders will be described. In
the same way, the Higgs spectrum of 
supersymmetric theories will be discussed
in the following section. Finally, the main 
features of strong $W$ interactions and their
analysis in $WW$ scattering experiments will
be presented in the last section.

Only the basic elements of electroweak symmetry
breaking and Higgs mechanism can be reviewed  
in this report. Other aspects may be traced back
from Ref. \cite{24b} and recent reports collected in Ref. \cite{24A}.\\[1cm]

\section{The Higgs Sector of the Standard Model}
\subsection{The Higgs Mechanism}
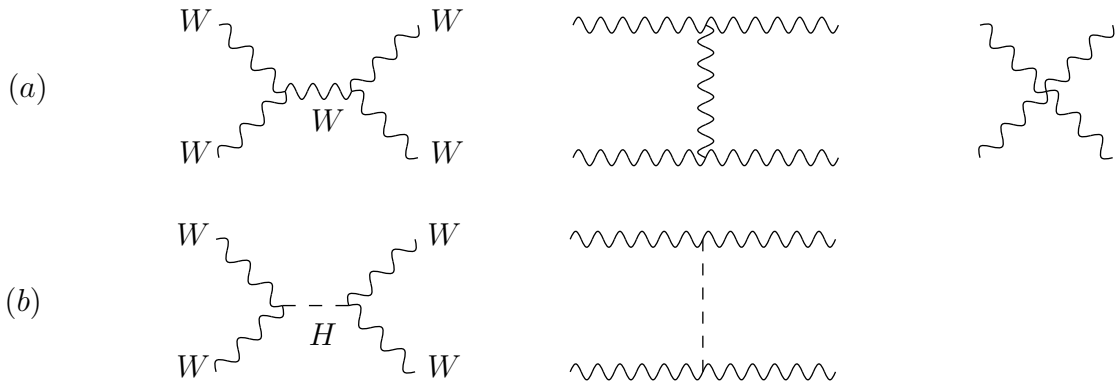
\begin{figure}[hbt]
\begin{center}
\begin{picture}(80,80)(50,-20)
\Photon(0,50)(25,25){3}{3}
\Photon(0,0)(25,25){3}{3}
\Photon(25,25)(50,25){3}{3}
\Photon(50,25)(75,50){3}{3}
\Photon(50,25)(75,0){3}{3}
\put(-80,23){$(a)$}
\put(-15,-2){$W$}
\put(-15,48){$W$}
\put(80,-2){$W$}
\put(80,48){$W$}
\put(35,10){$W$}
\end{picture}
\begin{picture}(60,80)(0,-20)
\Photon(0,50)(100,50){3}{12}
\Photon(0,0)(100,0){3}{12}
\Photon(50,50)(50,0){3}{6}
\end{picture}
\begin{picture}(60,80)(-90,-20)
\Photon(0,50)(25,25){3}{3}
\Photon(0,0)(25,25){3}{3}
\Photon(25,25)(50,50){3}{3}
\Photon(25,25)(50,0){3}{3}
\end{picture} \\
\begin{picture}(80,60)(83,0)
\Photon(0,50)(25,25){3}{3}
\Photon(0,0)(25,25){3}{3}
\DashLine(25,25)(50,25){6}
\Photon(50,25)(75,50){3}{3}
\Photon(50,25)(75,0){3}{3}
\put(-80,23){$(b)$}
\put(-15,-2){$W$}
\put(-15,48){$W$}
\put(80,-2){$W$}
\put(80,48){$W$}
\put(35,10){$H$}
\end{picture}
\begin{picture}(60,60)(33,0)
\Photon(0,50)(100,50){3}{12}
\Photon(0,0)(100,0){3}{12}
\DashLine(50,50)(50,0){5}
\end{picture}  \\
\end{center}
\caption[]{\it \label{fg:wwtoww} Generic diagrams of elastic $WW$ scattering:
(a) pure gauge-boson dynamics, and (b) Higgs-boson exchange.}
\end{figure}
\noindent
At high energies, the amplitude for the elastic 
scattering of massive $W$ bosons, $WW \to WW$, 
grows indefinitely with energy for longitudinally 
polarized particles, Fig. \ref{fg:wwtoww}a. This is a consequence
of the linear rise of the longitudinal $W_L$ wave function, 
$\epsilon_L = (p,0,0,E)/M_W$,
with the energy of the particle. Even though the term of 
the scattering amplitude rising as the fourth power in the energy
is cancelled by virtue of the non-Abelian 
gauge symmetry, the amplitude remains quadratically
divergent in the energy. On the other hand, 
unitarity requires elastic-scattering 
amplitudes of partial waves $J$ to be bounded by
$\Re e A_J \leq 1/2$.
Applied to the asymptotic $S$-wave amplitude
$A_0 = G_F s/8\pi\sqrt{2}$ of the isospin-zero channel
$2W_L^+W_L^- + Z_L Z_L$,  the bound  \cite{25}
\begin{equation}
s \leq 4\pi\sqrt{2}/G_F \sim (1.2~\mbox{TeV})^2
\end{equation}
on the c.m. energy $\sqrt{s}$ can be derived for 
the validity of a theory of weakly 
coupled massive gauge bosons.

However, the quadratic rise in the energy can be damped by 
exchanging a new scalar particle, Fig. \ref{fg:wwtoww}b. 
To achieve the 
cancellation, the size of the coupling must be given 
by the product of the gauge coupling with the gauge
boson mass. For high energies, the
amplitude $A'_0 = -G_F s/8\pi\sqrt{2}$
cancels exactly the quadratic divergence of the pure 
gauge-boson amplitude $A_0$.
Thus, unitarity can be restored by introducing a weakly
coupled \underline{\it Higgs particle}.

In the same way, the linear divergence of the amplitude 
$A(f\bar f\to W_L W_L)\sim gm_f\sqrt{s}$
for the annihilation of a fermion--antifermion pair to 
a pair of longitudinally polarized gauge bosons,
can be damped by adding the Higgs exchange to 
the gauge-boson exchange. In this case
the Higgs particle must couple proportionally
to the mass $m_f$ of the fermion $f$.

These observations can be summarized in a theorem:
{\it A theory of massive gauge bosons and fermions
that are weakly coupled up to very high 
energies, requires, by unitarity, the existence 
of a Higgs particle; the Higgs particle is a
scalar $0^+$ particle that couples to other particles 
proportionally to the masses of the particles.}

The assumption that the couplings of the 
fundamental particles are weak up to very
high energies is qualitatively supported
by the perturbative renormalization of the
electroweak mixing angle $\sin^2\theta_W$
from the symmetry value 3/8 at the GUT scale
down to $\sim 0.2$, 
which is close to the experimentally observed
value at low energies.\\

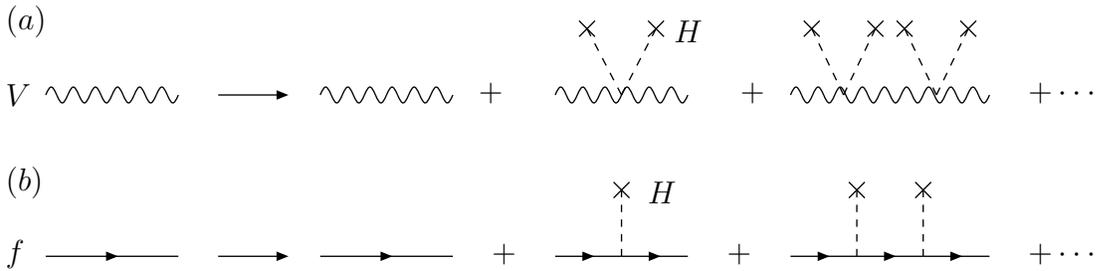
\begin{figure}[hbt]
\begin{center}
\begin{picture}(60,10)(80,40)
\Photon(0,25)(50,25){3}{6}
\LongArrow(65,25)(90,25)
\put(-15,21){$V$}
\put(-15,50){$(a)$}
\end{picture}
\begin{picture}(60,10)(40,40)
\Photon(0,25)(50,25){3}{6}
\put(60,23){$+$}
\end{picture}
\begin{picture}(60,10)(15,40)
\Photon(0,25)(50,25){3}{6}
\DashLine(25,25)(12,50){3}
\DashLine(25,25)(38,50){3}
\Line(9,53)(15,47)
\Line(9,47)(15,53)
\Line(35,53)(41,47)
\Line(35,47)(41,53)
\put(45,45){$H$}
\put(70,23){$+$}
\end{picture}
\begin{picture}(60,10)(-10,40)
\Photon(0,25)(75,25){3}{9}
\DashLine(20,25)(8,50){3}
\DashLine(20,25)(32,50){3}
\DashLine(55,25)(43,50){3}
\DashLine(55,25)(67,50){3}
\Line(5,53)(11,47)
\Line(5,47)(11,53)
\Line(29,53)(35,47)
\Line(29,47)(35,53)
\Line(40,53)(46,47)
\Line(40,47)(46,53)
\Line(64,53)(70,47)
\Line(64,47)(70,53)
\put(90,23){$+ \cdots$}
\end{picture} \\
\begin{picture}(60,80)(80,20)
\ArrowLine(0,25)(50,25)
\LongArrow(65,25)(90,25)
\put(-15,23){$f$}
\put(-15,50){$(b)$}
\end{picture}
\begin{picture}(60,80)(40,20)
\ArrowLine(0,25)(50,25)
\put(65,23){$+$}
\end{picture}
\begin{picture}(60,80)(15,20)
\ArrowLine(0,25)(25,25)
\ArrowLine(25,25)(50,25)
\DashLine(25,25)(25,50){3}
\Line(22,53)(28,47)
\Line(22,47)(28,53)
\put(35,45){$H$}
\put(65,23){$+$}
\end{picture}
\begin{picture}(60,80)(-10,20)
\ArrowLine(0,25)(25,25)
\ArrowLine(25,25)(50,25)
\ArrowLine(50,25)(75,25)
\DashLine(25,25)(25,50){3}
\DashLine(50,25)(50,50){3}
\Line(22,53)(28,47)
\Line(22,47)(28,53)
\Line(47,53)(53,47)
\Line(47,47)(53,53)
\put(90,23){$+ \cdots$}
\end{picture}  \\
\end{center}
\caption[]{\it \label{fg:massgen} Generating (a) gauge boson and (b)
fermion masses through interactions with the scalar background field.}
\end{figure}

These ideas can be cast into an elegant
mathematical form by interpreting the electroweak
interactions as a gauge theory with spontaneous
symmetry breaking in the scalar sector. 
Such a theory consists of fermion fields, gauge
fields and a scalar field coupled by the 
standard gauge interactions and Yukawa interactions
to the other fields. Moreover, a self-interaction
\begin{equation}
V = \frac{\lambda}{2} \left[ |\phi|^2 - \frac{v^2}{2} \right]^2
\label{eq:potential}
\end{equation}
is introduced in the scalar sector, which leads to 
a non-zero ground-state value $v/\sqrt{2}$
of the scalar field. By fixing the phase of the 
vacuum amplitude to  zero, the gauge symmetry
is spontaneously broken in the scalar sector. Interactions of the 
gauge fields with the scalar background field,
Fig. \ref{fg:massgen}a, and Yukawa interactions of the 
fermion fields with the background field, Fig. \ref{fg:massgen}b, 
shift the masses of these fields from 
zero to non-zero values:
\begin{equation}
\begin{array}{lrclclcl}
\displaystyle
(a) \hspace*{2.0cm} &
\displaystyle
\frac{1}{q^2} & \to & \displaystyle \frac{1}{q^2} + \sum_j \frac{1}{q^2}
\left[ \left( \frac{gv}{\sqrt{2}} \right)^2 \frac{1}{q^2} \right]^j & = &
\displaystyle \frac{1}{q^2-M^2} & : & \displaystyle M^2 = g^2 \frac{v^2}{2}
\\ \\
(b) &
\displaystyle
\frac{1}{\not \! q} & \to & \displaystyle \frac{1}{\not \! q} +
\sum_j \frac{1}{\not \! q} \left[ \frac{g_fv}{\sqrt{2}} \frac{1}{\not
\! q} \right]^j & = & \displaystyle \frac{1}{\not \! q-m_f} & : &
\displaystyle m_f = g_f \frac{v}{\sqrt{2}} 
\end{array}
\end{equation}
Thus, in theories with gauge and Yukawa interactions,
in which the scalar field acquires a non-zero
ground-state value, the couplings are naturally
proportional to the masses. This ensures the
unitarity of the theory as discussed before.
These theories are renormalizable (as a result
of the gauge invariance, which is only disguised 
in the unitary formulation adopted so far), and 
thus they are well-defined and mathematically 
consistent.

\subsection{The Higgs Mechanism in the Standard Model}
Besides the Yang--Mills and the fermion parts, the
electroweak $SU_2 \times U_1$
Lagrangian includes a scalar isodoublet field
$\phi$, coupled to itself in the potential $V$,
cf. eq. (\ref{eq:potential}),
to the gauge fields through the covariant derivative
$iD = i\partial - g \vec{I} \vec{W} - g'YB$,
and to the up and down fermion fields $u,d$
through Yukawa interactions:
\begin{equation}
{\cal L}_0 = |D\phi|^2 - \frac{\lambda}{2} \left[ |\phi|^2
- \frac{v^2}{2} \right]^2 - g_d \bar d_L \phi d_R - g_u \bar u_L
\phi_c u_R + {\rm hc} ~.
\end{equation}
In the unitary gauge, the isodoublet $\phi$
is replaced by the physical Higgs field
$H$, $\phi\to [0,(v+H)/\sqrt{2}]$,
which describes the fluctuation of the $I_3=-1/2$
component of the isodoublet field about the ground-state 
value $v/\sqrt{2}$. The scale $v$
of the electroweak symmetry breaking is fixed
by the $W$ mass, which in turn can be reexpressed by the
Fermi coupling, $v = 1/\sqrt{\sqrt{2}G_F} \approx 246$ GeV.
The quartic coupling $\lambda$
and the Yukawa couplings $g_f$
can be reexpressed in terms of the physical
Higgs mass $M_H$ and the fermion masses $m_f$: 
\begin{eqnarray}
M_H^2 & = & \lambda v^2 \nonumber \\
m_f & = & g_f v / \sqrt{2}
\end{eqnarray}
respectively.

Since the couplings of the Higgs particle to gauge particles, 
to fermions and to itself are given by the gauge couplings
and the masses of the particles, the only unknown
parameter in the Higgs sector (apart from the CKM 
mixing matrix) is the Higgs mass. When this mass is fixed,
all properties of the Higgs particle can be predicted,
i.e. the lifetime and decay branching ratios, as 
well as the production mechanisms and the corresponding
cross sections.

\subsubsection{The SM Higgs Mass}
Even though the mass of the Higgs boson cannot be 
predicted in the Standard Model, stringent upper
and lower bounds can nevertheless be derived from
internal consistency conditions and extrapolations
of the model to high energies.

The Higgs boson has been introduced as a fundamental
particle to render 2--2 scattering amplitudes involving
longitudinally polarized $W$
bosons compatible with unitarity. Based on the general
principle of time-energy uncertainty, particles must
decouple from a physical system if their mass grows
indefinitely. The mass of the Higgs particle must 
therefore be bounded to restore unitarity in the 
perturbative regime. From the asymptotic expansion of 
the elastic $W_L W_L$ $S$-wave scattering amplitude including 
$W$ and Higgs exchanges, $A(W_L W_L \to W_L W_L) \to -G_F M_H^2/4\sqrt{2}\pi$,
it follows \cite{25} that
\begin{equation}
M_H^2 \leq 2\sqrt{2}\pi/G_F \sim (850~\mbox{GeV})^2 ~.
\end{equation}
Within the canonical formulation of the Standard Model,
consistency conditions  therefore require a Higgs mass below 1 TeV.\\

\begin{figure}[hbt]
\vspace*{-0.5cm}

\begin{center}
\begin{picture}(90,80)(60,-10)
\DashLine(0,50)(25,25){3}
\DashLine(0,0)(25,25){3}
\DashLine(50,50)(25,25){3}
\DashLine(50,0)(25,25){3}
\put(-15,45){$H$}
\put(-15,-5){$H$}
\put(55,-5){$H$}
\put(55,45){$H$}
\end{picture}
\begin{picture}(90,80)(10,-10)
\DashLine(0,50)(25,25){3}
\DashLine(0,0)(25,25){3}
\DashLine(75,50)(50,25){3}
\DashLine(75,0)(50,25){3}
\DashCArc(37.5,25)(12.5,0,360){3}
\put(-15,45){$H$}
\put(-15,-5){$H$}
\put(35,40){$H$}
\put(80,-5){$H$}
\put(80,45){$H$}
\end{picture}
\begin{picture}(50,80)(-40,2.5)
\DashLine(0,0)(25,25){3}
\DashLine(0,75)(25,50){3}
\DashLine(50,50)(75,75){3}
\DashLine(50,25)(75,0){3}
\ArrowLine(25,25)(50,25)
\ArrowLine(50,25)(50,50)
\ArrowLine(50,50)(25,50)
\ArrowLine(25,50)(25,25)
\put(-15,70){$H$}
\put(-15,-5){$H$}
\put(35,55){$t$}
\put(80,-5){$H$}
\put(80,70){$H$}
\end{picture}  \\
\setlength{\unitlength}{1pt}
\caption[]{\label{fg:lambda} \it Diagrams generating the evolution of
the Higgs self-interaction $\lambda$.}
\end{center}
\end{figure}
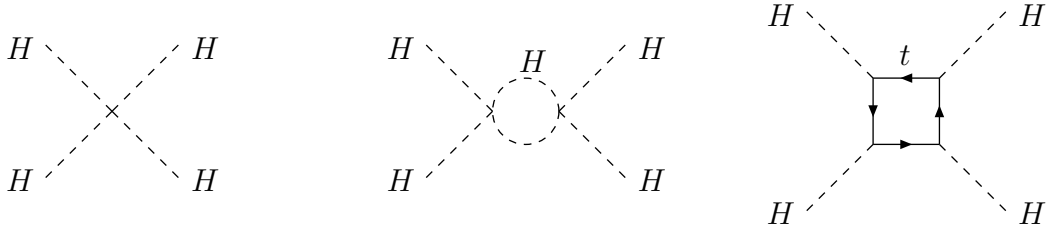

\begin{figure}[hbtp]

\vspace*{0.8cm}

\hspace*{2.0cm}
\begin{turn}{90}%
\epsfxsize=8.5cm \epsfbox{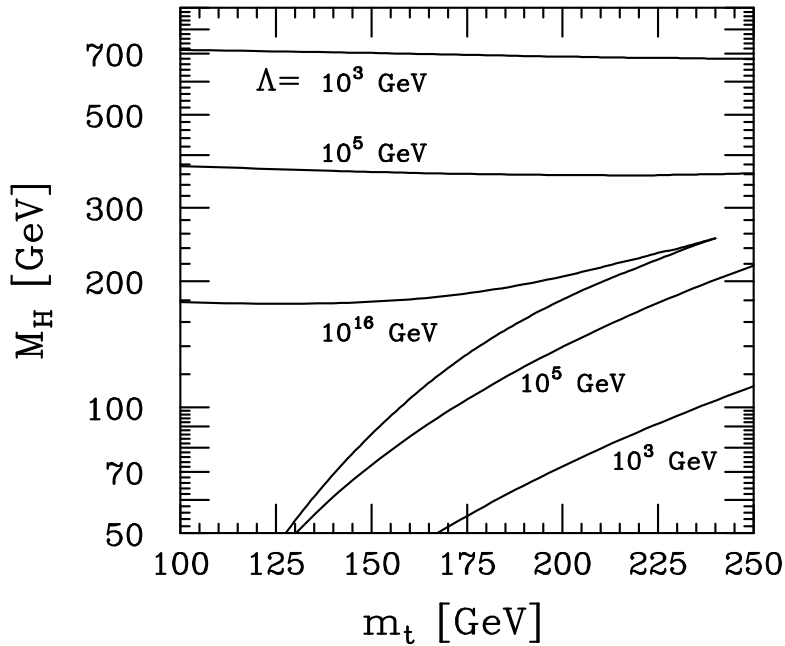}
\end{turn}
\vspace*{-0.2cm}

\caption[]{\label{fg:triviality} \it Bounds on the mass of the Higgs boson in
the SM. Here $\Lambda$ denotes the energy scale at which the Higgs-boson system of
the SM would become strongly interacting (upper bound); the lower bound follows
from the requirement of vacuum stability. (Refs. \cite{4,5}.)}
\vspace*{-0.3cm}
\end{figure}

Quite restrictive bounds on the value of the SM Higgs
mass follow from hypotheses on the 
energy scale $\Lambda$
up to which the Standard Model can be extended before
new physical phenomena emerge, which would be associated 
with new strong 
interactions between the fundamental particles. The 
key to these bounds is the evolution of the quartic
coupling $\lambda$
with the energy (i.e. the field strength)
due to quantum fluctuations \cite{4}. The basic
contributions are depicted in Fig. \ref{fg:lambda}. The
Higgs loop itself gives rise to an indefinite increase
of the coupling while the fermionic top-quark
loop, with increasing top mass, drives the coupling
to smaller values, finally even to values below zero.
The variation of the quartic Higgs coupling
$\lambda$ and the top-Higgs Yukawa coupling $g_t$
with energy, parametrized by $t=\log \mu^2/v^2$,
may be written as \cite{4}
\begin{equation}
\begin{array}{rclcl}
\displaystyle \frac{d\lambda}{dt} & = & \displaystyle \frac{3}{8\pi^2}
\left[ \lambda^2 + \lambda g_t^2 - g_t^4 \right] & : & \displaystyle
\lambda(v^2) = M_H^2/v^2 \\ \\
\displaystyle \frac{d g_t}{dt} & = & \displaystyle \frac{1}{32\pi^2}
\left[ \frac{9}{2} g_t^3 - 8 g_t g_s^2 \right] & : & \displaystyle
g_t(v^2) = \sqrt{2}~m_f/v ~.
\end{array}
\end{equation}
Only the leading contributions from Higgs, top
and QCD loops are taken into account.\\

For moderate top masses, the quartic coupling $\lambda$
rises indefinitely, $\partial \lambda / \partial t \sim + \lambda^2$,
and the coupling becomes strong shortly before 
reaching the Landau pole:
\begin{equation}
\lambda (\mu^2) = \frac{\lambda(v^2)}{1- \frac{3\lambda(v^2)}{8\pi^2} \log
\frac{\mu^2}{v^2}} ~.
\end{equation}
Reexpressing the initial value of $\lambda$
by the Higgs mass, the condition $\lambda (\Lambda) < \infty$,
can be translated to an \underline{upper bound} on the Higgs
mass: 
\begin{equation}
M_H^2 \leq \frac{8\pi^2 v^2}{3\log \frac{\Lambda^2}{v^2}} ~.
\end{equation}
This mass bound is related logarithmically to the energy $\Lambda$
up to which the Standard Model is assumed to be valid.
The maximal value of $M_H$ for the minimal cut-off $\Lambda \sim $ 1~TeV
is given by $\sim 750$ GeV. This bound is close to the estimate
of $\sim 700$ GeV
in lattice calculations for $\Lambda \sim 1$ TeV,
which allow  proper control of non-perturbative 
effects near the boundary \cite{6}.\\

\begin{table}[hbt]
\renewcommand{\arraystretch}{1.5}
\begin{center}
\begin{tabular}{|l||l|} \hline
$\Lambda$ & $M_H$ \\ \hline \hline
1 TeV & 55 GeV $\lsim M_H \lsim 700$ GeV \\
$10^{19}$ GeV & 130 GeV $\lsim M_H \lsim 190$ GeV \\ \hline
\end{tabular}
\renewcommand{\arraystretch}{1.2}
\caption[]{\label{tb:triviality}
\it Higgs mass bounds for two values of the cut-off $\Lambda$.}
\end{center}
\end{table}
A \underline{lower bound} on the Higgs mass can be derived from 
the requirement of vacuum stability \cite{4,5}. Since
top-loop corrections reduce $\lambda$
for increasing top-Yukawa coupling, $\lambda$
becomes negative if the top mass becomes too large. 
In such a case, the self-energy potential would 
become deep negative and the ground state would no longer 
be stable. To avoid the instability, the Higgs
mass must exceed a minimal value for a given top 
mass. This lower bound depends on the cut-off value $\Lambda$.\\

For any given $\Lambda$ the allowed values of $(M_t,M_H)$
pairs are shown in Fig. \ref{fg:triviality}. For a central top mass $M_t =$ 175
GeV, the allowed Higgs mass values are collected in 
Table \ref{tb:triviality} for two specific cut-off values $\Lambda$.
If the Standard Model is assumed to be valid up 
to the scale of grand unification, the Higgs
mass is restricted to a narrow window between
130 and 190~GeV. The observation of a Higgs mass
above or below this window would demand a new
physics scale below the GUT scale.

\subsubsection{Decays of the Higgs Particle}
The profile of the Higgs particle is uniquely
determined if the Higgs mass is fixed. The
strength of the Yukawa couplings of the Higgs
boson to fermions is set by the fermion masses $m_j$,
and the coupling to the electroweak gauge bosons
$V=W,Z$ by their masses $M_V$:
\begin{eqnarray}
g_{ffH} & = & \left[ \sqrt{2} G_F \right]^{1/2} m_f \\
g_{VVH} & = & 2 \left[ \sqrt{2} G_F \right]^{1/2} M_V^2 ~. \nonumber
\end{eqnarray}
The total decay width and lifetime, as well as the
 branching ratios for specific decay channels, are 
determined by these parameters. The measurement
of the decay characteristics can therefore
by exploited to establish, experimentally, 
that Higgs couplings grow with the masses
of the particles, a direct consequence of
the Higgs mechanism {\it sui generis}.

For Higgs particles in the intermediate mass range
${\cal O}(M_Z) \leq M_H \leq 2M_Z$, 
the main decay modes are decays into 
$b\bar b$ pairs and $WW,ZZ$
pairs,  one of the gauge bosons being virtual
below the respective threshold. Above the
$WW,ZZ$ pair thresholds, the Higgs particles decay almost
exclusively into these two channels, with a small
admixture of top decays near the $t\bar t$ threshold.
Below 140 GeV, the decays $H\to \tau^+\tau^-, c\bar c$
and $gg$ are also important besides the dominating
$b\bar b$ channel; $\gamma\gamma$
decays, though suppressed in rate, nevertheless provide 
 a clear 2-body signature for the
formation of Higgs particles in this mass range.

\paragraph{(a) Higgs decays to fermions} ~\\[0.5cm]
The partial width of Higgs decays to lepton
and quark pairs is given by \cite{26}
\begin{equation}
\Gamma (H\to f\bar f) = {\cal N}_c \frac{G_F}{4\sqrt{2}\pi} m_f^2(M_H^2) M_H
~,
\end{equation}
${\cal N}_c = 1$ or 3 being the colour factor. Near threshold the partial
width is suppressed by an additional factor $\beta_f^3$, where $\beta_f$
is the fermion velocity. Asymptotically, the
fermionic width grows only linearly with the
Higgs mass.
The bulk of QCD radiative corrections can be 
mapped into the scale dependence of the quark mass,
evaluated at the Higgs mass. For $M_H\sim 100$ GeV
the relevant parameters are $m_b (M_H^2) \simeq 3$ GeV and
$m_c (M_H^2) \simeq$ 0.6~GeV.
The reduction of the effective $c$-quark mass
overcompensates the colour factor in the ratio 
between charm and $\tau$
decays of Higgs bosons. The residual QCD corrections,
$\sim 5.7 \times (\alpha_s/\pi)$, modify the widths only slightly. 

\paragraph{(b) Higgs decays to $WW$ and $ZZ$ boson pairs} ~\\[0.5cm]
Above the $WW$ and $ZZ$ decay thresholds, the partial widths for these 
channels may be written as \cite{27}
\begin{equation}
\Gamma (H\to VV) = \delta_V \frac{G_F}{16\sqrt{2}\pi} M_H^3 (1-4x+12x^2)
\beta_V ~,
\end{equation}
where $x=M_V^2/M_H^2$ and $\delta_V = 2$ and 1 for $V=W$ and $Z$, 
respectively. For large Higgs masses, the vector bosons
are longitudinally polarized. Since the wave functions
of these states are linear in the energy, the widths
grow as the third power of the Higgs mass. Below the
threshold for two real bosons, the Higgs particle can
decay into $VV^*$
pairs, one of the vector bosons being virtual. The partial 
width is given in this case \cite{28} by
\begin{equation}
\Gamma(H\to VV^*) = \frac{3G^2_F M_V^4}{16\pi^3}~M_H  R(x)~\delta'_V ~,
\end{equation}
where $\delta'_W = 1$, $\delta'_Z = 7/12 - 10\sin^2\theta_W/9 + 40
\sin^4\theta_W/27$ and
\begin{displaymath}
R(x) = \frac{3(1-8x+20x^2)}{(4x-1)^{1/2}}\arccos\left(\frac{3x-1}{2x^{3/2}}
\right) - \frac{1-x}{2x} (2-13x+47x^2)
- \frac{3}{2} (1-6x+4x^2) \log x ~.
\end{displaymath}
The $ZZ^*$ channel becomes relevant for Higgs masses beyond 
$\sim 140$ GeV.  Above the threshold, the 4-lepton channel 
$H\to ZZ \to 4 \ell^\pm$
provides a very clear signal for Higgs bosons.

\paragraph{(c) Higgs decays to $gg$ and $\gamma\gamma$ pairs} ~\\[0.5cm]
In the Standard Model, gluonic Higgs decays are 
mediated by top- and bottom-quark loops, photonic decays in addition by 
$W$ loops. Since these decay modes are significant
only far below the top and $W$
thresholds, they are described by the approximate
expressions \cite{29,30}
\begin{eqnarray}
\Gamma (H\to gg) & = & \frac{G_F \alpha_s^2(M_H^2)}{36\sqrt{2}\pi^3}M_H^3
\left[ 1+ \left(\frac{95}{4} - \frac{7N_F}{6} \right) \frac{\alpha_s}{\pi}
\right] \label{eq:htogg} \\ \nonumber \\
\Gamma (H\to \gamma\gamma) & = & \frac{G_F \alpha^2}{128\sqrt{2}\pi^3}M_H^3
\left|  \frac{4}{3} {\cal N}_C e_t^2 - 7 \right|^2 ~,
\end{eqnarray}
which are valid in the limit $M_H^2 \ll 4M_W^2, 4M_t^2$.
The QCD radiative corrections, which include the $ggg$ and $gq\bar q$
final states in (\ref{eq:htogg}), are very important; they
increase the partial width by about 65\%. Even
though photonic Higgs decays are very rare, 
they nevertheless offer a simple and 
attractive signature for Higgs particles
by leading to just two stable particles 
in the final state.

\paragraph{\underline{Digression:}} Loop-mediated Higgs couplings can 
easily be calculated in the limit in which the  Higgs
mass is small compared to the loop mass, by exploiting a low-energy 
theorem \citer{29,32} for the external Higgs amplitude ${\cal A} (XH)$:
\begin{equation}
\lim_{p_H\to 0} {\cal A}(XH) = \frac{1}{v} \frac{\partial {\cal A}(X)}{\partial
 \log m} ~.
\end{equation}
The theorem can be derived by observing that the
insertion of an external zero-energy Higgs line into a 
fermionic propagator, for instance, is equivalent
to the substitution
\begin{displaymath}
\frac{1}{\not\! p-m} \to \frac{1}{\not\! p-m} \frac{m}{v} \frac{1}{\not\! p-m}
= \frac{1}{v} \frac{\partial}{\partial \log m} \frac{1}{\not\! p-m} ~.
\end{displaymath}
The amplitudes for processes including an
external Higgs line can therefore be obtained
from the amplitude without the external Higgs
line by taking the logarithmic derivative.
If applied to the gluon propagator at $Q^2=0$, $\Pi_{gg} \sim 
\frac{\alpha_s}{12\pi}
GG \log m$, the $Hgg$ amplitude can easily be derived as
${\cal A}(Hgg) = GG \frac{\alpha_s}{12\pi} \frac{1}{v}$. If higher 
orders are included, the parameter $m$ must be interpreted
as bare mass.

\begin{figure}[hbtp]

\vspace*{0.5cm}
\hspace*{1.0cm}
\begin{turn}{-90}%
\epsfxsize=8.5cm \epsfbox{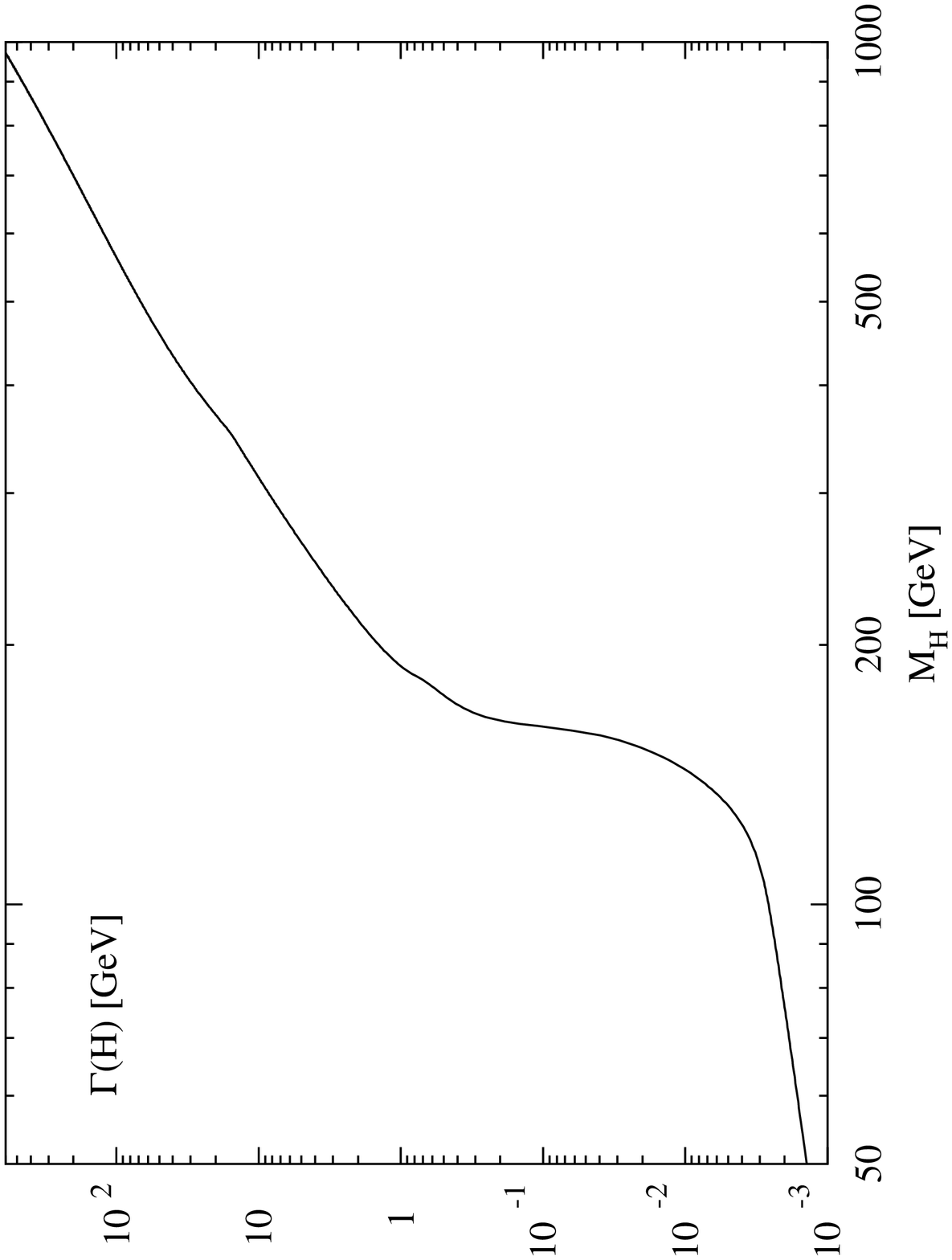}
\end{turn}

\vspace*{0.5cm}
\hspace*{1.0cm}
\begin{turn}{-90}%
\epsfxsize=8.5cm \epsfbox{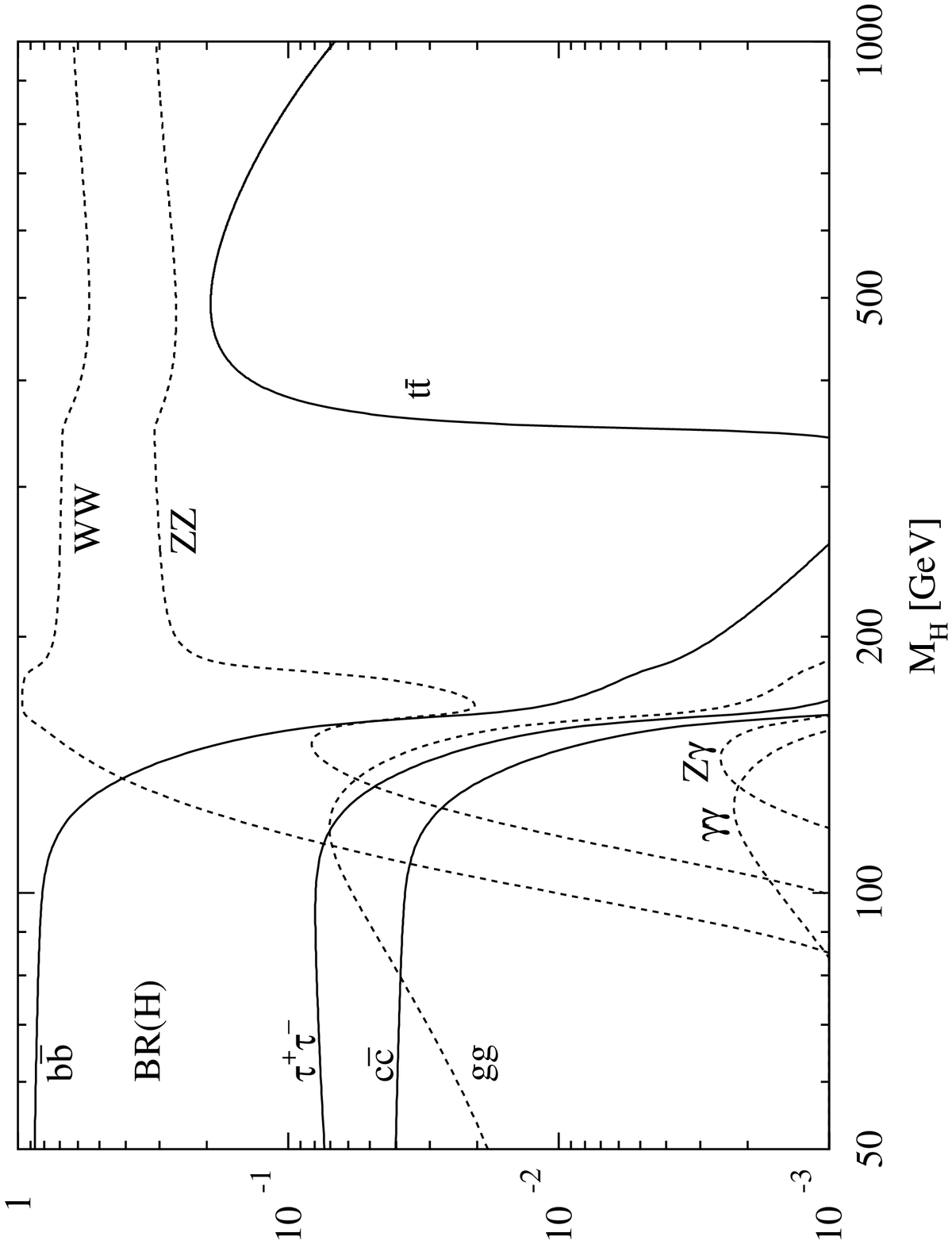}
\end{turn}
\vspace*{-0.0cm}

\caption[]{\label{fg:wtotbr} \it (a) Total decay width (in GeV) of the SM
Higgs boson as a function of its mass. (b) Branching ratios of the
dominant decay modes of the SM Higgs particle. All relevant higher-order
corrections are taken into account.}
\end{figure}

\paragraph{(d) Summary} ~\\[0.5cm]
By adding up all possible decay channels, we obtain 
the total width shown in Fig. \ref{fg:wtotbr}a. Up to masses of 
140 GeV, the Higgs particle is very narrow, $\Gamma(H) \leq 10$ MeV.
After opening up the real and virtual gauge-boson
channels, the state rapidly becomes  wider,
reaching a width of $\sim 1$ GeV at the $ZZ$
threshold. The width cannot be measured directly
in the intermediate mass region at the LHC or $e^+ e^-$
colliders; however, it could be measured directly at muon
colliders \cite{32A}. Above a mass of $\sim 250$ GeV,
the state becomes wide enough to be resolved experimentally in general.

The branching ratios of the main decay modes are
displayed in Fig. \ref{fg:wtotbr}b. A large variety of channels
will be accessible for Higgs masses below 140 GeV. 
The dominant mode is $b\bar b$ decays, yet
$c\bar c, \tau^+\tau^-$ and $gg$ decays 
still occur at a level of several per cent.
At $M_H=$ 120~GeV for instance, the branching ratios are 68\% for
$b\bar b$, 3.1\% for $c\bar c$, 6.9\% for $\tau^+\tau^-$ and 7\% for $gg$.
$\gamma\gamma$
decays occur at a level of 1 per mille. Above this mass value, 
the Higgs boson decay into $W$'s
becomes dominant, overwhelming all other channels if 
the decay mode into two real $W$'s is kinematically possible.
For Higgs masses far above the thresholds, 
$ZZ$ and $WW$ decays occur at a ratio of 1:2, slightly modified only
just above the $t\bar t$ threshold. Since the width grows as the third
power of the mass, the Higgs particle becomes very wide, $\Gamma(H) \sim
\frac{1}{2} M_H^3$ [TeV]. In fact, for $M_H\sim 1$ TeV,
the width reaches $\sim \frac{1}{2}$ TeV.

\subsection{Electroweak Precision Data: Estimate of the Higgs Mass}  
Indirect evidence for a light Higgs 
boson can be derived from the
high-precision measurements of electroweak 
observables at LEP and elsewhere. Indeed, the
fact that the Standard Model is renormalizable only 
after including the top and Higgs particles in 
the loop corrections, indicates that the
electroweak observables are sensitive to the
masses of these particles.

The Fermi coupling can be rewritten in terms of 
the weak coupling and the $W$ mass; at lowest order, 
$G_F/\sqrt{2} = g^2/8M_W^2$.
After substituting the electromagnetic coupling, 
the electroweak mixing angle and the $Z$
mass for the weak coupling, and the $W$
mass, this relation can be rewritten as
\begin{equation}
\frac{G_F}{\sqrt{2}} = \frac{2\pi\alpha}{\sin^2 2\theta_W M_Z^2}
[1+\Delta r_\alpha + \Delta r_t + \Delta r_H ] ~.
\end{equation}
The $\Delta$ terms take account of the radiative corrections:
$\Delta r_\alpha$ describes the shift in the electromagnetic coupling 
if evaluated at the scale $M_Z^2$ instead of zero-momentum;
$\Delta r_t$ denotes the top (and bottom) quark contributions to the
$W$ and $Z$ masses, which are quadratic in the top mass. Finally,
$\Delta r_H$ accounts for the virtual Higgs contributions to the masses;
this term depends only logarithmically \cite{7} on
the Higgs mass at leading order: 
\begin{equation}
\Delta r_H = \frac{G_F M_W^2}{8\sqrt{2}\pi^2} \frac{11}{3} \left[
\log \frac{M_H^2}{M_W^2} - \frac{5}{6} \right] \hspace{1cm} (M_H^2 \gg M_W^2)
~.
\end{equation}
The screening effect reflects the role of the Higgs
field as a regulator that renders the electroweak theory
renormalizable.

Although the sensitivity on the Higgs mass is 
only logarithmic, the increasing precision in the 
measurement of the electroweak observables allows 
us to derive interesting estimates and constraints
on the Higgs mass \cite{8}:
\begin{eqnarray}
M_H & = & 115^{+116}_{-66}~\mbox{GeV} \\
    & < & 420 ~\mbox{GeV~~~(95\% CL)}  ~. \nonumber
\end{eqnarray}
It may be concluded from these numbers that the 
canonical formulation of the Standard Model,  
which includes the existence of a Higgs boson
with a mass below $\sim 700$ GeV,
is compatible with the electroweak data. However,
alternative mechanisms cannot be ruled out.

\subsection{Higgs Production Channels at $e^+e^-$ Colliders}
The first process that was used to search
directly for Higgs bosons over a large mass range,
was the Bjorken process, $Z\to Z^* H, Z^* \to f\bar f$ \cite{34}.
By exploring this production channel,
Higgs bosons with masses less than 65.4 GeV 
were excluded by the LEP1 experiments.
The search now continues by reversing the
role of the real and virtual $Z$ bosons in the 
$e^+e^-$ continuum at LEP2.

The  main production mechanisms for Higgs
bosons in $e^+e^-$ collisions are
\begin{eqnarray}
\mbox{Higgs-strahlung} & : & e^+e^- \to Z^* \to ZH \\ \nonumber \\
\mbox{$WW$ fusion}     & : & e^+e^- \to \bar \nu_e \nu_e (WW) \to \bar \nu_e
\nu_e H
\label{eq:wwfusion}
\end{eqnarray}
In Higgs-strahlung \cite{30,34,35} the Higgs boson is emitted
from the $Z$-boson line, while $WW$ fusion is a formation
process of Higgs bosons in the collision of two quasi-real 
$W$ bosons radiated off the electron and positron beams \cite{36}.

As evident from the subsequent analyses, LEP2 can cover
the SM Higgs mass range up to about 100 GeV
\cite{10}. The high-energy $e^+e^-$
linear colliders can cover the entire Higgs
mass range in the second phase of the machines in which they
will reach a total energy of about 2~TeV \cite{13}.

\paragraph{(a) Higgs-strahlung} ~\\[0.5cm]
The cross section for Higgs-strahlung can be
written in a compact form as 
\begin{equation}
\sigma (e^+e^- \to ZH) = \frac{G_F^2 M_Z^4}{96\pi s} \left[ v_e^2 + a_e^2
\right] \lambda^{1/2} \frac{\lambda + 12 M_Z^2/s}{\left[ 1- M_Z^2/s \right]^2}
~, 
\end{equation}
where $v_e = -1 + 4 \sin^2 \theta_W$ and $a_e=-1$
are the vector and axial-vector $Z$
charges of the electron and $\lambda = [1-(M_H+M_Z)^2/s] [1-(M_H-M_Z)^2/s]$
is the usual two-particle phase-space
function. The cross section is of the size $\sigma \sim \alpha_W^2/s$,
i.e. of second order in the weak coupling, and 
it scales in the squared energy.

\begin{figure}[hbt]

\vspace*{-5.0cm}
\hspace*{-2.0cm}
\epsfxsize=20cm \epsfbox{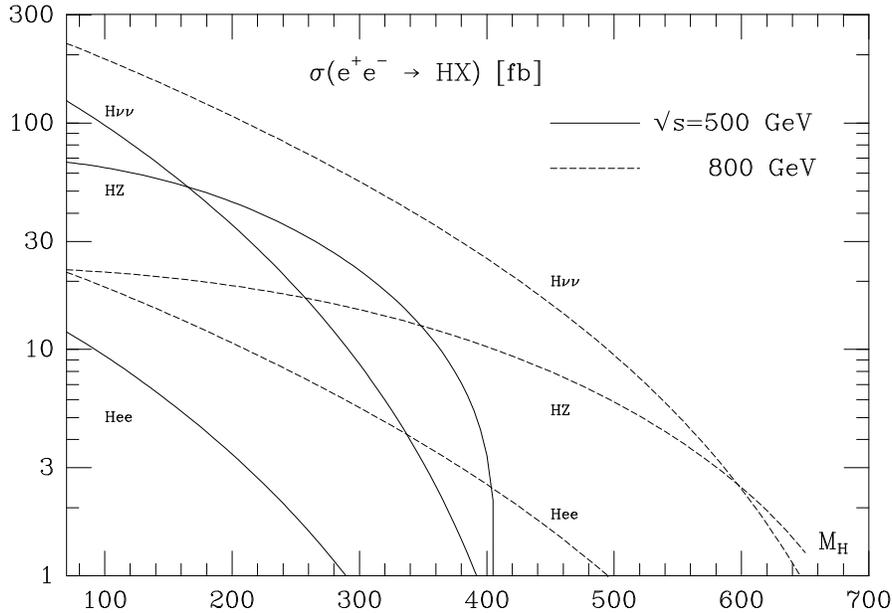}

\caption[]{\label{fg:eehx} \it The cross section for the production of SM
Higgs bosons in Higgs-strahlung $e^+e^-\to ZH$ and $WW/ZZ$ fusion $e^+e^- \to
\bar \nu_e \nu_e/e^+e^- H$; solid curves: $\sqrt{s}=500$ GeV, dashed curves:
$\sqrt{s}=800$ GeV.}
\end{figure}
\begin{figure}[hbt]

\vspace*{-0.5cm}
\hspace*{3.0cm}
\epsfxsize=10cm \epsfbox{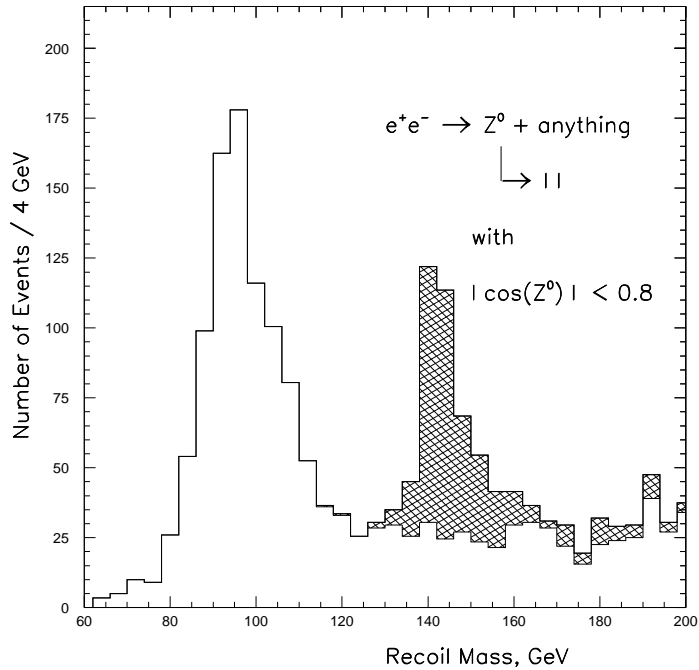}
\vspace*{-0.4cm}

\caption[]{\label{fg:zrecoil} \it Dilepton recoil mass analysis of Higgs-strahlung
$e^+e^- \to ZH\to \ell^+\ell^- +$ anything in the intermediate Higgs mass range
for $M_H=140$ GeV. The c.m.~energy is $\sqrt{s}=$ 360~GeV and the integrated
luminosity $\int {\cal L} = 50 fb^{-1}$. Ref. \cite{37}.}
\end{figure}

Since the cross section vanishes for asymptotic 
energies, the Higgs-strahlung process is most
useful for searching Higgs bosons in the
range where the collider energy is of the 
same order as the Higgs mass, $\sqrt{s} \gsim {\cal O} (M_H)$.
The size of the cross section is illustrated 
in Fig. \ref{fg:eehx} for the energy $\sqrt{s}=500$ GeV of 
$e^+e^-$ linear colliders as a function of the Higgs mass.
Since the recoiling $Z$ mass in the two-body reaction
$e^+e^- \to ZH$
is mono-energetic, the mass of the Higgs boson
can be reconstructed from the energy of the
$Z$ boson, $M_H^2 = s -2\sqrt{s}E_Z + M_Z^2$,
without any need  to analyse the decay products
of the Higgs boson. For leptonic $Z$
decays, missing-mass techniques provide a 
very clear signal, as demonstrated in Fig. \ref{fg:zrecoil}.

\paragraph{(b) $WW$ fusion} ~\\[0.5cm]
Also the cross section for the fusion process (\ref{eq:wwfusion})
can be cast implicitly into a compact form:
\begin{eqnarray}
\sigma (e^+e^-\to\bar \nu_e \nu_e H) & = & \frac{G_F^3 M_W^4}{4\sqrt{2}\pi^3}
\int_{\kappa_H}^1\int_x^1\frac{dx~dy}{[1+(y-x)/\kappa_W ]^2}f(x,y)
\\ \nonumber \\
f(x,y) & = & \left( \frac{2x}{y^3} - \frac{1+3x}{y^2} + \frac{2+x}{y} -1 \right)
\left[ \frac{z}{1+z} - \log (1+z) \right] + \frac{x}{y^3} \frac{z^2(1-y)}{1+z}
~, 
\nonumber
\end{eqnarray}
with $\kappa_H=M_H^2/s$, $\kappa_W=M_W^2/s$ and $z=y(x-\kappa_H)/(\kappa_Wx)$.

Since the fusion process is a 
$t$-channel exchange process, the size is set by the 
$W$ Compton wavelength, suppressed however with 
respect to Higgs-strahlung by the third power
of the electroweak coupling, $\sigma \sim \alpha_W^3/M_W^2$.
As a result, $W$
fusion becomes the leading production process
for Higgs particles at high energies. At
asymptotic energies the cross section simplifies to
\begin{equation}
\sigma (e^+e^- \to \bar \nu_e \nu_e H) \to \frac{G_F^3 M_W^4}{4\sqrt{2}\pi^3}
\left[ \log\frac{s}{M_H^2} - 2 \right] ~.
\end{equation}
In this limit, $W$
fusion to Higgs bosons can be interpreted as a
two-step process: the $W$
bosons are radiated as quasi-real particles 
from electrons and positrons, $e \to \nu W$, 
with the Higgs bosons formed subsequently in the 
colliding $W$ beams.

The size of the 
fusion cross section is compared with Higgs-strahlung
in Fig. \ref{fg:eehx}. At $\sqrt{s}=500$ GeV
the two cross sections are of the same order, yet the
fusion process becomes increasingly important with 
rising energy.

\paragraph{(c) $\gamma\gamma$ fusion} ~\\[0.5cm]
The production of Higgs bosons in $\gamma\gamma$
collisions \cite{38} can be exploited to determine 
important properties of these particles, in particular
the two-photon decay width. The $H\gamma\gamma$
coupling is mediated by loops of charged particles.
If the mass of the loop particle is generated through 
the Higgs mechanism, the decoupling of the
heavy particles is lifted and the $\gamma\gamma$
width reflects the spectrum of these states with
masses possibly far above the Higgs mass.

The two-photon width is related to the production cross section
for polarized $\gamma$ beams  by
\begin{equation}
\sigma (\gamma\gamma \to H) = \frac{16 \pi^2 \Gamma(H\to \gamma\gamma)}
{M_H} \times BW ~, 
\end{equation}
where $BW$ denotes the Breit--Wigner resonance factor in terms
of the $\gamma \gamma$  energy squared. For narrow Higgs bosons the observed 
cross section is found by folding the parton cross
section with the invariant $\gamma\gamma$ energy flux
$\tau d{\cal L}^{\gamma\gamma} / d\tau$ for $J_z^{\gamma\gamma}=0$ at
$\tau=M_H^2/s_{ee}$.

\begin{figure}

\vspace*{-0.0cm}
\hspace*{2.0cm}
\begin{turn}{-1.98}%
\epsfxsize=9.5cm \epsfbox{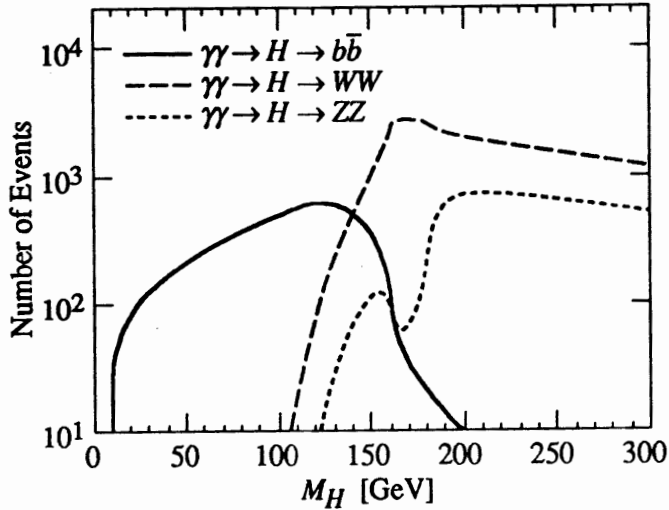}
\end{turn}
\vspace*{-0.4cm}

\caption[]{\label{fg:gagah} \it Production rate of Standard Model Higgs bosons
for three exclusive final states relevant for the intermediate- and heavy-mass
regions in $\gamma\gamma$ collisions. A value of \linebreak 
$4 \times 10^{-2} fb^{-1}/$GeV
is assumed for $d{\cal L}^{\gamma\gamma}/d W_{\gamma\gamma}$. Ref. \cite{38}.}
\end{figure}
The event rate for the production of Higgs bosons in 
$\gamma\gamma$ collisions of Weizs\"acker--Williams photons is 
too small to play a role in practice. However, the
rate is sufficiently large if the photon spectra are 
generated by Compton back-scattering of laser
light, Fig. \ref{fg:gagah}. The $\gamma\gamma$
invariant energy in such a Compton collider \cite{38A} is nearly 
of the same size as the parent $e^+e^-$
energy and the luminosity is
expected to be only slightly smaller than the 
luminosity in $e^+e^-$ collisions. In the Higgs mass range between
100 and 150 GeV, the final state consists primarily of $b\bar b$
pairs. The large $\gamma\gamma$
continuum background is suppressed in the
$J_z^{\gamma\gamma}=0$ polarization state. For Higgs masses above 
150 GeV, $WW$ final states become dominant, supplemented in 
the ratio 1:2 by $ZZ$ final states above the $ZZ$
decay threshold. While the continuum $WW$ background in 
$\gamma\gamma$ collisions is very large, the $ZZ$
background appears under control for masses up 
to order 300 GeV \cite{38B}.

\subsection{Higgs Production at Hadron Colliders}
Several processes can be exploited to produce 
Higgs particles in hadron colliders \cite{32,39}: \\[0.5cm]
\begin{tabular}{llll}
\hspace*{21mm} & gluon fusion & :              & $gg\to H$ \\ \\
& $WW,ZZ$ fusion           & :    & $W^+ W^-, ZZ \to H$ \\ \\
& Higgs-strahlung off $W,Z$ & :   & $q\bar q \to W,Z \to W,Z + H$ \\ \\
& Higgs bremsstrahlung off top & : & $q\bar q, gg \to t\bar t + H$
\end{tabular} \\[0.5cm]
While gluon fusion plays a dominant role 
throughout the entire Higgs mass range of the
Standard Model, the $WW/ZZ$
fusion process becomes increasingly important
with rising Higgs mass. The last two radiation 
processes are relevant only for light Higgs masses.

The production cross sections at hadron colliders, at
the LHC in particular, are quite sizeable so that a 
large sample of SM Higgs particles can be produced
in this machine. Experimental difficulties 
arise from the huge number  of background events
that come along with the Higgs signal events.
This problem will be tackled by either
triggering on leptonic decays of $W,Z$ and $t$
in the radiation processes or by exploiting
the resonance character of the Higgs decays
$H\to \gamma\gamma$ and $H\to ZZ \to 4\ell^\pm$.
In this way, the Tevatron is expected to 
search for Higgs particles in the mass range
above that of LEP2 up to about 110 to 120 GeV \cite{11}. 
The LHC is expected to cover the entire canonical
Higgs mass range $M_H \lsim 700$ GeV
of the Standard Model \cite{12}.

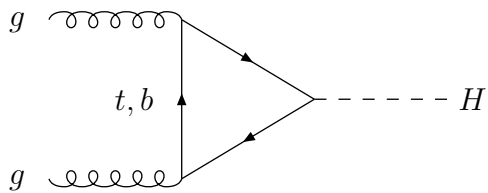
\begin{figure}[hbt]
\begin{center}
\setlength{\unitlength}{1pt}
\begin{picture}(180,90)(0,0)

\Gluon(0,20)(50,20){-3}{5}
\Gluon(0,80)(50,80){3}{5}
\ArrowLine(50,20)(50,80)
\ArrowLine(50,80)(100,50)
\ArrowLine(100,50)(50,20)
\DashLine(100,50)(150,50){5}
\put(155,46){$H$}
\put(25,46){$t,b$}
\put(-15,18){$g$}
\put(-15,78){$g$}

\end{picture}  \\
\setlength{\unitlength}{1pt}
\caption[ ]{\label{fg:gghlodia} \it Diagram contributing to the
formation of Higgs bosons in gluon-gluon collisions
at lowest order.}
\end{center}
\end{figure}

\paragraph{(a) Gluon fusion} ~\\[0.5cm]
The gluon-fusion mechanism \cite{29,32,39A}
\begin{displaymath}
pp \to gg \to H
\end{displaymath}
provides the dominant production mechanism of Higgs 
bosons at the LHC in the entire relevant Higgs
mass range up to about 1 TeV. The gluon coupling
to the Higgs boson in the SM is mediated by
triangular loops of top and bottom quarks, 
cf. Fig. \ref{fg:gghlodia}.
Since the Yukawa coupling of the Higgs particle
to heavy quarks grows with the quark mass, thus
balancing the decrease of the amplitude, the 
form factor approaches a non-zero value for large 
loop-quark masses. [If the masses of heavy quarks
beyond the third generation were generated solely
by the Higgs mechanism, these particles
would add the same amount to the form factor as
the top quark in the asymptotic heavy-quark limit.]
\\

The partonic cross section, Fig. \ref{fg:gghlodia}, can be expressed 
by the gluonic width of the Higgs boson at 
lowest order \cite{32}:
\begin{eqnarray}
\hat \sigma_{LO} (gg\to H) & = & \sigma_0 M_H^2 \delta(\hat s - M_H^2) \\
\sigma_0 & = & \frac{\pi^2}{8M_H^2} \Gamma_{LO} (H\to gg) =
\frac{G_F\alpha_s^2}{288\sqrt{2}\pi} \left| \sum_Q A_Q^H (\tau_Q) \right|^2 ~,
\nonumber
\end{eqnarray}
where the scaling variable is defined as $\tau_Q = 4M_Q^2/M_H^2$ and $\hat s$
denotes the partonic c.m. energy squared. The
form factor can easily be evaluated:
\begin{eqnarray}
A_Q^H (\tau_Q) & = & \frac{3}{2} \tau_Q \left[ 1+(1-\tau_Q) f(\tau_Q)
\right] \label{eq:ftau} \\
f(\tau_Q) & = & \left\{ \begin{array}{ll}
\displaystyle \arcsin^2 \frac{1}{\sqrt{\tau_Q}} & \tau_Q \ge 1 \\
\displaystyle - \frac{1}{4} \left[ \log \frac{1+\sqrt{1-\tau_Q}}
{1-\sqrt{1-\tau_Q}} - i\pi \right]^2 & \tau_Q < 1
\end{array} \right. \nonumber
\end{eqnarray}
For small loop masses the form factor vanishes, $A_Q^H(\tau_Q) \sim -3/8 \tau_Q
[\log (\tau_Q/4)+i\pi]^2$,
while for large loop masses it approaches a non-zero value,
$A_Q^H (\tau_Q) \to 1$.

In the narrow-width approximation, the hadronic
cross section can be cast into the form
\begin{equation}
\sigma_{LO} (pp\to H) = \sigma_0 \tau_H \frac{d{\cal L}^{gg}}{d\tau_H} ~,
\end{equation}
with $d{\cal L}^{gg}/d\tau_H$ denoting the $gg$ luminosity of the 
$pp$ collider, evaluated for the Drell--Yan variable
$\tau_H = M_H^2/s$, where $s$ is the total hadronic energy squared. \\

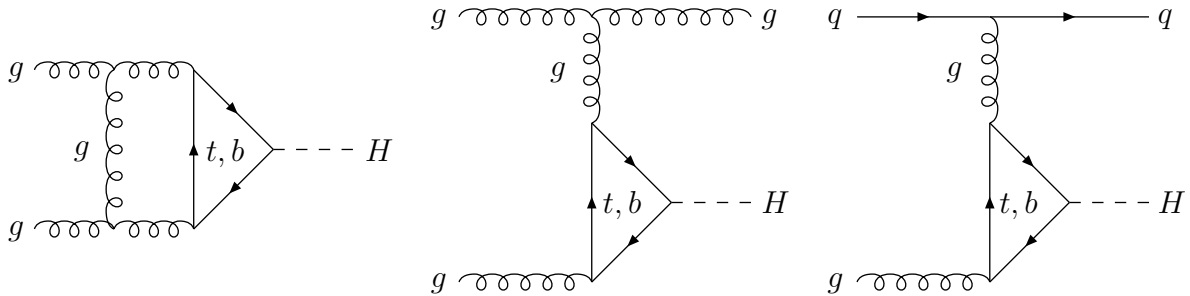
\begin{figure}[hbt]
\begin{center}
\setlength{\unitlength}{1pt}
\begin{picture}(450,100)(-10,0)

\Gluon(0,20)(30,20){3}{3}
\Gluon(0,80)(30,80){3}{3}
\Gluon(30,20)(60,20){3}{3}
\Gluon(30,80)(60,80){3}{3}
\Gluon(30,20)(30,80){3}{5}
\ArrowLine(60,20)(60,80)
\ArrowLine(60,80)(90,50)
\ArrowLine(90,50)(60,20)
\DashLine(90,50)(120,50){5}
\put(125,46){$H$}
\put(65,46){$t,b$}
\put(-10,18){$g$}
\put(-10,78){$g$}
\put(15,48){$g$}

\Gluon(160,100)(210,100){3}{5}
\Gluon(210,100)(270,100){3}{6}
\Gluon(160,0)(210,0){3}{5}
\Gluon(210,100)(210,60){3}{4}
\ArrowLine(210,0)(210,60)
\ArrowLine(210,60)(240,30)
\ArrowLine(240,30)(210,0)
\DashLine(240,30)(270,30){5}
\put(275,26){$H$}
\put(215,26){$t,b$}
\put(150,-2){$g$}
\put(150,98){$g$}
\put(195,78){$g$}
\put(275,98){$g$}

\ArrowLine(310,100)(360,100)
\ArrowLine(360,100)(420,100)
\Gluon(310,0)(360,0){3}{5}
\Gluon(360,100)(360,60){3}{4}
\ArrowLine(360,0)(360,60)
\ArrowLine(360,60)(390,30)
\ArrowLine(390,30)(360,0)
\DashLine(390,30)(420,30){5}
\put(425,26){$H$}
\put(365,26){$t,b$}
\put(300,-2){$g$}
\put(300,98){$q$}
\put(425,98){$q$}
\put(345,78){$g$}

\end{picture}  \\
\setlength{\unitlength}{1pt}
\caption[ ]{\label{fg:gghqcddia} \it Typical diagrams contributing to the
virtual/real QCD corrections to $gg\to H$.}
\end{center}
\end{figure}

The QCD corrections to the gluon fusion 
process \cite{32} are very important. They
stabilize the theoretical predictions for the 
cross section when the renormalization and
factorization scales are varied. Moreover,
they are large and positive, thus increasing the
production cross section for Higgs bosons. 
The QCD corrections consist of virtual 
corrections to the basic process $gg\to H$, 
and of real corrections due to the associated
production of the Higgs boson with massless
partons, $gg\to Hg$ and $gq\to Hq,\, q\bar q\to Hg$.
These subprocesses contribute to Higgs production
at ${\cal O}(\alpha_s^3)$.
The virtual corrections rescale the lowest-order
fusion cross section with a coefficient that depends
only on the ratios of the Higgs and quark masses. 
Gluon radiation leads to two-parton final states
with invariant energy $\hat s \geq M_H^2$ in the 
$gg, gq$ and $q\bar q$ channels.\\

The final result  for the hadronic cross section 
can be split into five parts:
\begin{equation}
\sigma(pp \rightarrow H+X) = \sigma_{0} \left[ 1+ C
\frac{\alpha_{s}}{\pi} \right] \tau_{H} \frac{d{\cal L}^{gg}}{d\tau_{H}} +
\Delta \sigma_{gg} + \Delta \sigma_{gq} + \Delta \sigma_{q\bar{q}} ~.
\label{eq:gghqcd}
\end{equation}
The calculation of the 
corrections has been performed in the $\overline{MS}$
scheme. The mass $M_Q$
is identified with the pole quark mass and the
renormalization scale in $\alpha_s$
and the factorization scale of the parton
densities can be fixed 
at the Higgs mass. [The general scale dependence
is also known.]

The coefficient $C(\tau_Q)$
denotes the finite part of the virtual two-loop
corrections. It splits into the infrared part $\pi^2$
and the finite piece, which depends on the quark mass:
\begin{equation}
C(\tau_Q) = \pi^{2}+ c(\tau_Q) ~.
\end{equation}
The finite parts of the hard contributions from gluon
radiation in $gg$ scattering, $gq$ scattering and
$q\bar q$ annihilation, may be written as
\begin{eqnarray}
\Delta \sigma_{gg} & = & \int_{\tau_{H}}^{1} d\tau \frac{d{\cal
L}^{gg}}{d\tau} \times \frac{\alpha_{s}}{\pi} \sigma_{0}  \Big\{  - z
P_{gg} (z) \log z + d_{gg} (z,\tau_Q) \non \\
& & \left. \hspace{3.7cm} + 12 \left[ \left(\frac{\log
(1-z)}{1-z} \right)_+ - z[2-z(1-z)] \log (1-z) \right]  \right\} \non \\ \non \\
\Delta \sigma_{gq} & = & \int_{\tau_{H}}^{1} d\tau \sum_{q,
\bar{q}} \frac{d{\cal L} ^{gq}}{d\tau} \times \frac{\alpha_{s}}{\pi}
\sigma_{0} \left\{ -\frac{z}{2} P_{gq}(z) \log\frac{z}{(1-z)^2}
+ d_{gq} (z,\tau_Q) \right\} \non \\ \non \\
\Delta \sigma_{q\bar{q}} & = & \int_{\tau_{H}}^{1} d\tau
\sum_{q} \frac{d{\cal L}^{q\bar{q}}}{d\tau} \times \frac{\alpha_{s}}{\pi}
\sigma_{0}~d_{q\bar q} (z,\tau_Q)
\end{eqnarray}
with $z=\tau_H/\tau=M_H^2/\hat s$; $P_{gg}$ and $P_{gq}$
are the standard Altarelli--Parisi splitting functions.
The coefficient functions $c(\tau_Q)$ and $d(z,\tau_Q)$
can be reduced analytically to one-dimensional 
integrals, which  must, in general, 
 be evaluated numerically. However, they can be
calculated analytically in the heavy-quark limit \cite{32,39B}:
\begin{eqnarray}
c(\tau_Q) & \displaystyle \to\frac{11}{2} \hspace{4.1cm} d_{gg}(z,\tau_Q) & \to
-\frac{11}{2} (1-z)^3 \non \\
d_{gq}(z,\tau_Q) & \displaystyle \to\frac{2}{3}z^2 - (1 - z)^2 \hspace{2cm}
d_{q\bar q}(z,\tau_Q) & \to \frac{32}{27} (1-z)^3 
\label{eq:gghqcdlim}
\end{eqnarray}
Thus, for light Higgs bosons the production cross 
section is available in complete analytic form, 
including the complicated QCD radiative corrections.

\begin{figure}[hbt]

\vspace*{0.4cm}
\hspace*{2.0cm}
\begin{turn}{-90}%
\epsfxsize=7cm \epsfbox{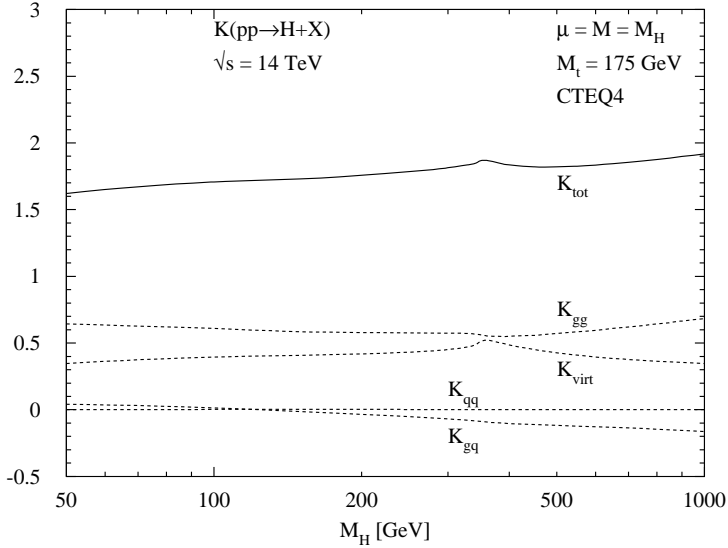}
\end{turn}
\vspace*{-0.2cm}

\caption[]{\label{fg:gghk} \it K factors of the QCD-corrected gluon-fusion
cross section $\sigma(pp \to H+X)$ at the LHC with c.m.~energy $\sqrt{s}=14$
TeV. The dashed lines show the individual contributions of the four terms of
the QCD corrections defined in eq. (\ref{eq:gghqcd}). The renormalization and
factorization scales have been identified with the Higgs mass,  
and  CTEQ4 parton densities have been adopted.}
\end{figure}
The size of the radiative corrections can be parametrized
by defining the $K$ factor as $K=\sigma_{NLO}/\sigma_{LO}$, in which
all quantities are evaluated in the  numerator and 
denominator in next-to-leading and leading order,
respectively. The results of this calculation are
shown in Fig. \ref{fg:gghk}. The virtual corrections 
$K_{virt}$ and the real corrections $K_{gg}$ for the $gg$ collisions 
 are apparently of the same size, and both are large
and positive; the corrections for $q\bar q$
collisions and the $gq$
inelastic Compton contributions are less important.
After including these higher-order QCD corrections,
the dependence of the cross section on the renormalization
and factorization scales is significantly reduced
from a level of ${\cal O}(100\%)$ down to a level of about 20\%.

\begin{figure}[hbt]

\vspace*{0.5cm}
\hspace*{2.0cm}
\begin{turn}{-90}%
\epsfxsize=7cm \epsfbox{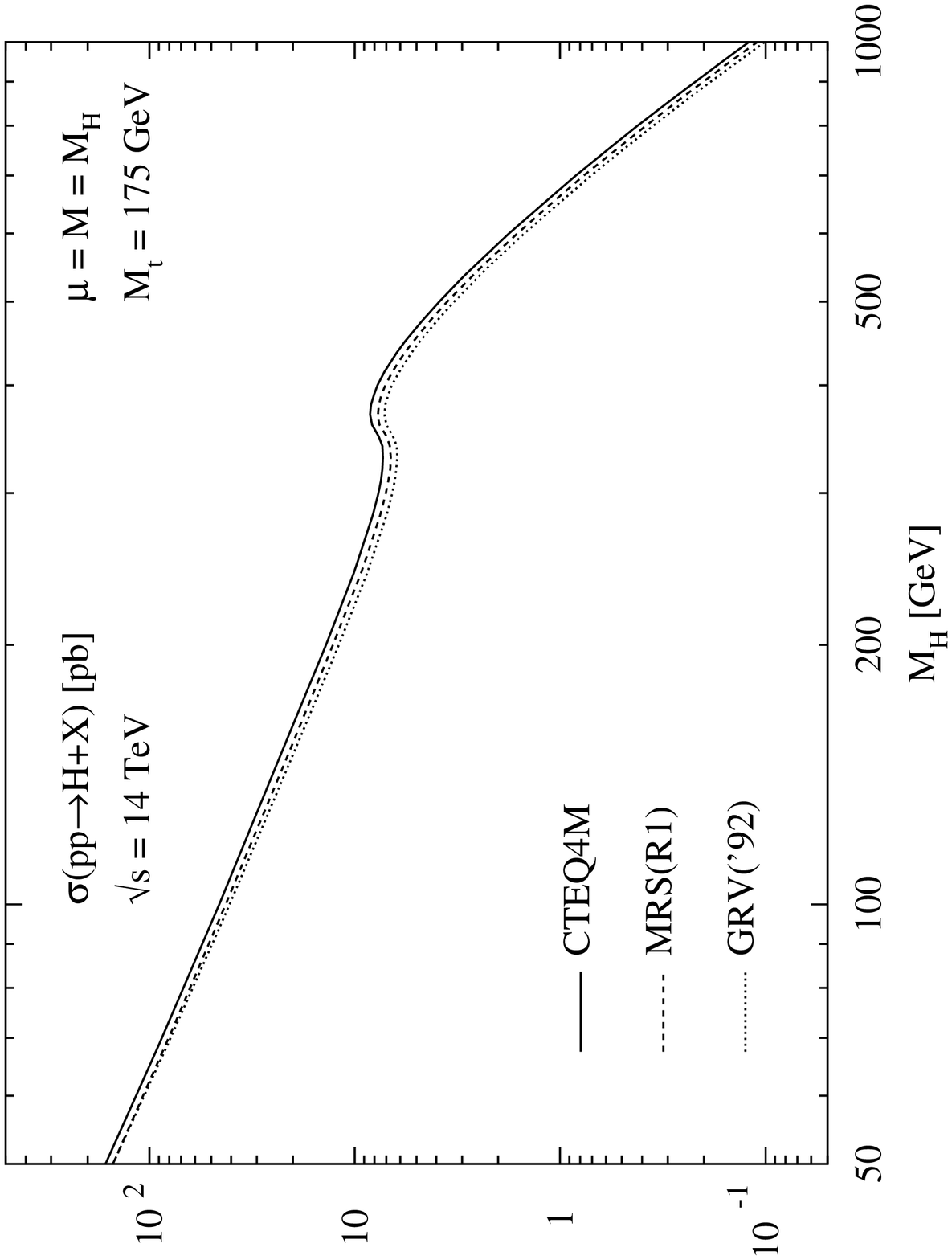}
\end{turn}
\vspace*{0.0cm}

\caption[]{\label{fg:gghparton} \it The cross 
section for the production of Higgs bosons;  three
different sets of parton densities are shown
[CTEQ4M, MRS(R1) and GRV('92)].}
\end{figure}
The theoretical prediction for the production cross 
section  of Higgs particles 
is presented in Fig. \ref{fg:gghparton} for the LHC as a 
function of the Higgs mass.
The cross section decreases with increasing Higgs mass.
This is, to a large extent, a consequence of the
sharply falling $gg$
luminosity for large invariant masses. The bump in 
the cross section is induced by the $t\bar t$
threshold
in the top triangle. The overall theoretical 
accuracy of this calculation is expected to be at
a level of 20 to 30\%.  

\paragraph{(b) Vector-boson fusion} ~\\[0.5cm]
The second important channel for Higgs production at the
LHC is vector-boson fusion, $W^+W^- \to H$
\cite{23,39}. For large Higgs masses this mechanism becomes
competitive to gluon fusion; for intermediate
masses the cross section is smaller by about an
order of magnitude.\\

For large Higgs masses, the two electroweak bosons $W,Z$
that form the Higgs boson are predominantly 
longitudinally polarized. At high energies, the 
equivalent particle spectra of the longitudinal $W,Z$
bosons in quark beams are given by
\begin{eqnarray}
f^W_L (x) & = & \frac{G_F M_W^2}{2\sqrt{2}\pi^2} \frac{1-x}{x} 
 \label{eq:xyz} \\ \non \\
f^Z_L (x) & = & \frac{G_F M_Z^2}{2\sqrt{2}\pi^2}
\left[(I_3^q - 2e_q \sin^2\theta_W)^2 + (I_3^q)^2\right] \frac{1-x}{x} ~, \non
\end{eqnarray}
where $x$ is the fraction of energy transferred from the quark
to the $W,Z$ boson in the splitting process
$q\to q +W/Z$. From these particle spectra, the $WW$ and $ZZ$
luminosities can easily be derived:
\begin{eqnarray}
\frac{d{\cal L}^{WW}}{d\tau_W} & = & \frac{G_F^2 M_W^4}{8\pi^4}
\left[ 2 - \frac{2}{\tau_W} -\frac{1+\tau_W}{\tau_W} \log \tau_W \right] \\
\non \\
\frac{d{\cal L}^{ZZ}}{d\tau_Z} & = & \frac{G_F^2 M_Z^4}{8\pi^4}
\left[(I_3^q - 2e_q \sin^2\theta_W)^2 + (I_3^q)^2\right]
\left[(I_3^{q'} - 2e_{q'} \sin^2\theta_W)^2 + (I_3^{q'})^2\right] \non \\
& & \hspace{1.5cm} \cdot \left[ 2 - \frac{2}{\tau_Z} -\frac{1+\tau_Z}{\tau_Z}
\log \tau_Z \right] \non
\end{eqnarray}
with the Drell--Yan variable defined as  $\tau_V = M_{VV}^2/s$.
Denoting the parton cross section for $WW,ZZ\to H$ by 
$\hat \sigma_0$ with 
\begin{eqnarray}
\hat \sigma_0(VV\to H) & = & \sigma_0 \delta\left(1-M_H^2/\hat s \right)
\\
\sigma_0 & = & \sqrt{2}~\pi G_F  ~, \non
\end{eqnarray}
the cross section for Higgs production in quark--quark
collisions is given by
\begin{equation}
\hat \sigma(qq\to qqH) = \frac{d{\cal L}^{VV}}{d\tau_V} \sigma_0 ~.
\label{eq:vvhpart}
\end{equation}
The hadronic cross section is finally obtained by
summing the parton cross section (\ref{eq:vvhpart}) 
over the flux of all possible pairs
of quark--quark and antiquark combinations
\begin{equation}
\sigma (qq'\to VV \to H) = \int_{M_H^2/s}^1 d\tau \sum_{qq'}
\frac{d{\cal L}^{qq'}}{d\tau} \hat \sigma (qq'\to qq'H; \hat s = \tau s) ~.
\end{equation}

Since to lowest order the proton remnants are
colour singlets in the $WW,ZZ$
fusion processes, no colour will be exchanged between the
two quark lines from which the two vector bosons are
radiated. As a result, the leading QCD corrections to
these processes are already accounted for
by the corrections to the quark parton densities.\\

\begin{figure}[hbt]

\vspace*{0.5cm}
\hspace*{0.0cm}
\begin{turn}{-90}%
\epsfxsize=10cm \epsfbox{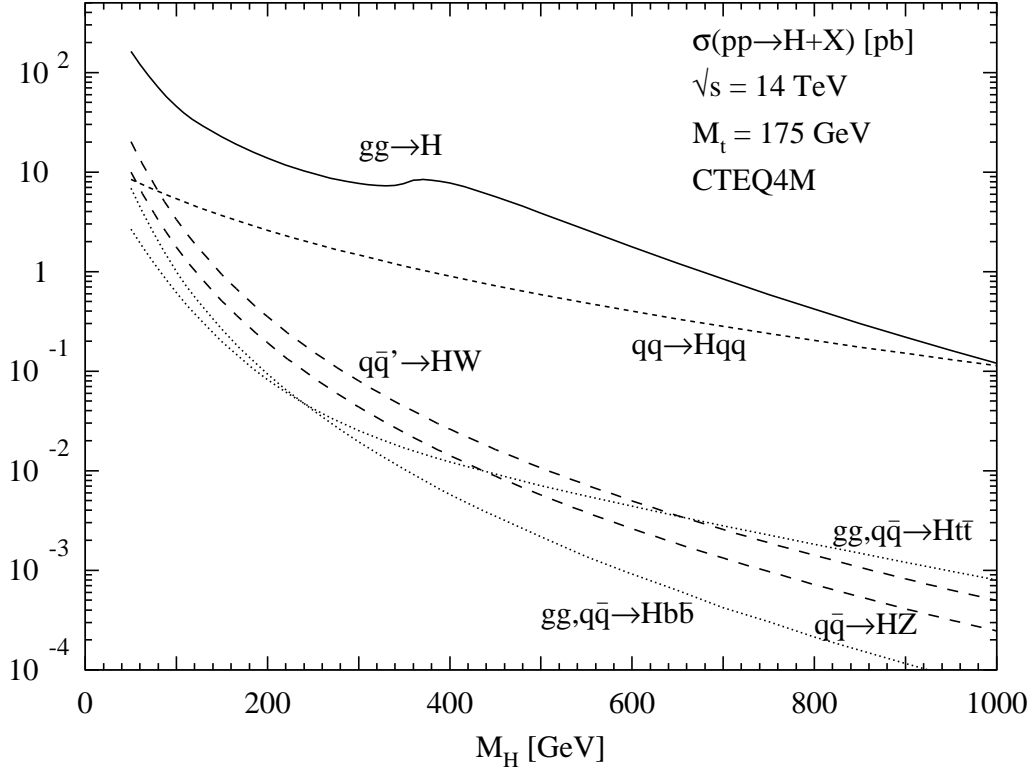}
\end{turn}
\vspace*{0.0cm}

\caption[]{\label{fg:lhcpro} \it Higgs production cross sections at the LHC
 for the various production mechanisms as a function of the
Higgs mass. The full QCD-corrected results for the gluon fusion $gg
\to H$, vector-boson fusion $qq\to VVqq \to Hqq$, vector-boson bremsstrahlung
$q\bar q \to V^* \to HV$ and associated production $gg,q\bar q \to Ht\bar t,
Hb\bar b$ are shown.
[The QCD corrections to the last processes are unknown.]}
\end{figure}

The $WW/ZZ$ fusion cross section for Higgs bosons at the LHC
is shown in Fig. \ref{fg:lhcpro}. The process is apparently
most important in the upper range of Higgs
masses, where the cross section approaches values
close to gluon fusion.

\paragraph{(c) Higgs-strahlung off vector bosons} ~\\[0.5cm]
Higgs-strahlung $q\bar q \to V^* \to VH~(V=W,Z)$
is a very important mechanism (Fig. \ref{fg:lhcpro}) for the 
search of light Higgs bosons at the hadron colliders
Tevatron and LHC. Though the cross section is 
smaller than for gluon fusion, leptonic decays
of the electroweak vector bosons are
extremely useful to filter Higgs signal events
out of the huge background. Since the dynamical
mechanism is the same as for $e^+e^-$
colliders, except for the  folding with
the quark--antiquark densities, intermediate steps of the
 calculation need not be noted, and merely  
the final values 
of the cross sections for the Tevatron and the 
LHC are recorded in Fig. \ref{fg:lhcpro}.

\paragraph{(d) Higgs bremsstrahlung off top quarks} ~\\[0.5cm]
Also the process $gg,q\bar q \to t\bar t H$
is relevant only for small Higgs masses, Fig. \ref{fg:lhcpro}.
The analytical expression for the parton cross
section, even at lowest order, is quite involved, 
so that just the final results for the LHC
cross section are shown in Fig. \ref{fg:lhcpro}.

 Higgs
bremsstrahlung off top quarks is also an interesting
process for measurements of the fundamental $Htt$
Yukawa coupling. The cross section $\sigma (pp\to t\bar t H)$
is directly proportional to the square of
this fundamental coupling.

\paragraph{\underline{Summary.}} An overview of the production cross
sections for Higgs particles at the LHC
is presented in Fig. \ref{fg:lhcpro}. Three classes
of channels can be distinguished. The gluon fusion of Higgs particles
is a universal process, dominant over the 
entire SM Higgs mass range. Higgs-strahlung
off electroweak $W,Z$
bosons or top quarks is prominent for light
Higgs bosons. The $WW/ZZ$
fusion channel, by contrast, becomes increasingly
important in the upper part of the SM Higgs
mass range.

The signatures for the search for Higgs particles are
dictated by the decay branching ratios. In the
lower part of the intermediate mass range, resonance
reconstruction in $\gamma\gamma$ final states and $b\bar b$
jets can be exploited. In the upper part of the
intermediate mass range, decays to $ZZ^*$ and $WW^*$
are important, with the two electroweak bosons 
decaying leptonically. In the mass range above
the on-shell $ZZ$ decay threshold, the charged-lepton decays
$H\to ZZ \to 4\ell^\pm$ provide  gold-plated signatures. Only at the
upper end of the classical SM Higgs mass range,
 decays to neutrinos and jets,  
generated in $W$ and $Z$ decays, complete the search techniques.

\subsection{The Profile of the SM Higgs Particle}
To establish the Higgs mechanism
experimentally, the nature of this particle
must be explored by measuring all its
characteristics, the mass and lifetime,
the external quantum numbers spin-parity,
the couplings to gauge bosons and fermions,
and last but not least, the Higgs self-couplings.
While part of this program
can be realized at the LHC, the complete
profile of the particle can be reconstructed
across the entire mass range in $e^+ e^-$ colliders.

\paragraph{(a) Mass} ~\\[0.5cm]
The mass of the Higgs particle can be
measured by collecting the decay products
of the particle at hadron and $e^+e^-$ colliders. Moreover, in
$e^+e^-$ collisions Higgs-strahlung can be exploited
to reconstruct the mass very precisely from
the $Z$ recoil energy in the two-body
process $e^+e^-\to ZH$.
 An overall
accuracy of about $\delta M_H \sim 100$ MeV can be expected.

\paragraph{(b) Width/lifetime} ~\\[0.5cm]
The width of the state, i.e. the lifetime of
the particle, can be directly measured
above the $ZZ$ decay threshold where the
width grows rapidly. In the lower part of
the intermediate mass range the width can be
measured indirectly, by combining the branching
ratio for $H\to \gamma\gamma$, 
accessible at the LHC, with the measurement
of the partial $\gamma\gamma$ width, accessible through
$\gamma\gamma$ production at a Compton collider:
$\Gamma_{tot} = \Gamma_{\gamma \gamma} / BR_{\gamma \gamma}$.
In the upper
part of the intermediate mass range, the
combination of the branching ratios for $H\to WW,ZZ$
decays with the production cross sections for
$WW$ fusion and Higgs-strahlung, which can be
expressed both through the partial Higgs decay
widths to $WW$ and $ZZ$ pairs, will allow us to
extract the width of the Higgs particle. Thus,
the total width of the Higgs particle can be
determined throughout the entire  mass
range when the experimental results from the LHC,
$e^+e^-$ and  $\gamma\gamma$
colliders can be combined. The direct measurement
of the width in the intermediate mass
range will be possible at muon colliders
in which Higgs bosons can be generated
as  $s$-channel resonances: $\mu^+\mu^- \to H \to f\bar f,VV$.
The energy resolution of the muon beams
is expected to be so high that the
Breit--Wigner excitation curve can be
reconstructed \cite{32A}.

\begin{figure}[hbt]

\vspace*{0.5cm}
\hspace*{0.0cm}
\epsfxsize=8cm \epsfbox{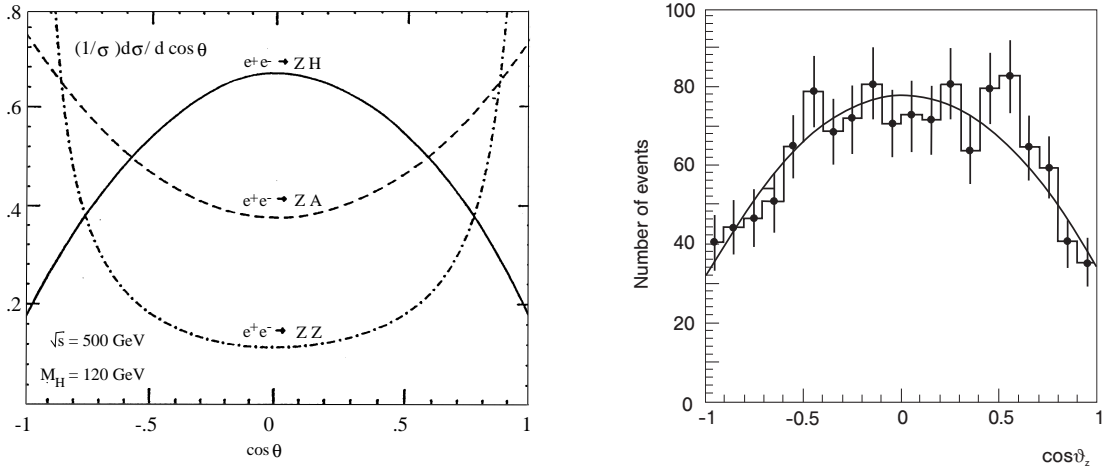}
\vspace*{0.0cm}

\vspace*{-6.35cm}
\hspace*{9.0cm}
\epsfxsize=6.25cm \epsfbox{higgs_sm_spin_exp.eps}
\vspace*{0.0cm}

\caption[]{\it \label{fg:spinpar} Left: Angular distribution of $Z/H$ bosons in
Higgs-strahlung, compared with the production of pseudoscalar
particles and the $ZZ$ background final states; Ref. \cite{41}.
Right: The same for the signal plus background in the experimental
simulation of Ref. \cite{42}.}
\end{figure}

\paragraph{(c) Spin-parity} ~\\[0.5cm]
The angular distribution of the $Z/H$ bosons
in the Higgs-strahlung process is 
sensitive to the spin and parity of the
Higgs particle \cite{13}. Since the production
amplitude is given by ${\cal A}(0^+) \sim \vec{\epsilon}_{Z^*} \cdot
\vec{\epsilon}_Z$, the $Z$ boson is produced in a state of
longitudinal polarization at high
energies -- in accordance  with the equivalence
theorem. As a result, the angular distribution
\begin{equation}
\frac{d\sigma}{d\cos\theta} \sim \sin^2 \theta + \frac{8M_Z^2}{\lambda s}
\end{equation}
approaches the spin-zero $\sin^2\theta$
law asymptotically. This may be contrasted
with the distribution $\sim 1 + \cos^2\theta$
for negative parity states, which follows
from the transverse polarization amplitude
${\cal A}(0^-) \sim \vec{\epsilon}_{Z^*} \times \vec{\epsilon}_Z \cdot
\vec{k}_Z$. It is also characteristically different
from the distribution of the background
process $e^+e^- \to ZZ$, which, as a result of $t/u$-channel $e$ exchange,
is strongly peaked in the forward/backward
direction, Fig. \ref{fg:spinpar}.\\

In a similar way, the zero-spin of the
Higgs particle can be determined from the
isotropic distribution of the decay
products. Moreover, the parity can be
measured by observing the spin correlations
of the decay products. According to the
equivalence theorem, the azimuthal angles
of the decay planes in $H\to ZZ\to (\mu^+\mu^-) (\mu^+\mu^-)$
are asymptotically uncorrelated, $d\Gamma^+/d\phi_* \to 0$,
for a $0^+$ particle; this is to be contrasted with 
$d\Gamma^-/d\phi_* \to 1-\frac{1}{4} \cos 2\phi_*$
for the distribution of the azimuthal angle
between the planes for the decay of a $0^-$
particle. The difference between the angular distributions
is a consequence of the different polarization
states of the vector bosons in the two
cases. While they approach states of 
longitudinal polarization for scalar Higgs
decays, they are transversely polarized
for pseudoscalar particle decays. 

\paragraph{(d) Higgs couplings} ~\\[0.5cm]
Since the fundamental particles acquires a
mass through the interaction with the
Higgs field, the strength of the Higgs
couplings to fermions and gauge bosons
is set by the masses of these particles.
It will therefore be a very important experimental  
task to measure these couplings, which
are uniquely predicted by the very
nature of the Higgs mechanism.\\

\begin{figure}[hbt]

\vspace*{0.0cm}
\hspace*{2.0cm}
\epsfxsize=11cm \epsfbox{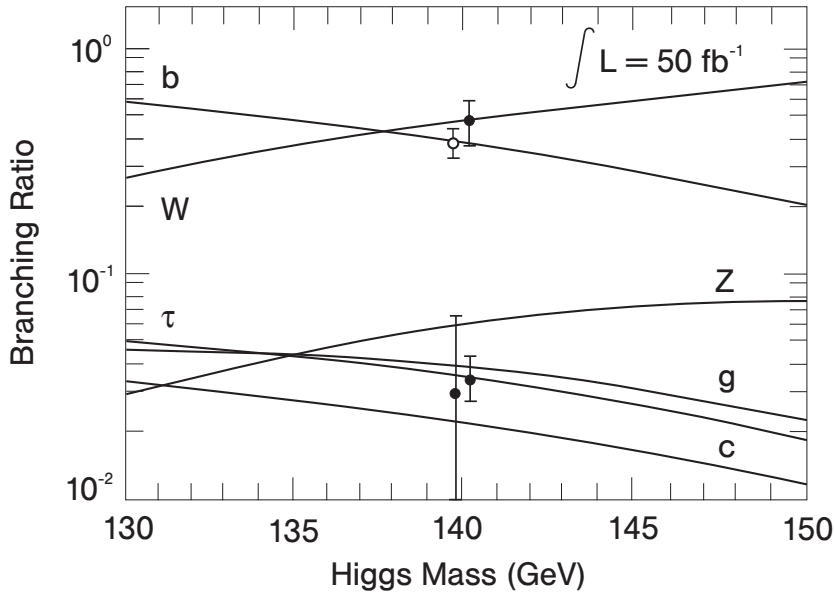}
\vspace*{-0.2cm}

\caption[]{\it \label{fg:brmeas} The measurement of decay branching ratios
of the SM Higgs boson for $M_H=140$ GeV.
In the bottom part of the figure the small
error bar belongs to the $\tau$
branching ratio, the large bar to the average
of the charm and gluon branching ratios, 
which were not separated in the simulation
of Ref. \cite{43}. In the upper part of the
figure the open circle denotes the $b$ branching
ratio, the full circle the $W$ branching ratio.}
\end{figure}

The Higgs couplings to massive gauge
bosons can be determined from the
production cross sections in
Higgs-strahlung and $WW,ZZ$ fusion, with the
accuracy expected at the per cent level.
For heavy enough Higgs bosons the decay
width can be exploited to determine the
coupling to electroweak gauge bosons.
For Higgs couplings to fermions the
branching ratios $H\to b\bar b, c\bar c, \tau^+\tau^-$
can be used in the lower part of the
intermediate mass range; these observables
allow the direct measurement of the Higgs
Yukawa couplings. This is exemplified
in Fig. \ref{fg:brmeas} for a Higgs mass of 140 GeV.\\

A particularly interesting coupling
is the Higgs coupling to top quarks.
Since the top quark is by far the
heaviest fermion in the Standard Model,
irregularities in the standard picture
of electroweak symmetry breaking through
a fundamental Higgs field may become
apparent first in this coupling. Thus 
the $Ht t$ Yukawa coupling may eventually provide
essential clues to the nature of the
mechanism breaking the electroweak
symmetries.

\begin{figure}[hbt]

\vspace*{0.0cm}
\hspace*{2.0cm}
\epsfxsize=12cm \epsfbox{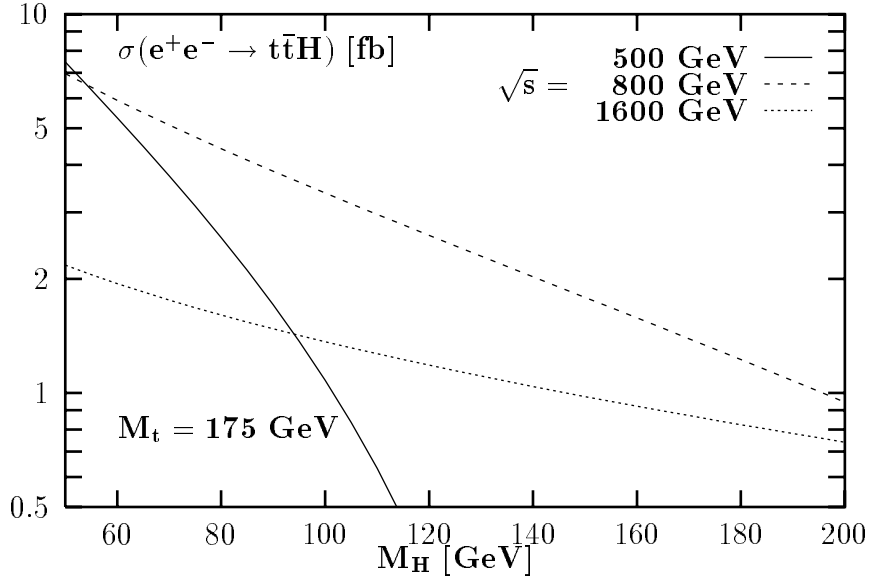}
\vspace*{-0.5cm}

\caption[]{\it \label{fg:eetth} The cross section for bremsstrahlung of SM
Higgs bosons off top quarks in the Yukawa
process $e^+e^-\to t\bar t H$.
[The amplitude for radiation off the
intermediate $Z$-boson line is small.] Ref. \cite{44}.}
\end{figure}

Top loops mediating the production
processes $gg\to H$ and $\gamma\gamma\to H$
(and the corresponding decay channels)
give rise to cross sections and partial
widths, which are proportional to the square of
the Higgs--top Yukawa coupling. This
Yukawa coupling can be measured directly,
for the lower part of the intermediate
mass range, in the bremsstrahlung
processes $pp\to t\bar t H$ and $e^+e^- \to t\bar t H$ \cite{44}.
The Higgs boson is radiated, in the first
process exclusively, in the second process
predominantly, from the heavy top quarks.
Even though these experiments are 
difficult because of  the small cross sections
[cf. Fig. \ref{fg:eetth} for $e^+e^-$ collisions] 
and of the complex topology of
the $b\bar bb\bar bW^+W^-$ final state, this process 
is an important tool for exploring the
mechanism of electroweak symmetry breaking.
For large Higgs masses above the $t\bar t$
threshold, the decay channel $H\to t\bar t$
can be studied; in $e^+e^-$ collisions the cross section of
$e^+e^- \to t\bar t Z$ increases through the reaction
$e^+e^- \to ZH (\to t\bar t)$ \cite{45}. Higgs exchange between
$t\bar t$ quarks also affects the excitation curve
near the threshold at a level of a few per cent.


\paragraph{(e) Higgs self-couplings} ~\\[0.5cm]
\begin{figure}[hbtp]
\vspace*{-2.3cm}

\hspace*{-2.0cm}
\epsfxsize=20cm \epsfbox{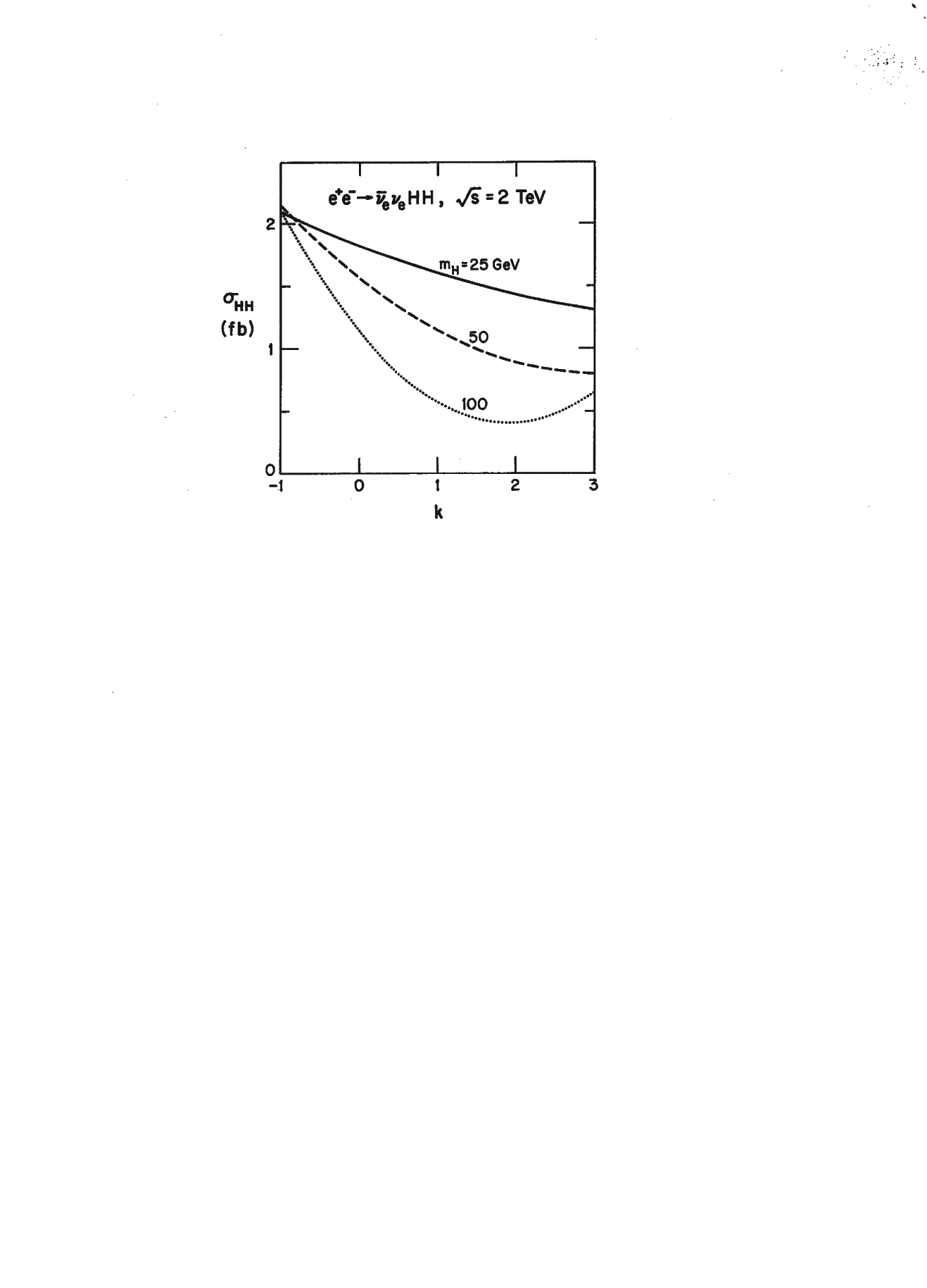}
\vspace*{-17.9cm}

\caption[]{\label{fg:wwtohh} \it Dependence of the cross section for 
Higgs-boson pair production via $W$ fusion on the self-coupling $k$ in 
units of the
Standard Model coupling $g_{H^3}$ at $e^+ e^-$ colliders. Ref.~\cite{46}.}
\end{figure}
The Higgs mechanism, based on a non-zero
value of the Higgs field in the vacuum, must
finally be made manifest experimentally by
reconstructing the interaction potential
that  generates the non-zero field in
the vacuum. This program can be carried out
by measuring the strength of the  trilinear
and quartic self-couplings of the Higgs
particles:
\begin{eqnarray}
g_{H^3} & = & 3 \sqrt{\sqrt{2} G_F} M_H^2 \\ \non \\
g_{H^4} & = & 3 \sqrt{2} G_F M_H^2 ~.
\end{eqnarray}
This is a  difficult task since the
processes to be exploited are suppressed
by small couplings and phase space.
Nevertheless, the problem can be solved
at the LHC and in the high-energy phase
of the $e^+e^-$ linear colliders for sufficiently high
luminosities \cite{46}. The best-suited reaction
for the measurement of the trilinear 
coupling for Higgs masses in the theoretically
preferred mass range of ${\cal O}(100~\mbox{GeV})$, is the
WW fusion process
\begin{equation}
pp, e^+e^- \to WW \to HH
\end{equation}
in which, among other mechanisms, the two-Higgs
final state is generated by the $s$-channel
exchange of a virtual Higgs particle so that this
process is sensitive to the trilinear $HHH$
coupling in the Higgs potential, Fig. \ref{fg:wwtohh}. Since
the cross section is only a fraction of 1 fb
at an energy of $\sim 1.6$ TeV, an integrated luminosity of
$\sim 1 ab^{-1}$ is needed to isolate the events at linear
colliders. The quartic coupling $H^4$
seems to be accessible only through loop
effects in the foreseeable future.\\

To sum up, the essential elements of the
Higgs mechanism can be established experimentally
at the LHC and TeV $e^+e^-$ linear colliders.\\

\vspace*{0.0cm}

\section{Higgs Bosons in Supersymmetric Theories}
Arguments  deeply rooted in the Higg sector 
play an eminent role in introducing
supersymmetry as a fundamental symmetry
of nature \cite{14}. This is the only symmetry
that correlates bosonic with fermionic
degrees of freedom.
\paragraph{(a)}
The cancellation between bosonic and
fermionic contributions to the radiative
corrections of the light Higgs masses
in supersymmetric theories provides a
solution of the hierarchy problem in
the Standard Model. If the Standard Model
is embedded in a grand-unified theory,
the large gap between the high grand-unification
scale and the low scale of
electroweak symmetry breaking can be
stabilized in a natural way in boson--fermion symmetric
theories \cite{15,601}. Denoting
the bare Higgs mass by $M_{H,0}^2$,
the radiative corrections due to vector-boson
loops in the Standard Model by $\delta M_{H,V}^2$, 
and the contributions of supersymmetric
fermionic gaugino partners by $\delta M_{\tilde H,\tilde V}^2$, 
the physical Higgs mass is given by the sum
$M_H^2 = M_{H,0}^2 + \delta M_{H,V}^2 + \delta M_{\tilde H,\tilde V}^2$.
The vector-boson correction is quadratically
divergent, $\delta M_{H,V}^2 \sim \alpha [\Lambda^2 - M^2]$, 
so that for a cut-off scale $\Lambda \sim \Lambda_{GUT}$
extreme fine-tuning between  the
intrinsic bare mass and the radiative quantum fluctuations 
would be needed to
generate a Higgs mass of order $M_W$.
However, owing  to Pauli's principle, the
additional fermionic gaugino contributions
in supersymmetric theories are just
opposite in sign, $\delta M_{\tilde H,\tilde V}^2\sim -\alpha
[\Lambda^2-\tilde M^2]$,
so that the divergent terms cancel. Since
$\delta M_H^2\sim\alpha [\tilde M^2-M^2]$,
any fine-tuning is avoided for supersymmetric
particle masses $\tilde M \lsim {\cal O}(1$ TeV).
Thus, within this symmetry scheme the Higgs 
sector is stable in the low-energy range $M_H\sim M_W$
even in the context of high-energy GUT scales.

\paragraph{(b)}
The concept of supersymmetry is strongly
supported by the successful prediction
of the electroweak mixing angle in
the minimal version of this theory \cite{16}.
There, the extended particle spectrum 
 drives the evolution of the
electroweak mixing angle from the
GUT value 3/8 down to $\sin^2\theta_W = 0.2336 \pm 0.0017$,
the error including unknown threshold
contributions at the low and the
high supersymmetric mass scales.
The prediction coincides with the
experimentally measured value $\sin^2\theta_W^{exp} = 0.2317 \pm 0.0003$
within the theoretical uncertainty
of less than 2 per mille.

\paragraph{(c)}
Conceptually very interesting is 
 the interpretation
of the Higgs mechanism in supersymmetric
theories as a quantum effect \cite{50A}. The
breaking of the electroweak symmetry $SU(2)_L \times U(1)_Y$
can be induced radiatively while
leaving the electromagnetic gauge
symmetry $U(1)_{EM}$
and the colour gauge symmetry $SU(3)_C$
unbroken for top-quark masses
between 150 and 200 GeV. Starting
with a set of universal scalar
masses at the high GUT scale, the
squared mass parameter of the Higgs
sector evolves to negative values
at the low electroweak scale, while
the squared squark and slepton
masses remain positive.

This fundamental mechanism can easily
be studied \cite{50A'} in a simplified model 
for the two stop fields $\tilde t_R$ and $\tilde t_L$,
and the Higgs field $H_2$.
The Yukawa terms in the renormalization group equations
\begin{equation}
4\pi^2 \frac{\partial}{\partial \log \mu^2/M_G^2} \left[
\begin{array}{c}
M_{H_2}^2 \\ \\
M_{\tilde t_R}^2 \\ \\
M_{\tilde t_L}^2
\end{array}
\right] = g_t^2 \left[
\begin{array}{ccc}
3 & 3 & 3 \\ \\
2 & 2 & 2 \\ \\
1 & 1 & 1
\end{array}
\right] \left[
\begin{array}{c}
M_{H_2}^2 \\ \\
M_{\tilde t_R}^2 \\ \\
M_{\tilde t_L}^2
\end{array}
\right] + g_t^2 A_t^2 \left[
\begin{array}{c}
3 \\ \\
2 \\ \\
1
\end{array}
\right] - \frac{16}{3} g_s^2 M_3^2 \left[
\begin{array}{c}
0 \\ \\
1 \\ \\
1
\end{array}
\right]
\end{equation}
drive the masses to smaller values in the evolution from the GUT scale down to
the electroweak scale $\mu^2=M_G^2 \to M_Z^2$. This force is strongest for the
Higgs mass and increases with the top Yukawa coupling. It is balanced by the
SUSY-QCD contribution for the squark masses, if the top Yukawa coupling is not
too large.
Solving the renormalization group equations with the initial condition
\begin{equation}
\hspace*{1.6cm} \mbox{GUT scale:} \hspace*{1.3cm}
M_{H_2}^2 = M_{\tilde t_R}^2 = M_{\tilde t_L}^2 = M_0^2 > 0 ~, 
\end{equation}
the masses evolve down to
\begin{eqnarray*}
\mbox{ELW scale:} \hspace*{1cm}
M^2_{H_2}        & = & -\frac{1}{2} M_0^2 < 0 \\ \\
M_{\tilde t_R}^2 & = & 0 \\ \\
M_{\tilde t_L}^2 & = & +\frac{1}{2} M_0^2 > 0
\end{eqnarray*}
at low energies in the limit of vanishing gauge and trilinear couplings. Both 
stop states preserve the normal particle 
character, while the negative mass squared
of the field $H_2$
generates the Higgs mechanism.\\

The Higgs sector of supersymmetric
theories differs in several aspects
from the Standard Model \cite{17}. To preserve
supersymmetry and gauge invariance,
at least two iso-doublet fields must
be introduced, leaving us with a
spectrum of five or more physical
Higgs particles. In the minimal
supersymmetric extension of the
Standard Model (MSSM) the Higgs
self-interactions are generated
by the scalar-gauge action, so that the
quartic couplings are related to
the gauge couplings in this scenario. This leads
to strong bounds of less than
about 130 GeV for the mass of 
the lightest Higgs boson \cite{19}. If the
system is assumed to remain
weakly interacting up to scales
of the order of the GUT or Planck
scale, the mass remains small, 
for reasons
quite analogous to those found in the Standard
Model,
even in more complex supersymmetric
theories involving additional Higgs fields and  
Yukawa interactions. The masses of the heavy 
Higgs bosons are expected to be
of the scale of electroweak symmetry
breaking up to order 1 TeV.

\subsection{The Higgs Sector of the MSSM}
The particle spectrum of the MSSM \cite{14} consists
of leptons, quarks and their scalar 
supersymmetric partners, and gauge
particles, Higgs particles and their
spin-1/2 partners. The matter and force fields are coupled
in supersymmetric and gauge-invariant
actions:
\begin{equation}
\begin{array}{lrcll}
S = S_V + S_\phi + S_W: \hspace*{1cm}
& S_V    & = & \frac{1}{4} \int d^6 z \hat W_\alpha \hat W_\alpha
\hspace*{1cm} & \mbox{gauge action} ~, \\ \\
& S_\phi & = & \int d^8 z \hat \phi^* e^{gV} \hat \phi
& \mbox{matter action} ~, \\ \\
& S_W    & = & \int d^6 z W[\hat \phi]
& \mbox{superpotential} ~.
\end{array}
\end{equation}
Decomposing the superfields into fermionic
and bosonic components, and carrying out
the integration over the Grassmann
variables in $z\to x$,
the following Lagrangians can be derived, which  
describe the interactions of the
gauge, matter and Higgs fields:
\begin{eqnarray*}
{\cal L}_V & = & -\frac{1}{4}F_{\mu\nu}F_{\mu\nu}+\ldots+\frac{1}{2}D^2 ~, \\ \\
{\cal L}_\phi & = & D_\mu \phi^* D_\mu \phi +\ldots+\frac{g}{2} D|\phi|^2  ~, \\ \\
{\cal L}_W & = & - \left| \frac{\partial W}{\partial \phi} \right|^2 ~. 
\end{eqnarray*}
The $D$ field is an auxiliary field that 
does not propagate in space-time and
 can be eliminated by applying  
the equations of motion: $D=-\frac{g}{2} |\phi|^2$.
Reinserted into the Lagrangian, the
quartic self-coupling of the scalar Higgs
fields is generated:
\begin{equation}
{\cal L} [\phi^4] = -\frac{g^2}{8} |\phi^2|^2 ~.
\end{equation}
Thus, the quartic coupling of the Higgs 
fields is given, in the minimal
supersymmetric theory, by the square
of the gauge coupling. Unlike the Standard
Model case, the quartic coupling is not a free parameter. Moreover,
this coupling is weak.\\

Two independent Higgs doublet fields $H_1$ and $H_2$
must be introduced into the superpotential: 
\begin{equation}
W = -\mu \epsilon_{ij} \hat H_1^i \hat H_2^j + \epsilon_{ij} [f_1 \hat H_1^i
\hat L^j \hat R + f_2 \hat H_1^i \hat Q^j \hat D +
f_2' \hat H_2^j \hat Q^i \hat U]
\end{equation}
to provide the down-type particles ($H_1$)
and the up-type particles ($H_2$) with a mass.
Unlike the Standard Model, the second Higgs
field cannot be identified with the
charge conjugate of the first Higgs field
since $W$ must be analytic to preserve
supersymmetry. Moreover, the Higgsino
fields associated with a single Higgs
field would generate triangle anomalies;
they cancel if the two conjugate doublets
are added up, and the classical gauge
invariance of the interactions is not
destroyed at the quantum level. 
Integrating the superpotential over
the Grassmann coordinates generates
the supersymmetric Higgs self-energy
$V_0 = |\mu|^2 (|H_1|^2 + |H_2|^2)$.
The breaking of supersymmetry can be
incorporated in the Higgs sector by
introducing bilinear mass terms $\mu_{ij} H_i H_j$.
Added to the supersymmetric self-energy part $H^2$
and the quartic part $H^4$
generated by the gauge action, they
lead to the following Higgs potential
\begin{eqnarray}
V & = & m_1^2 H_1^{*i} H_1^i + m_2^2 H_2^{*i} H_2^i - m_{12}^2 (\epsilon_{ij}
H_1^i H_2^j + hc) \non \\ \non \\
& & + \frac{1}{8} (g^2 + g'^2) [H_1^{*i} H_1^i -
H_2^{*i} H_2^i]^2 + \frac{1}{2} |H_1^{*i} H_2^{*i}|^2 ~. 
\end{eqnarray}
The Higgs potential includes three
bilinear mass terms, while the strength
of the quartic couplings is set by the
$SU(2)_L$ and $U(1)_Y$
gauge couplings squared. The three mass
terms are free parameters.

The potential develops a stable minimum
for $H_1 \to [0,v_1]$ and $H_2\to [v_2,0]$,
if the following conditions are met:
\begin{equation}
m_1^2 +  m_2^2 >  2 | m^2_{12} |  \hspace*{0.5cm} \mbox{and}  \hspace*{0.5cm}
m_1^2    m_2^2  <  | m^2_{12} |^2 ~.
\end{equation}
Expanding the fields about the ground-state 
values $v_1$ and $v_2$,
\begin{equation}
\begin{array}{rclcl}
H_1^1 & = & & & H^+ \cos \beta + G^+ \sin \beta \\ \\
H_1^2 & = & v_1 & + & [H^0 \cos \alpha - h^0 \sin \alpha + i A^0 \sin \beta - i G^0
\cos \beta ]/\sqrt{2}
\end{array}
\end{equation}
and
\begin{equation}
\begin{array}{rclcl}
H_2^1 & = & v_2 & + & [H^0 \sin \alpha + h^0 \cos \alpha + i A^0 \cos \beta + i G^0
\sin \beta ]/\sqrt{2} \\ \\
H_2^2 & = & & & H^- \sin \beta - G^- \cos \beta ~, 
\end{array}
\end{equation}
the mass eigenstates are given by the
neutral states $h^0,H^0$ and $A^0$,
which are even and odd under ${\cal CP}$
transformations, and by the charged states $H^\pm$;
the $G$ states correspond to the Goldstone
modes, which are absorbed by the gauge
fields to build up the longitudinal
components. After introducing the three
parameters
\begin{eqnarray}
M_Z^2 & = & \frac{1}{2} (g^2 + g'^2) (v_1^2 + v_2^2) \non \\ \non \\
M_A^2 & = & m_{12}^2 \frac{v_1^2 + v_2^2}{v_1v_2} \non \\ \non \\
\tgb  & = & \frac{v_2}{v_1} ~, 
\end{eqnarray}
the mass matrix can be decomposed into
three $2\times 2$ blocks, which are easy to
diagonalize:
\begin{displaymath}
\begin{array}{lrcl}
\mbox{\bf charged matrix:} & M_\pm^2 & = & \sin 2\beta
(M_A^2+M_W^2) \left[
\begin{array}{cc}
\tgb & 1 \\ \\
1 & \ctgb
\end{array}
\right] \\ \\
& & & \mbox{\underline{charged mass:}}~M_{H^\pm}^2 = M_A^2 + M_W^2 \\ \\ \\
\mbox{\bf pseudoscalar matrix:} & M_a^2 & = & \sin 2\beta
M_A^2 \left[
\begin{array}{cc}
\tgb & 1 \\ \\
1 & \ctgb
\end{array}
\right] \\ \\
& & & \mbox{\underline{pseudoscalar mass:}}~M_A^2 \\ \\ \\
\mbox{\bf scalar matrix:} & M_s^2 & = & \displaystyle \sin 2\beta
\left( \frac{M_A^2}{2} \left[
\begin{array}{cc}
\tgb & -1 \\ \\
-1 & \ctgb
\end{array}
\right]
+ \frac{M_Z^2}{2} \left[
\begin{array}{cc}
\ctgb & -1 \\ \\
-1 & \tgb
\end{array}
\right] \right) \\ \\
& & & \mbox{\underline{scalar masses:}} \\ \\
& & & M_{h,H}^2 = \frac{1}{2} \left[ M_A^2 + M_Z^2 \mp \sqrt{(M_A^2+M_Z^2)^2
- 4M_A^2M_Z^2 \cos^2 2\beta} \right] \\ \\
& & & \displaystyle \tg 2\alpha = \tg 2\beta \frac{M_A^2 + M_Z^2}{M_A^2 - M_Z^2}
\hspace*{0.5cm} \mbox{with} \hspace*{0.5cm} -\frac{\pi}{2} < \alpha < 0
\nonumber
\end{array}
\end{displaymath}
The three zero-mass Goldstone eigenvalues
of the charged and pseudoscalar mass
matrices are not denoted explicitly.\\

From the mass formulae, two important
inequalities can readily be derived,
\begin{eqnarray}
M_h & \leq & M_Z, M_A \leq M_H \\ \nonumber \\
M_W & \leq & M_{H^\pm} ~, 
\end{eqnarray}
which, by construction, are valid in
the tree approximation. As a result,
the lightest of the scalar Higgs masses
is predicted to be bounded by the $Z$ mass,
{\it modulo} radiative corrections. These bounds
follow from the fact that the quartic
coupling of the Higgs fields is determined
in the MSSM by the size of the gauge
couplings squared. \\


\noindent
\underline{\bf SUSY Radiative Corrections} \\[0.5cm]
The tree-level relations between the
Higgs masses are strongly modified
by radiative corrections that involve
the supersymmetric particle spectrum
of the top sector \cite{50B}. These effects
are proportional to the fourth power
of the top mass and to the logarithm
of the stop mass. Their origin are
incomplete cancellations between virtual
top and stop loops, reflecting the
breaking of supersymmetry. Moreover,
the mass relations are affected by 
the potentially large mixing between
$\tilde t_L$ and $\tilde t_R$
due to the top Yukawa coupling.\\

To leading order in $M_t^4$
the radiative corrections can be
summarized in the parameter
\begin{equation}
\epsilon = \frac{3G_F}{\sqrt{2}\pi^2}\frac{M_t^4}{\sin^2\beta}\log
\frac{M_{\tilde t_1}M_{\tilde t_2}}{M_t^2} ~.
\end{equation}
In this approximation the light Higgs mass $M_h$ 
can be expressed by $M_A$ and $\tgb$
in the following compact form:
\begin{eqnarray*}
M^2_h & = & \frac{1}{2} \left[ M_A^2 + M_Z^2 + \epsilon \right.
\non \\
& & \left. - \sqrt{(M_A^2+M_Z^2+\epsilon)^2
-4 M_A^2M_Z^2 \cos^2 2\beta
-4\epsilon (M_A^2 \sin^2\beta + M_Z^2 \cos^2\beta)} \right]
\end{eqnarray*}
The heavy Higgs masses $M_H$ and $M_{H^\pm}$
follow from the sum rules
\begin{eqnarray*}
M_H^2 & = & M_A^2 + M_Z^2 - M_h^2 + \epsilon \non \\
M_{H^\pm}^2 & = & M_A^2 + M_W^2 ~.
\end{eqnarray*}
Finally, the mixing parameter $\alpha$,  
which diagonalizes the ${\cal CP}$-even mass
matrix, is given by the radiatively
improved relation:
\begin{equation}
\tg 2 \alpha = \tg 2\beta \frac{M_A^2 + M_Z^2}{M_A^2 - M_Z^2 +
\epsilon/\cos 2\beta} ~. 
\label{eq:mssmalpha}
\end{equation}

\begin{figure}[hbtp]

\vspace*{-0.9cm}
\hspace*{-5.5cm}
\begin{turn}{-90}%
\epsfxsize=12cm \epsfbox{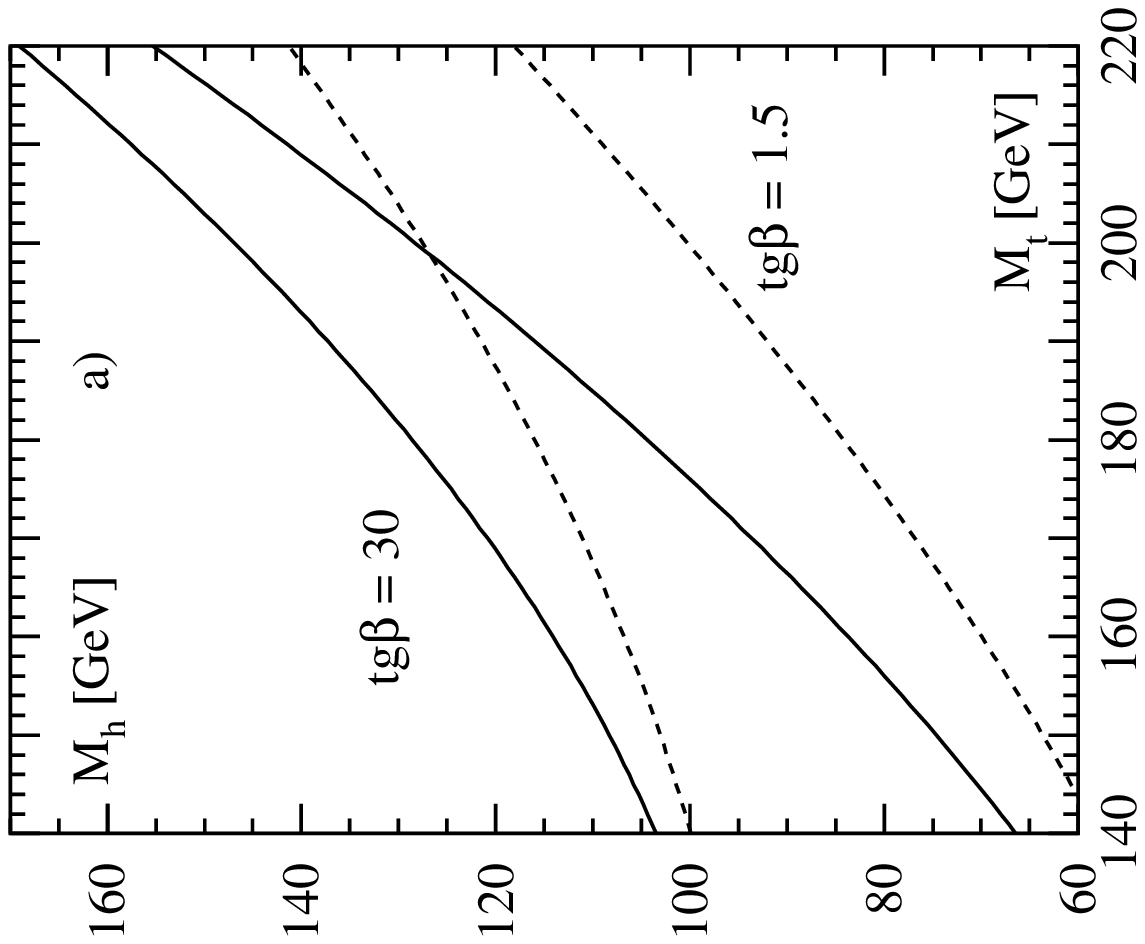}
\end{turn}
\vspace*{-12.05cm}

\hspace*{2.5cm}
\begin{turn}{-90}%
\epsfxsize=12cm \epsfbox{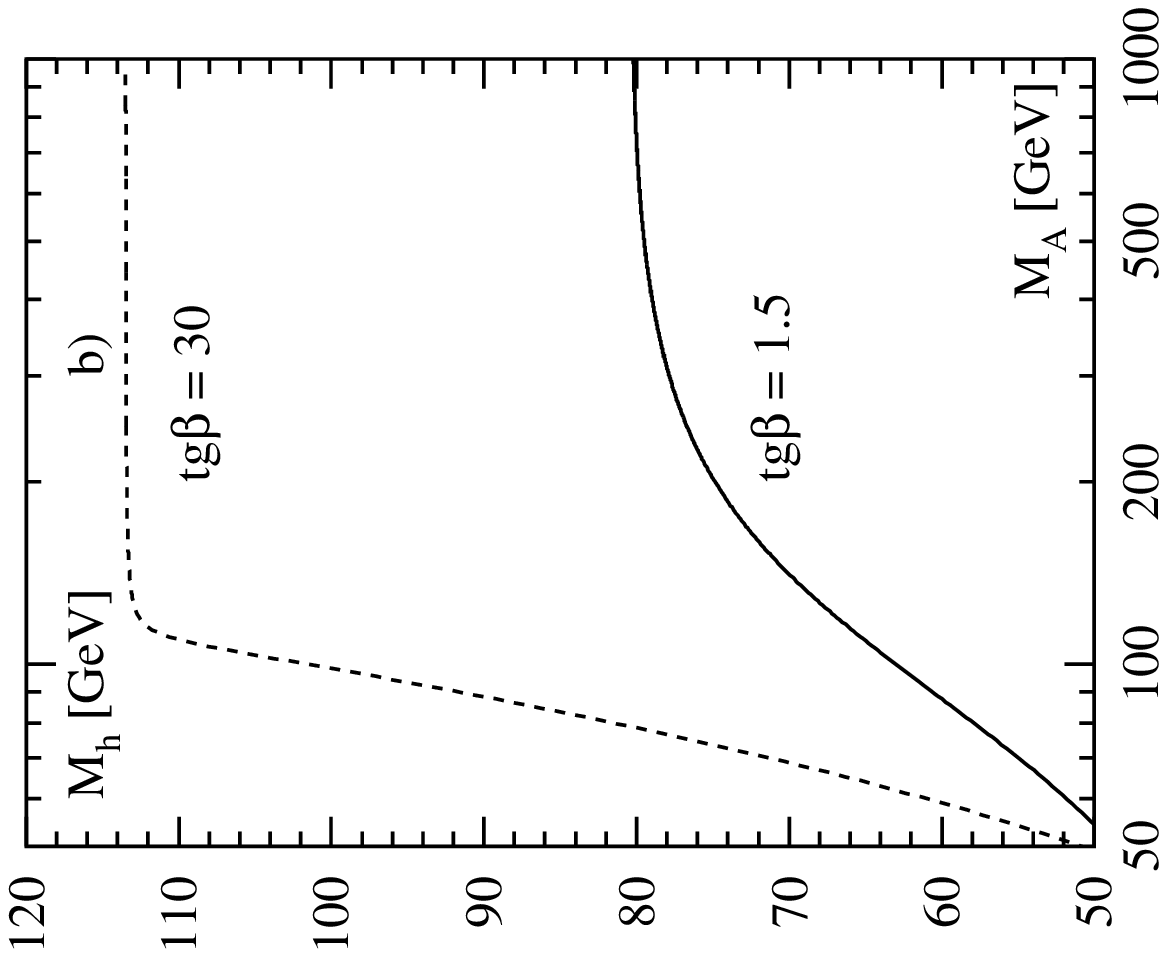}
\end{turn}
\vspace*{-3.5cm}

\hspace*{-5.5cm}
\begin{turn}{-90}%
\epsfxsize=12cm \epsfbox{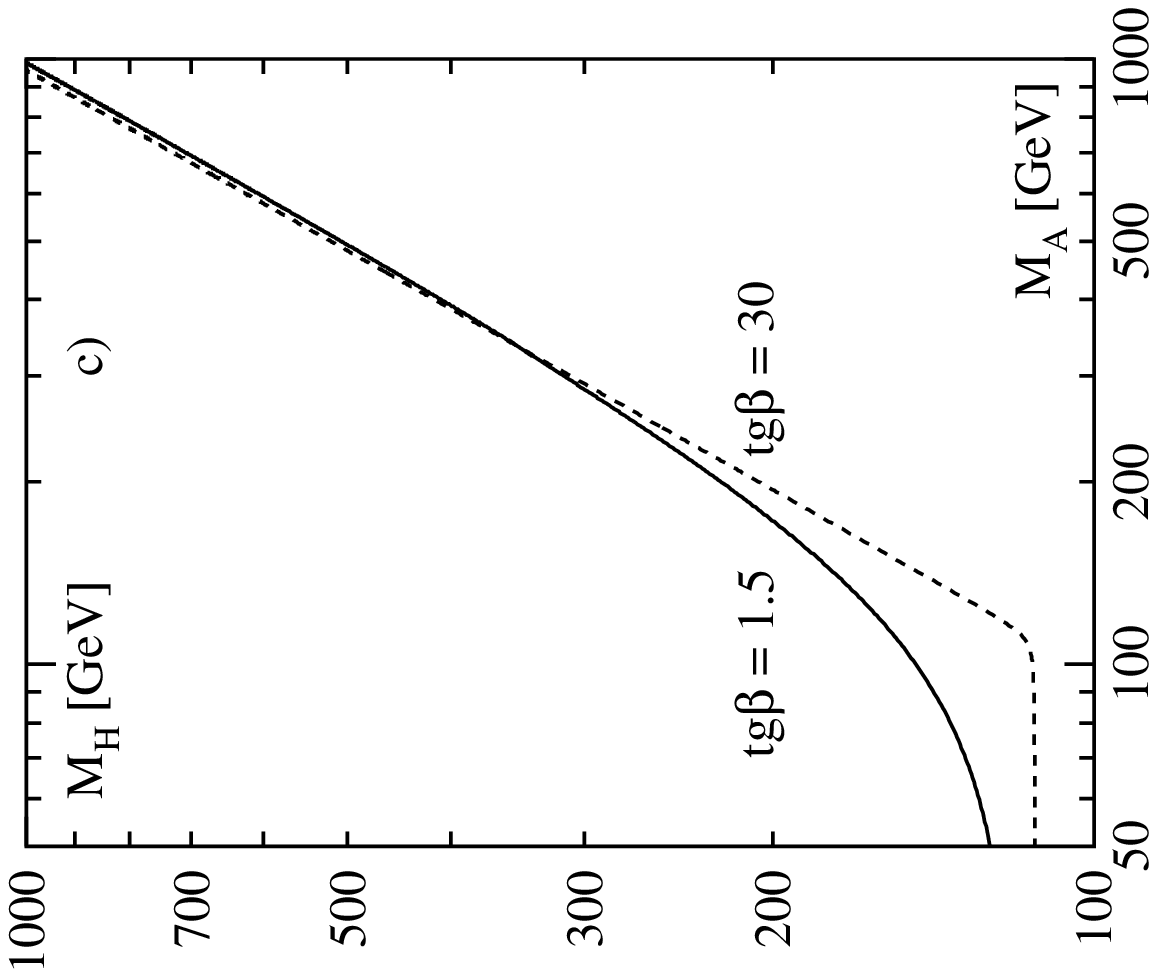}
\end{turn}
\vspace*{-12.05cm}

\hspace*{2.5cm}
\begin{turn}{-90}%
\epsfxsize=12cm \epsfbox{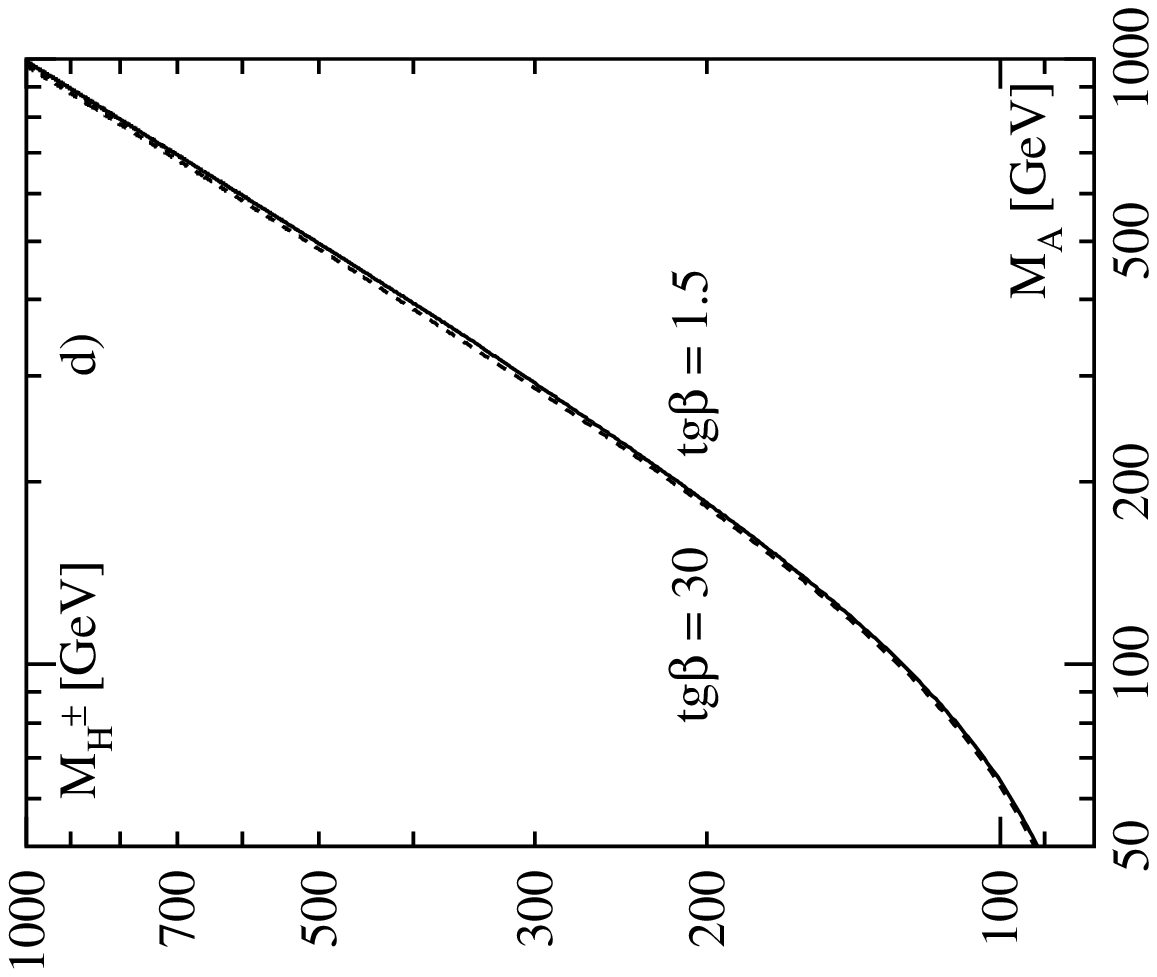}
\end{turn}
\vspace*{-2.5cm}

\caption[ ]{\label{fg:mssmhiggs} \it (a) The upper limit on the light scalar
Higgs pole mass in the MSSM as a function of the top quark mass for two values
of $\tgb=1.5,30$; the
common squark mass has been chosen 
as $M_S=1$ TeV. The full lines correspond to the case of maximal
mixing  [$A_t=\sqrt{6} M_S$, $A_b=\mu=0$] and the dashed lines to vanishing
mixing. The pole masses of the other Higgs bosons, $H,A,H^\pm$, are shown as a
function of the pseudoscalar mass in (b--d) for two values of $\tgb=1.5, 30$, 
 vanishing mixing and $M_t=175$ GeV.}
\end{figure}

The spectrum of the Higgs masses $M_h,M_H$ and $M_{H^\pm}$
is displayed as a function of the pseudoscalar mass $M_A$ 
in Fig. \ref{fg:mssmhiggs} for two representative values of 
$\tgb = 1.5$ and 30. For large $A$  mass, the
masses of the heavy Higgs particles
coincide approximately, $M_A\simeq M_H \simeq M_{H^\pm}$,
while the light Higgs mass approaches
a small asymptotic value. The spectrum for large
values of $\tgb$ is quite  regular: for small $M_A$ one finds
$\{ M_h\simeq M_A; M_H  \simeq \mbox{const} \}$; 
for large $M_A$ the opposite relationship
$\{ M_h\simeq \mbox{const}, M_H \simeq M_{H^\pm}\simeq M_A \}$.\\

While the non-leading effects of
mixing on the Higgs mass relations are
quite involved, the impact on the upper
bound of the light Higgs mass $M_h$ 
can be summarized in a simple way:
\begin{equation}
M_h^2 \leq M_Z^2 \cos^2 2\beta + \delta M_t^2 + \delta M_X^2 ~.
\end{equation}
The leading top contribution is related to
the parameter $\epsilon$,
\begin{equation}
\delta M_t^2 = \epsilon \sin^2\beta ~. 
\end{equation}
The second contribution
\begin{equation}
\delta M_X^2 = \frac{3G_F}{2\sqrt{2}\pi^2} X_t \left[ 2 h(M_{\tilde t_1}^2,
M_{\tilde t_2}^2) + X_t~g(M_{\tilde t_1}^2, M_{\tilde t_2}^2) \right]
\end{equation}
depends on the mixing parameter
\begin{equation}
M_tX_t = M_t \left[A_t - \mu~\ctgb \right] ~, 
\end{equation}
which couples left- and right-chirality
states in the stop mass matrix; $h,g$ are
functions of the stop masses:
\begin{equation}
h = \frac{1}{a-b} \log \frac{a}{b} \hspace*{0.5cm} \mbox{and} \hspace*{0.5cm}
g = \frac{1}{(a-b)^2} \left[ 2 - \frac{a+b}{a-b} \log \frac{a}{b} \right] ~.
\end{equation}
Subdominant contributions can essentially
be reduced to higher-order QCD effects.
They can effectively be incorporated by
interpreting the top mass parameter
$M_t \to M_t(\mu_t)$ as the $\overline{\rm MS}$
top mass evaluated at the geometric mean
between top and stop masses, $\mu_t^2 = M_t M_{\tilde t}$.\\

Upper bounds on the light Higgs mass are
shown in Fig. \ref{fg:mssmhiggs}a for two representative
values of  $\tgb=1.5$ and 30. The curves are the results of 
calculations with and without the mixing effects. It turns out that
$M_h$ is bounded by about $M_h \lsim 100$ GeV 
for moderate values of $\tgb$ 
while the general upper bound is given  by $M_h\lsim 130$ GeV, 
including large values of $\tgb$.  
The light Higgs sector can therefore
be entirely covered, for small $\tgb$, 
 by the LEP2 experiments -- a most
exciting prospect of the search for this
Higgs particle in the next few years.\\

The two ranges of $\tgb$  near $\tgb\sim 1.7$ and
$\tgb \sim M_t/M_b \sim 30$ to 50 are theoretically preferred in
the MSSM, if the model is embedded in a
grand-unified scenario \cite{52A}. Given the
experimentally observed top quark mass,
universal $\tau$ and $b$ 
masses at the unification scale
can be evolved down to the experimental
mass values at low energies within these
two bands of $\tgb$.
Qualitative support for small $\tgb$ 
follows from the observation that in
this scenario the top mass can be
interpreted as a fixed-point of the
evolution down from the unification 
scale \cite{52B}. Moreover, the small $\tgb$ 
range is  slightly preferred, as
radiative corrections that reduce
the light Higgs mass extracted
from the high-precision electroweak
observables, are minimized  \cite{52C}. By contrast, tuning 
problems in adjusting the $\tau/b$ 
mass ratio are more severe for the
large $\tgb$ solution. Nevertheless, this solution
is attractive as the $SO(10)$ 
symmetry relation between $\tau/b/t$ 
masses can be accommodated in this scenario.

\subsection{SUSY Higgs Couplings to SM Particles}
The size of MSSM Higgs couplings to quarks,
leptons and gauge bosons is similar to
the Standard Model, yet modified by the
mixing angles $\alpha$ and $\beta$.
Normalized to the SM values, they are
listed in Table \ref{tb:hcoup}. The pseudoscalar Higgs
boson $A$ does not couple to gauge bosons
at the tree level, but the coupling, 
compatible with  ${\cal CP}$ symmetry, can be
generated by higher-order loops.
The charged Higgs bosons couple to up   
and down fermions with the left- and
right-chiral amplitudes $g_\pm = -\frac{1}{\sqrt{2}}
\left[ g_t (1 \mp \gamma_5) + g_b (1 \pm \gamma_5) \right]$, where
$g_{t,b} = (\sqrt{2} G_F)^{1/2} m_{t,b}$.
\begin{table}[hbt]
\renewcommand{\arraystretch}{1.5}
\begin{center}
\begin{tabular}{|lc||ccc|} \hline
\multicolumn{2}{|c||}{$\Phi$} & $g^\Phi_u$ & $g^\Phi_d$ &  $g^\Phi_V$ \\
\hline \hline
SM~ & $H$ & 1 & 1 & 1 \\ \hline
MSSM~ & $h$ & $\cos\alpha/\sin\beta$ & $-\sin\alpha/\cos\beta$ &
$\sin(\beta-\alpha)$ \\
& $H$ & $\sin\alpha/\sin\beta$ & $\cos\alpha/\cos\beta$ &
$\cos(\beta-\alpha)$ \\
& $A$ & $ 1/\tg\beta$ & $\tg\beta$ & 0 \\ \hline
\end{tabular}
\renewcommand{\arraystretch}{1.2}
\caption[]{\label{tb:hcoup}
\it Higgs couplings in the MSSM to fermions and gauge bosons [$V=W,Z$]
relative to SM couplings.}
\end{center}
\end{table}

\begin{figure}[hbtp]
\vspace*{1.5cm}
\hspace*{-7.0cm}
\begin{turn}{-90}%
\epsfxsize=19cm \epsfbox{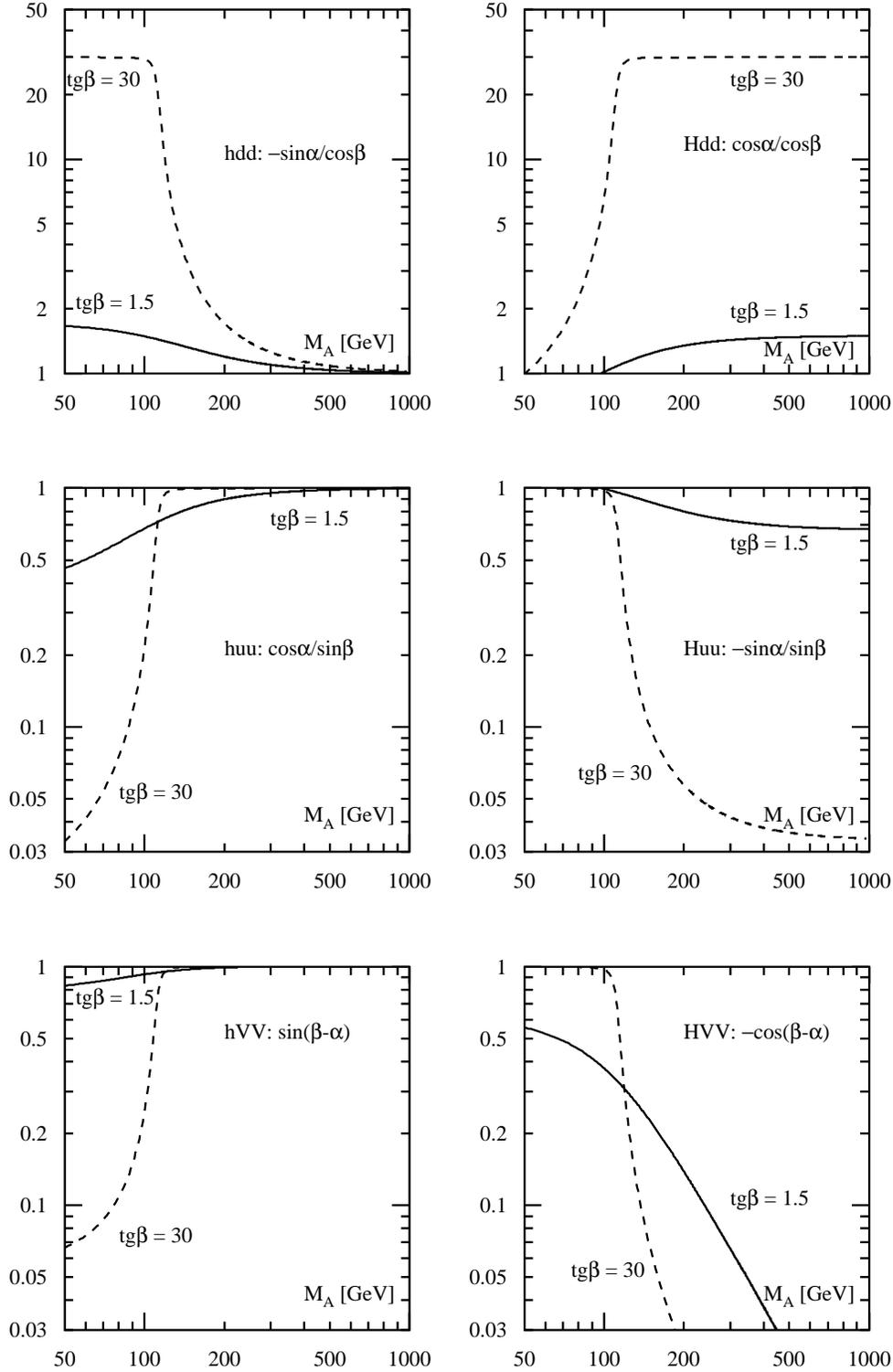}
\end{turn}
\vspace*{-0.5cm}
\caption[ ]{\label{fg:mssmcoup} \it The coupling parameters of the neutral MSSM
Higgs bosons as a function of the pseudoscalar mass $M_A$ for two values of
$\tgb =1.5,30$ and vanishing mixing. The couplings 
 are defined in Table \ref{tb:hcoup}.}
\end{figure}

The modified couplings incorporate the
renormalization due to SUSY radiative
corrections, to leading order in $M_t$, 
if the mixing angle $\alpha$ is related to
$\beta$ and $M_A$
through the corrected formula Eq.~(\ref{eq:mssmalpha}).
The behaviour of the couplings as a
function of mass $M_A$ is exemplified in Fig. \ref{fg:mssmcoup}.\\

For large $M_A$, in practice $M_A\gsim 200$ GeV, 
the couplings of the light Higgs boson
$h$ to the fermions and gauge bosons
approach the SM values asymptotically.
This is the essence of the \underline{decoupling theorem}: 
Particles with large masses
must decouple from the light-particle
system as a consequence of the
quantum-mechanical uncertainty principle.

\subsection{Decays of Higgs Particles}
The lightest \underline{\it neutral Higgs boson} $h$ 
will decay mainly into fermion pairs
since the mass is smaller than $\sim 130$
GeV, Fig. \ref{fg:mssmbr}a (cf. \cite{613A} for a
comprehensive summary). This is, in general,
also the dominant decay mode of the
pseudoscalar boson $A$. For values of $\tgb$
larger than unity and for masses less than
$\sim 140$ GeV, the main decay modes of the neutral
Higgs bosons are decays into $b\bar b$ and $\tau^+\tau^-$
pairs; the branching ratios are of order $\sim 90\%$ and $8\%$,
respectively. The decays into $c\bar c$
pairs and gluons are suppressed, especially
for large $\tgb$.  
For large masses, the top decay channels
$H,A \to t\bar t$ open up; yet for large $\tgb$
this mode remains suppressed and the
neutral Higgs bosons decay almost 
exclusively into $b\bar b$ and $\tau^+\tau^-$
pairs. If the mass is large enough, the
heavy ${\cal CP}$-even Higgs boson $H$
can in principle decay into weak gauge
bosons, $H\to WW,ZZ$.
Since the partial widths are proportional
to $\cos^2(\beta - \alpha)$,
they are strongly suppressed in general,
and the gold-plated $ZZ$ signal of the
heavy Higgs boson in the Standard Model
is lost in the supersymmetric extension.
As a result, the total widths of the Higgs
bosons are much smaller in supersymmetric
theories than in the Standard Model.
\begin{figure}[hbtp]

\vspace*{-2.5cm}
\hspace*{-4.5cm}
\begin{turn}{-90}%
\epsfxsize=16cm \epsfbox{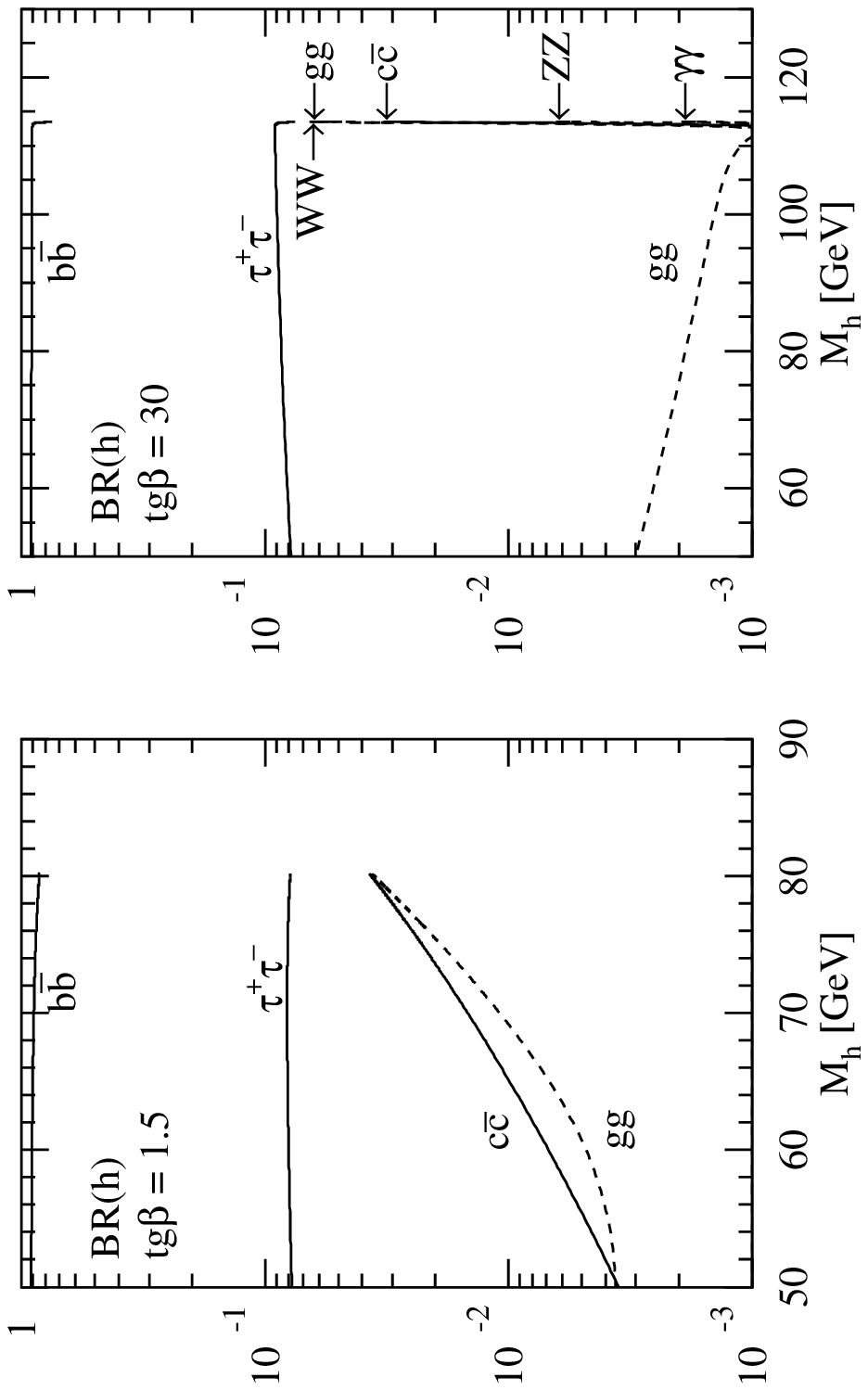}
\end{turn}
\vspace*{-4.2cm}

\centerline{\bf Fig. \ref{fg:mssmbr}a}

\vspace*{-2.5cm}
\hspace*{-4.5cm}
\begin{turn}{-90}%
\epsfxsize=16cm \epsfbox{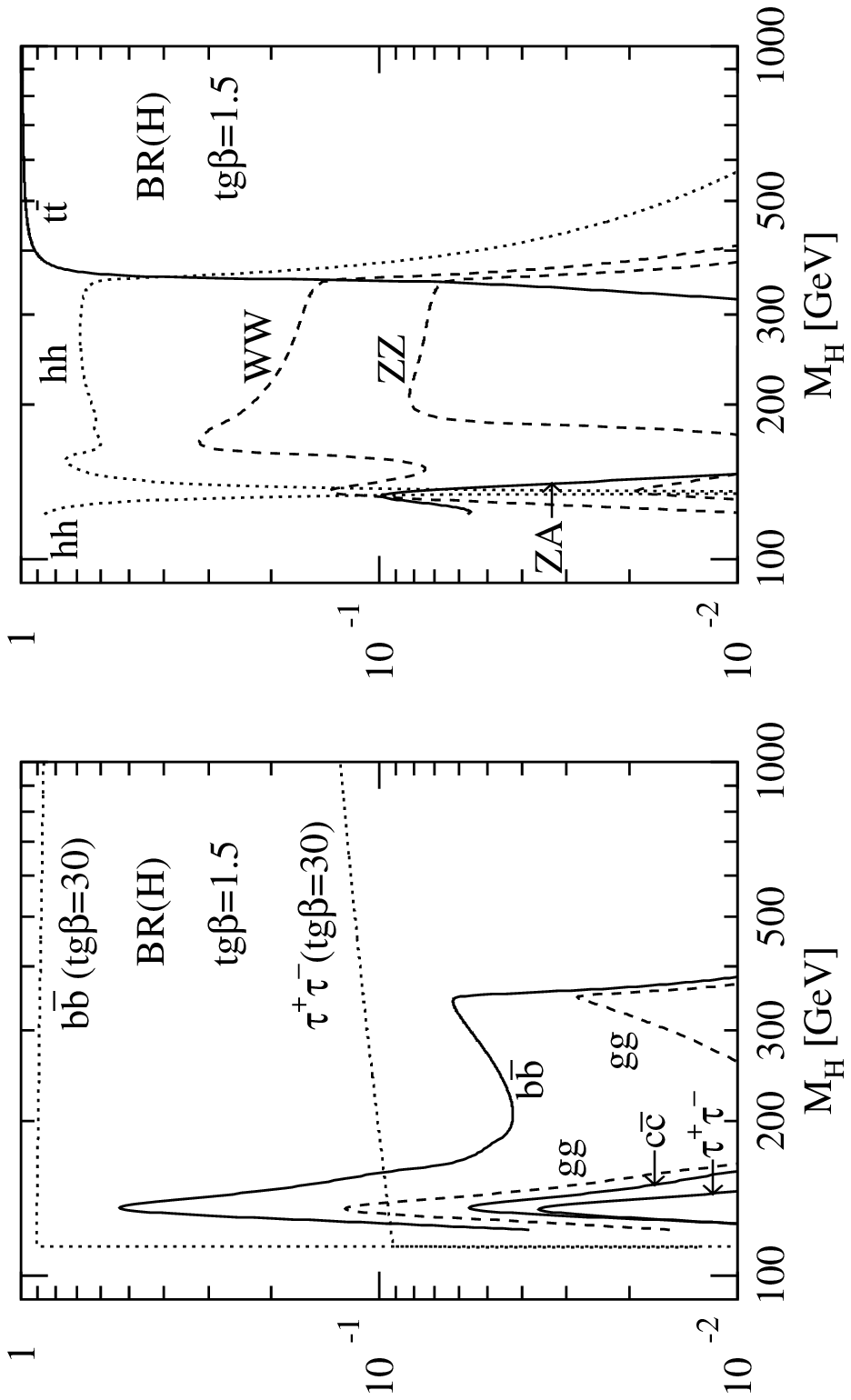}
\end{turn}
\vspace*{-4.2cm}

\centerline{\bf Fig. \ref{fg:mssmbr}b}

\caption[]{\label{fg:mssmbr} \it Branching ratios of the MSSM Higgs bosons $h,
 H, A, H^\pm$ for non-SUSY decay modes as a function of the
masses for two values of $\tgb=1.5, 30$ and vanishing mixing. The common squark
mass has been chosen as $M_S=1$ TeV.}
\end{figure}
\addtocounter{figure}{-1}
\begin{figure}[hbtp]

\vspace*{-2.5cm}
\hspace*{-4.5cm}
\begin{turn}{-90}%
\epsfxsize=16cm \epsfbox{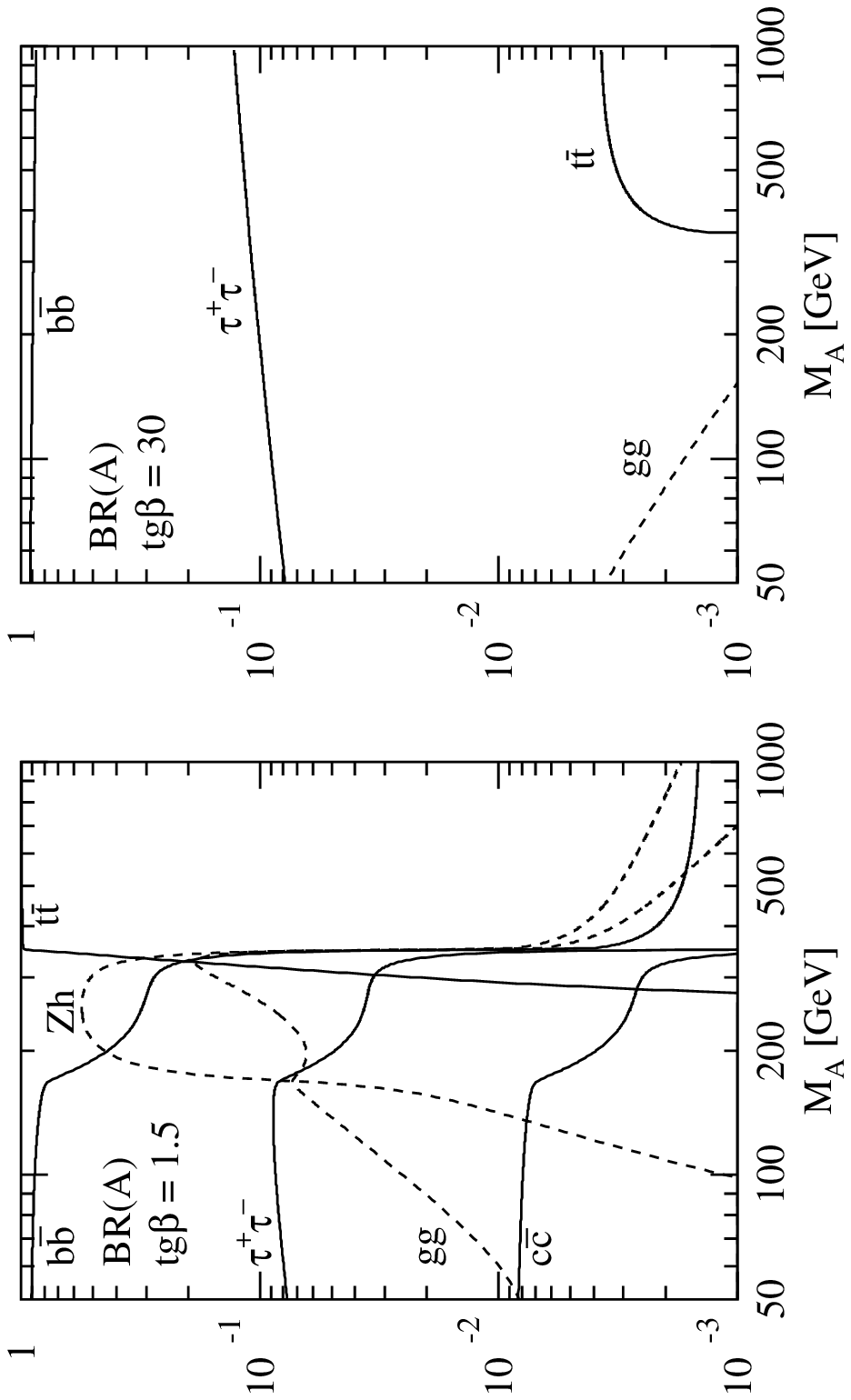}
\end{turn}
\vspace*{-4.2cm}

\centerline{\bf Fig. \ref{fg:mssmbr}c}

\vspace*{-2.5cm}
\hspace*{-4.5cm}
\begin{turn}{-90}%
\epsfxsize=16cm \epsfbox{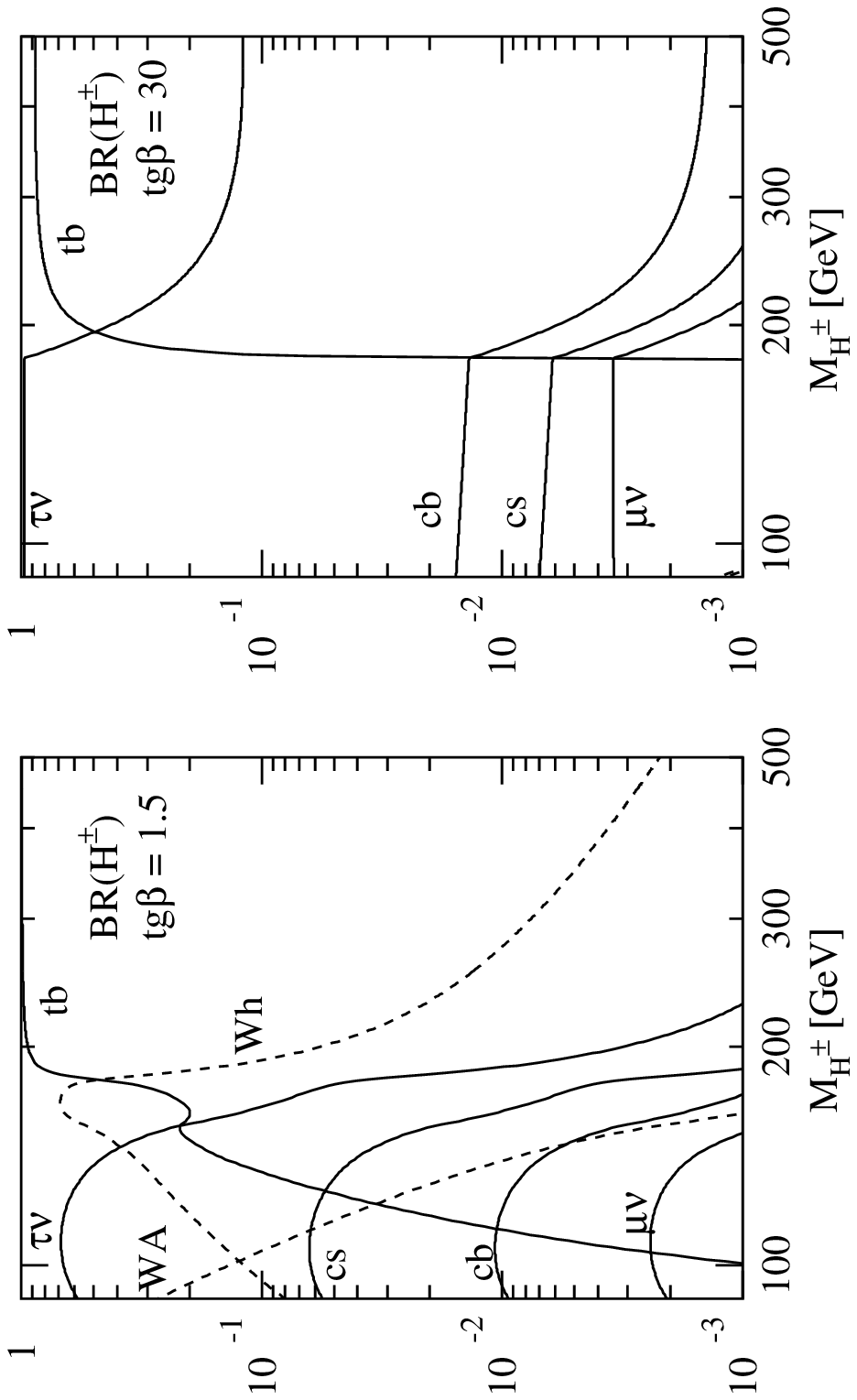}
\end{turn}
\vspace*{-4.2cm}

\centerline{\bf Fig. \ref{fg:mssmbr}d}

\caption[]{\it Continued.}
\end{figure}

\begin{figure}[hbt]

\vspace*{-2.5cm}
\hspace*{-4.5cm}
\begin{turn}{-90}%
\epsfxsize=16cm \epsfbox{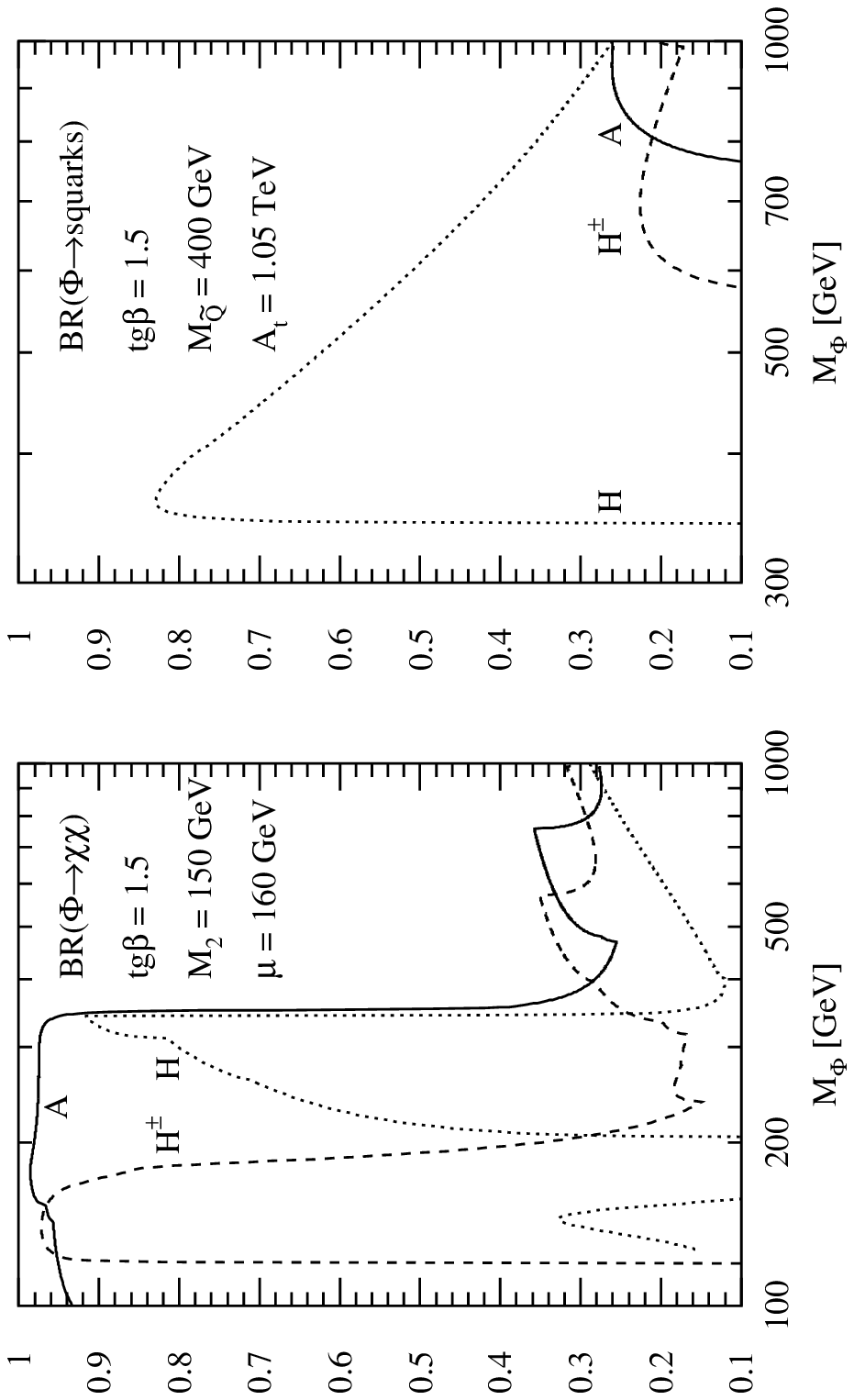}
\end{turn}
\vspace*{-4.2cm}

\caption[]{\label{fg:hcharneutsq} \it Branching ratios of the MSSM Higgs boson
$H,A,H^\pm$ decays into charginos/neutralinos and squarks as a function of their
masses for $\tgb=1.5$. The mixing parameters have been chosen as $\mu=160$ GeV,
$A_t=1.05$ TeV, $A_b=0$ and the squark masses of the first two generations as
$M_{\widetilde{Q}}=400$ GeV. The gaugino mass parameter has been set to
$M_2=150$ GeV.}
\end{figure}

The heavy neutral Higgs boson $H$
can also decay into two lighter Higgs bosons.
Other possible channels are Higgs cascade
decays and decays into supersymmetric
particles \citer{614,616}, Fig. \ref{fg:hcharneutsq}. In addition to light
sfermions, Higgs boson decays into charginos
and neutralinos could eventually be important.
These new channels are kinematically accessible, 
at least for the heavy Higgs bosons $H,A$ and $H^\pm$;
in fact, the branching fractions can be very
large and they can even become dominant in some
regions of the MSSM parameter space. Decays of $h$
into the lightest neutralinos (LSP) are also
important, exceeding 50\% in some parts of
the parameter space. These decays
 strongly affect experimental search techniques.\\

The \underline{\it charged Higgs particles} decay into
fermions, but also, if allowed kinematically,
into the lightest neutral Higgs and a
$W$ boson. Below the $tb$ and $Wh$ thresholds,
the charged Higgs particles will 
decay mostly into $\tau \nu_\tau$ and $cs$
pairs, the former being dominant for $\tgb>1$.
For large $M_{H^\pm}$ values, the top--bottom decay mode
$H^+\to t\bar b$ becomes dominant. In some parts of
the SUSY parameter space, decays into
supersymmetric particles may exceed
50 \%.\\

Adding up the various decay modes,
the width of all five Higgs bosons
remains very narrow, being of order
10 GeV even for large masses.

\subsection{The Production of SUSY Higgs Particles in $e^+e^-$ Collisions}
The search for the neutral SUSY Higgs bosons at \ee linear colliders will be
a straight\-forward ex\-tension of the search now being 
performed at LEP2, which
is expected to cover the mass range up to $\sim
100$~GeV for neutral Higgs bosons.  Higher
energies, $\sqrt{s}$ in excess of $250$~GeV, are required to sweep the
entire parameter space of the MSSM for moderate to large values
of $\tgb$.

\GS The main production mechanisms of \underline{\it neutral Higgs bosons} at
\ee colliders \cite{19, 615, 617} are the \Hs process and associated
pair production, as well as the fusion processes:
\begin{eqnarray}
(a) \ \ \mbox{Higgs--strahlung:} \hspace{1.4cm} \epem &
\stackrel{Z}{\longrightarrow} & Z+h/H \hspace{5cm}
\nonumber  \\
(b) \ \ {\rm Pair \ production:} \hspace{13.6mm} \epem &
\stackrel{Z}{\longrightarrow} & A+h/H 
\nonumber \\
(c) \ \ {\rm Fusion \ processes:} \hspace{10.7mm} \ \epem &
\stackrel{WW}{\longrightarrow} & \overline{\nu_e} \ \nu_e \ + h/H 
\hspace{3.3cm} \nonumber  \\
\epem & 
\stackrel{ZZ}{\longrightarrow} &  \epem + h/H  \nonumber
\end{eqnarray}
The ${\cal CP}$-odd Higgs boson $A$ cannot be produced in fusion
processes to leading order.  The cross sections for the four \Hs and
pair production processes can be expressed as
\begin{eqnarray}
\sigma(\epem \ra Z + h/H) & =& \sin^2/\cos^2(\beta-\alpha) \ \sigma_{SM}
\nonumber \\
\sigma(\epem \ra A + h/H) & =& \cos^2/\sin^2(\beta-\alpha) \
\bar{\lambda} \  \sigma_{SM} ~, 
\end{eqnarray}
where $\sigma_{SM}$ is the SM cross section for \Hs and the coefficient
$\bar{\lambda} \sim \lambda^{3/2}_{Aj} / \lambda^{\demi}_{Zj}$ accounts 
for the suppression of the $P$-wave 
$Ah/H$ cross sections near the threshold.

\STS The cross sections for  Higgs-strahlung and for  pair
production, much as those for the production of the
light and the heavy neutral Higgs bosons $h$ and $H$, are 
complementary, coming either with coefficients
$\sin^2(\beta-\alpha)$ or $\cos^2(\beta-\alpha)$.  As a result, since
$\sigma_{SM}$ is large, at least the lightest ${\cal CP}$-even Higgs
boson must be detected in $e^+e^-$ experiments.

\begin{figure}[hbtp]
\begin{center}
\hspace*{5mm}
\epsfig{file=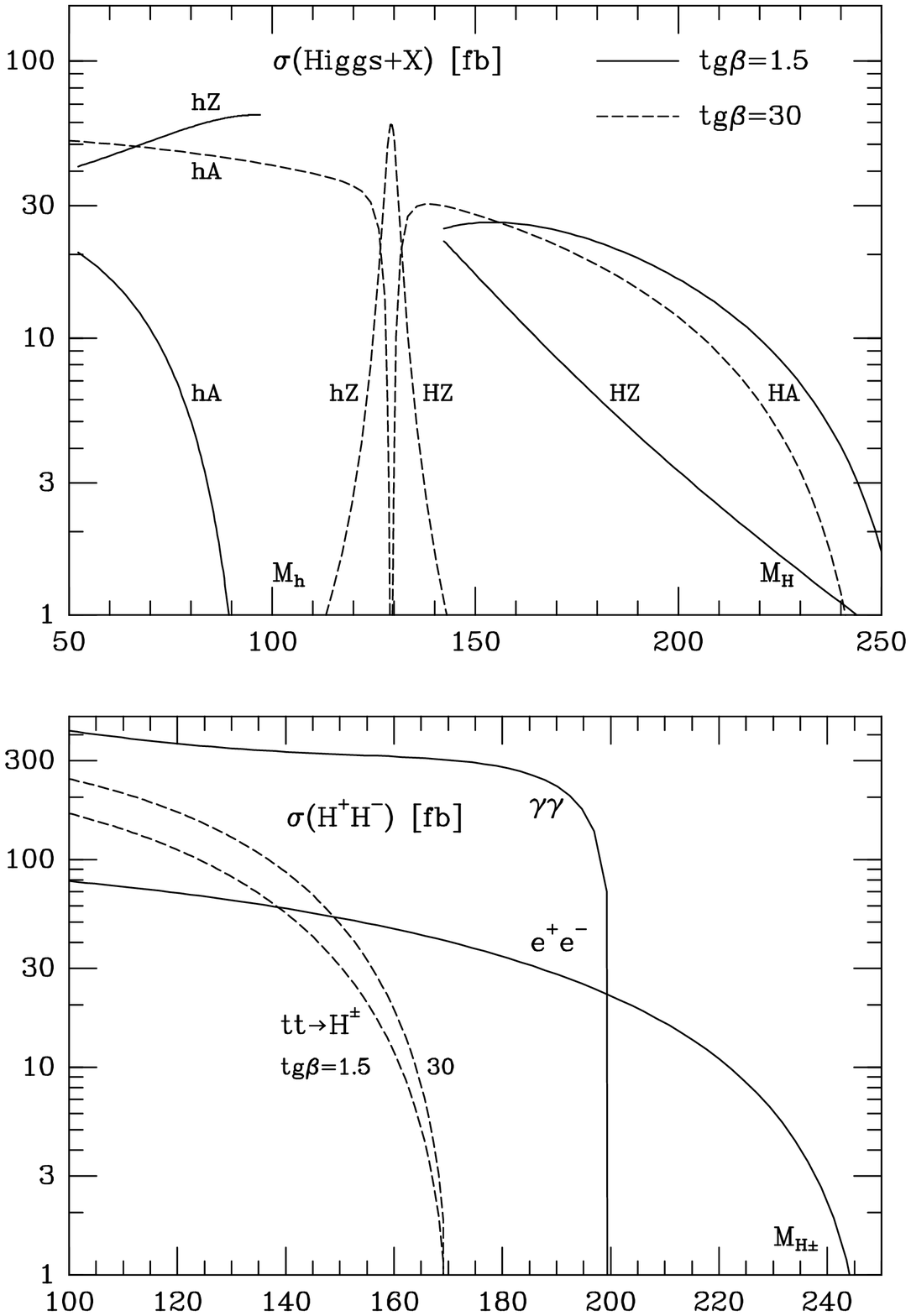,width= 15cm}
\end{center}
\vspace{-2.5cm}
\caption[]{\it 
  Production cross sections of MSSM Higgs bosons at $\sqrt{s} =
  500$~GeV: Higgs-strahlung and pair production; upper part: neutral
  Higgs bosons, lower part: charged Higgs bosons.
  Ref. \protect\cite{613A}.  \protect\label{f604}\label{prodcs}}
\end{figure}
\STS Representative examples of the cross sections for the production
mechanisms of the neutral Higgs bosons are shown in Fig. \ref{f604}, 
as a function of the Higgs masses,  
 for $\tgb= 1.5$ and 30.  The cross
section for $hZ$ is large for $M_h$ near the maximum value allowed
for $\tgb$; it is of order 50~fb, corresponding to $\sim$ 2,500
events for an integrated luminosity of 50 fb$^{-1}$.  By contrast, the
cross section for $HZ$ is large if $M_h$ is sufficiently below the
maximum value  [implying small $M_H$].  For $h$
and for the light $H$, the signals consist of a $Z$ boson accompanied by
a $b\bar{b}$ or $\tau^+ \tau^-$ pair.  These signals are easy to separate
from the background,  which comes mainly from $ZZ$ production if the
Higgs mass is close to $M_Z$.  For the associated channels $\epem \to
Ah$ and $AH$, the situation is opposite to the previous case: the
cross section for $Ah$ is large for light $h$, whereas $AH$ pair
production is the dominant mechanism in the complementary region for
heavy $H$ and $A$ bosons.  The sum of the two cross sections
decreases from $\sim 50$ to 10~fb if $M_A$ increases from $\sim 50$ to
200~GeV at $\sqrt{s} = 500$~GeV.  In major parts of the parameter
space, the signals consist of four $b$ quarks in the final state,
requiring provisions for efficient $b$-quark tagging.  Mass
constraints will help to eliminate the backgrounds from QCD jets and
$ZZ$ final states.  For the $WW$ fusion mechanism, the cross sections
are larger than for Higgs-strahlung, if the Higgs mass is moderately small -- less
than 160~GeV at $\sqrt{s} = 500$ GeV.  However, since the final state
cannot be fully reconstructed, the signal is more difficult to
extract.  As in the case of the \Hs processes, the production of light
$h$ and heavy $H$ Higgs bosons complement each other in $WW$
fusion, too.

\GS The \underline{\it charged Higgs bosons}, if lighter than the top 
quark, can
be produced in top decays, $t \ra b + H^+$, with a branching ratio
varying between $2\%$ and $20\%$ in the kinematically allowed region.
Since the cross section for top-pair production is of order 0.5 pb at
$\sqrt{s} = 500$~GeV, this corresponds to 1,000 to 10,000 charged
Higgs bosons at a luminosity of 50~fb$^{-1}$.  Since, for $\tgb$ larger
than unity, the charged Higgs bosons will decay mainly into $\tau
\nu_\tau$, there is  a surplus of $\tau$ final states over $e,
\mu$ final states in $t$ decays, an apparent breaking of lepton
universality.  For large Higgs masses the dominant decay mode is the
top decay $H^+ \to t \overline{b}$.  In this case the charged Higgs
particles must be pair-produced in \ee colliders:
\[
              \epem \to H^+H^- ~.
\]
The cross section depends only on the charged Higgs mass.  It is of
order 100 fb for small Higgs masses at $\sqrt{s} = 500$~GeV, but it
drops very quickly due to the $P$-wave suppression $\sim \beta^3$
near the threshold.  For $M_{H^{\pm}} = 230$~GeV, the cross section
falls to a level of $\simeq 5\,$~fb, which corresponds, for an integrated
luminosity of $50\,{\rm fb}^{-1}$, to 250 events.  The \cs
is considerably larger for $\gamma \gamma$ collisions.

\begin{figure}[hbtp]
\begin{center}
\hspace*{-1cm}
\epsfig{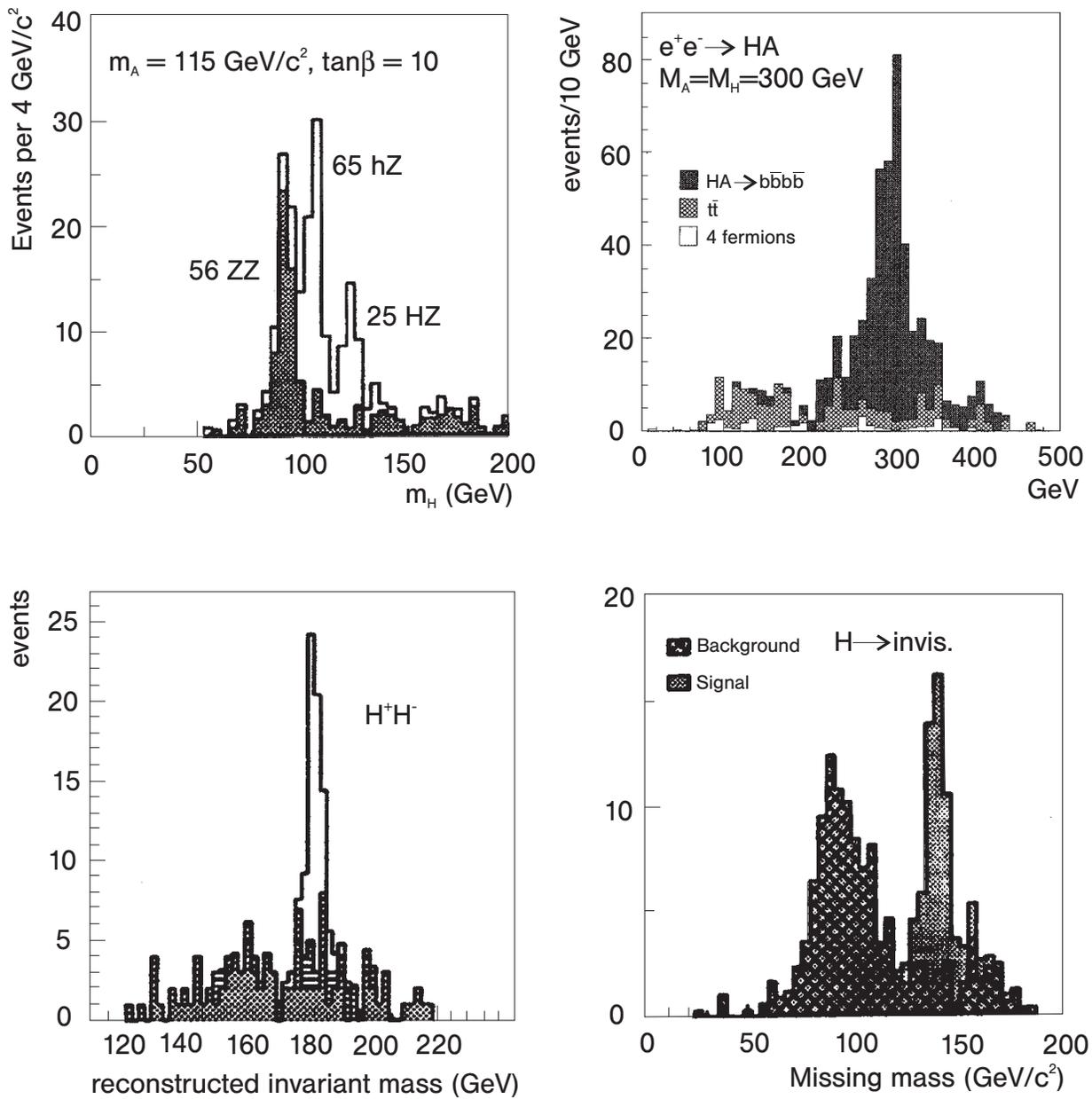}
\end{center}
\vspace{-.5cm}
\caption[]{\it 
  Experimental simulations of the search for MSSM Higgs bosons in
  Higgs-strahlung $hZ / HZ$, heavy-pair production $HA$, charged-Higgs 
production $H^+ H^-$, and neutral invisible Higgs decays in
  Higgs-strahlung.  Refs. \protect\citer{620,620A}.
  \protect\label{f605}}
\end{figure}

\GS \noindent
\underline{\bf Experimental Search Strategies} \\[0.5cm]
Search strategies have been summarized for neutral
Higgs bosons in Refs. \cite{620,620AA} and for charged Higgs bosons in
Ref. \cite{620A}.  Examples of the results for \Hs $Zh, ZH$ and pair
production $Ah$, $AH$ and $H^+H^-$ are given in Fig. \ref{f605}.
Visible as well as invisible decays are under experimental control
already for an integrated luminosity of 10~fb$^{-1}$.
The overall experimental situation
can be summarized as the following two points:

\STS
\noindent
{\bf (i)} The lightest ${\cal CP}$-even Higgs particle $h$ can be
detected in the entire range of the MSSM parameter space, either
via Higgs-strahlung  $\epem \to hZ$ or via pair
production $\epem \rightarrow hA$.  This conclusion holds true even at
a c.m. energy of 250 GeV, independently of the squark mass values; it
is also valid if decays to invisible neutralinos and other SUSY \ps
are realized in the Higgs sector.

\STS
\noindent
{\bf (ii)} The area in the parameter space where {\it all SUSY Higgs
bosons} can be discovered at \ee colliders is characterized by $M_H,
M_A \lessim \frac{1}{2} \sqrt{s}$, independently of $\tgb$.  The $h,
H$ Higgs bosons can be produced either via \Hs or in $Ah, AH$
associated production; charged Higgs bosons will be produced in
$H^+H^-$ pairs. \\

The search for the lightest neutral SUSY
Higgs boson $h$ is one of the most important experimental
tasks at LEP2. 
Up to the present time, mass values of the
pseudoscalar boson $A$ of less than 75 GeV have 
been excluded, independently of $\tgb$. In MSSM scenarios
without mixing effects, the entire mass range
of the lightest Higgs particle $h$ has already
been covered for $\tgb$ less than about 1.6; however,
this conclusion does not hold true  for scenarios
with strong mixing effects \cite{9}.
With a final energy close
to 200 GeV, the Higgs boson $h$ could be
discovered within the theoretically
allowed mass range if the mixing
parameter were realized below $\tgb \lsim 2.4$.
This range covers one of the two
solutions singled out by $\tau/b$
mass unification; moreover, it
corresponds to the area predicted by 
the fixed-point solution of the top-quark mass.
  
\subsection{The Production of SUSY Higgs Particles in Hadron \\ Collisions}
The basic production processes of SUSY
Higgs particles at hadron colliders \cite{620B,32}
are essentially the same as in the
Standard Model. Important differences
are  nevertheless generated by the
modified couplings, the extended particle
spectrum, and the negative parity of
the $A$ boson. For large $\tgb$ 
the coupling $hb\bar b$ is enhanced so that
the bottom-quark loop becomes competitive 
to the top-quark loop in the effective
$hgg$ coupling. Moreover squark loops
will contribute to this coupling \cite{sqloop}.\\

The partonic cross section $\sigma(gg\to \Phi)$
for the gluon fusion of Higgs particles
can be expressed by couplings $g$, in units
of the corresponding SM couplings, and
form factors $A$; to lowest order \cite{32,sqloopqcd}:
\begin{eqnarray}
\hat\sigma^\Phi_{LO} (gg\to \Phi) & = & \sigma^\Phi_0 \delta
\left(1 - \frac{M_\Phi^2}{\hat s}\right) \\
\sigma^{h/H}_0 & = & \frac{G_{F}\alpha_{s}^{2}(\mu)}{128 \sqrt{2}\pi} \
\left| \sum_{Q} g_Q^{h/H} A_Q^{h/H} (\tau_{Q})
+ \sum_{\widetilde{Q}} g_{\widetilde{Q}}^{h/H} A_{\widetilde{Q}}^{h/H}
(\tau_{\widetilde{Q}}) \right|^{2} \nonumber \\
\sigma^A_0 & = & \frac{G_{F}\alpha_{s}^{2}(\mu)}{128 \sqrt{2}\pi} \
\left| \sum_{Q} g_Q^A A_Q^A (\tau_{Q}) \right|^{2} \nonumber
\end{eqnarray}
While the quark couplings have been
defined in Table \ref{tb:hcoup}, the couplings of 
the Higgs particles to squarks are given by
\begin{eqnarray}
g_{\tilde Q_{L,R}}^{h} & = & \frac{M_Q^2}{M_{\tilde Q}^2} g_Q^{h}
\mp \frac{M_Z^2}{M_{\tilde Q}^2} (I_3^Q - e_Q \sin^2 \theta_W)
\sin(\alpha + \beta) \nonumber \\ \nonumber \\
g_{\tilde Q_{L,R}}^{H} & = & \frac{M_Q^2}{M_{\tilde Q}^2} g_Q^{H}
\pm \frac{M_Z^2}{M_{\tilde Q}^2} (I_3^Q - e_Q \sin^2 \theta_W)
\cos(\alpha + \beta)
\end{eqnarray}
${\cal CP}$ invariance only allows for non-zero
squark couplings to the pseudoscalar $A$
boson. 
The form factors can be expressed
in terms of the scaling function $f(\tau_i=4M_i^2/M_\Phi^2)$,
cf. Eq. (\ref{eq:ftau}):
\begin{eqnarray}
A_Q^{h/H} (\tau) & = & \tau [1+(1-\tau) f(\tau)] \nonumber \\
A_Q^A (\tau) & = & \tau f(\tau) \nonumber \\
A_{\tilde Q}^{h/H} (\tau) & = & -\frac{1}{2}\tau [1-\tau f(\tau)] ~.
\end{eqnarray}
For small $\tgb$ the contribution of the top loop is
dominant, while for large $\tgb$
the bottom loop is strongly enhanced.
The squark loops can be significant
for squark masses below $\sim 400$ GeV \cite{sqloopqcd}.\\

The limits of both large and small
loop masses are interesting for SUSY
Higgs particles. The contribution of
the top loop to the $hgg$ coupling can
be calculated approximately in the limit
of large loop masses, while the bottom
contributions to the $\Phi gg$ 
couplings can be calculated in the
approximation of small $b$ masses.

The limits of large loop masses for the
${\cal CP}$-even $h,H$ Higgs bosons are the same
as in the Standard Model,
\begin{eqnarray}
A_Q^{h/H} \to 2/3
\end{eqnarray}
while the corresponding limit for the
${\cal CP}$-odd $A$ Higgs boson reads:
\begin{eqnarray}
A_Q^A \to 1 ~. 
\end{eqnarray}
As a result of the non-renormalization
of the axial anomaly, the $Agg$ coupling
is not altered by QCD radiative 
corrections for large loop masses.

In the opposite limit in which the
quark-loop mass is much smaller than
the Higgs mass, the amplitudes are
the same for scalar and pseudoscalar
Higgs bosons:
\begin{eqnarray}
A_Q^\Phi \to -\frac{\tau_Q}{4} \left( \log \frac{\tau_Q}{4} - i\pi \right)^2 ~.
\end{eqnarray}
This result follows from the restoration
of chiral symmetry in the limit of
vanishing quark masses. \\

Other production mechanisms for SUSY
Higgs bosons, vector boson fusion,
Higgs-strahlung off $W,Z$ bosons and
Higgs-bremsstrahlung off top and bottom
quarks, can be treated in analogy to
the corresponding SM processes.\\
\begin{figure}[hbtp]

\vspace*{0.3cm}
\hspace*{1.0cm}
\begin{turn}{-90}%
\epsfxsize=8.5cm \epsfbox{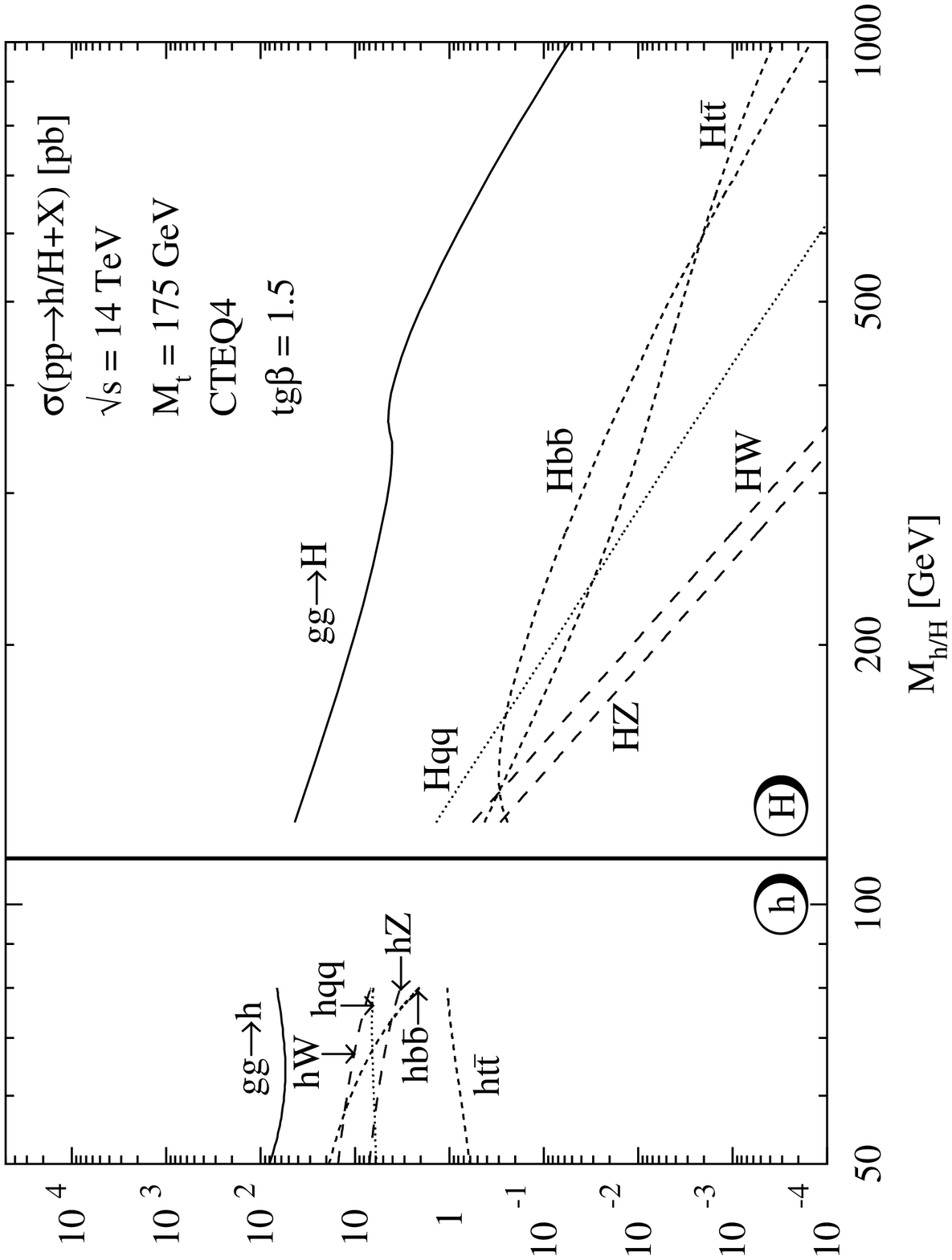}
\end{turn}
\vspace*{0.3cm}

\centerline{\bf Fig. \ref{fg:mssmprohiggs}a}

\vspace*{0.2cm}
\hspace*{1.0cm}
\begin{turn}{-90}%
\epsfxsize=8.5cm \epsfbox{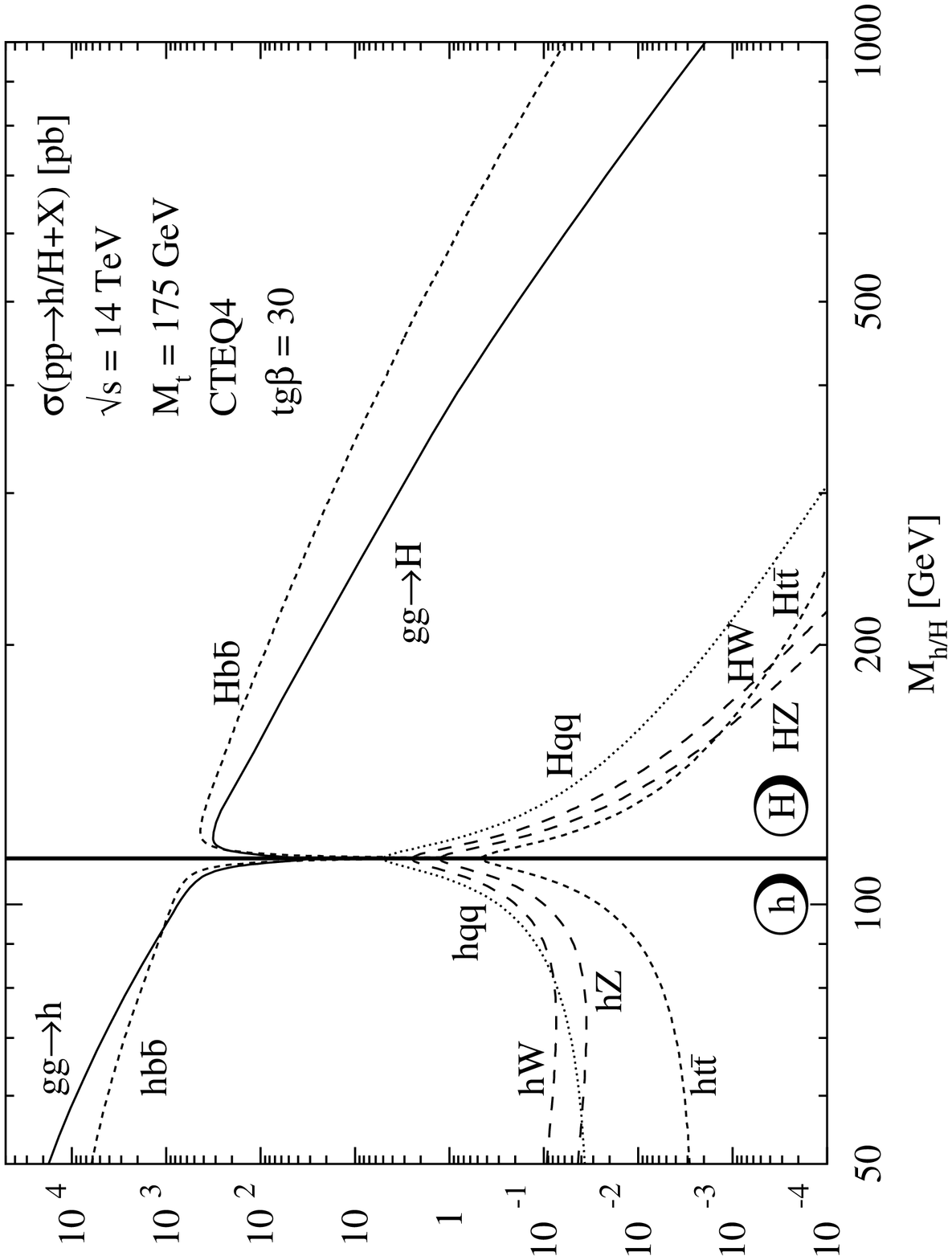}
\end{turn}
\vspace*{0.3cm}

\centerline{\bf Fig. \ref{fg:mssmprohiggs}b}

\caption[]{\label{fg:mssmprohiggs} \it Neutral MSSM Higgs production cross
sections at the LHC  for gluon fusion $gg\to \Phi$,
vector-boson fusion $qq\to qqVV \to qqh/
qqH$, vector-boson bremsstrahlung $q\bar q\to V^* \to hV/HV$ and the associated
production $gg,q\bar q \to  b\bar b \Phi/ t\bar t \Phi$, including all known
QCD corrections. (a) $h,H$ production for $\tgb=1.5$, (b) $h,H$ production for
$\tgb=30$, (c) $A$ production for $\tgb=1.5$, (d) $A$ production for $\tgb=30$.}
\end{figure}
\addtocounter{figure}{-1}
\begin{figure}[hbtp]

\vspace*{0.3cm}
\hspace*{1.0cm}
\begin{turn}{-90}%
\epsfxsize=8.5cm \epsfbox{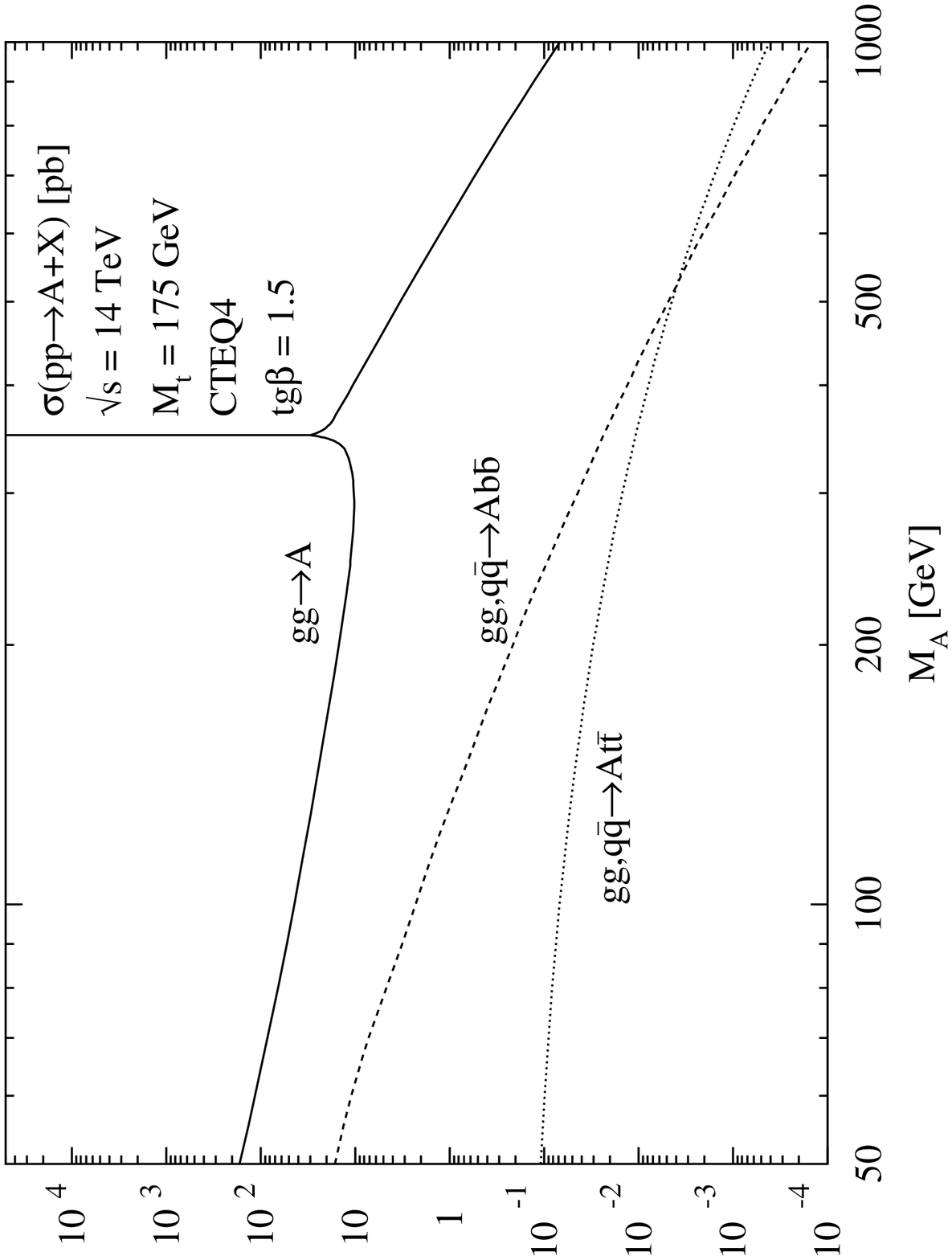}
\end{turn}
\vspace*{0.3cm}

\centerline{\bf Fig. \ref{fg:mssmprohiggs}c}

\vspace*{0.2cm}
\hspace*{1.0cm}
\begin{turn}{-90}%
\epsfxsize=8.5cm \epsfbox{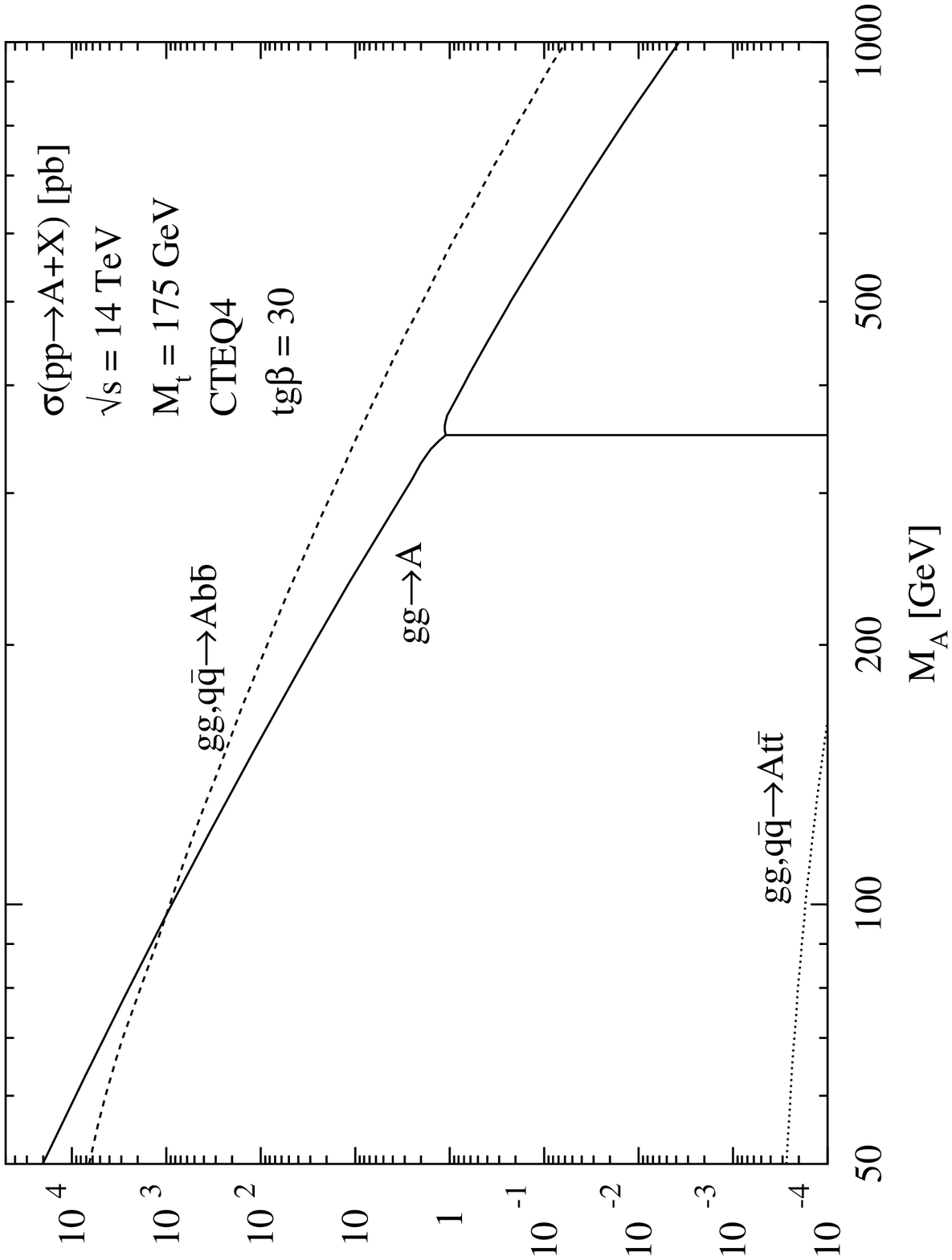}
\end{turn}
\vspace*{0.3cm}

\centerline{\bf Fig. \ref{fg:mssmprohiggs}d}

\caption[]{\it Continued.}
\end{figure}

Data from the Tevatron in the channel $p \bar p \to b \bar b \tau^+ \tau^-$
 have been exploited \cite{63A} to exclude part of the supersymmetric Higgs
parameter space in the $[ M_A, \tgb]$ plane. In the
interesting range of $\tgb$ between 30 and 50, 
pseudoscalar masses $M_A$ of up to 150 to 190 GeV appear 
to be excluded.\\

The cross sections of the various MSSM Higgs production mechanisms at the LHC
are shown in Figs. \ref{fg:mssmprohiggs}a--d for two representative values of
$\tgb = 1.5$ and 30, as a function of the corresponding Higgs mass.
The CTEQ4M
parton densities have been adopted with $\alpha_s(M_Z)=0.116$; the top and
bottom masses have been set to $M_t=175$ GeV and $M_b=5$ GeV. For the
Higgs bremsstrahlung off $t,b$ quarks, $pp \to Q\bar Q \Phi +X$,  the
leading-order CTEQ4L parton densities have been used.
For small and moderate values of $\tgb\lsim 10$
the gluon-fusion cross section provides the dominant production cross section
for the entire Higgs mass region up to $M_\Phi\sim 1$ TeV. However, for large
$\tgb$, Higgs bremsstrahlung off bottom quarks, $pp\to b\bar b \Phi+X$, 
dominates
over the gluon-fusion mechanism since  the  bottom Yukawa
couplings are strongly enhanced in this case.\\

The MSSM Higgs search at the LHC will be more involved than the SM Higgs
search. The basic features can be summarized as follows.\\
 
\noindent
{\bf(i)} For large pseudoscalar Higgs masses, $M_A \gsim 200$ GeV, the light
scalar Higgs boson $h$ can only be found in the photonic decay mode $h\to
\gamma \gamma$. In a significant part of this MSSM parameter region, especially
for moderate values of $\tgb$, no other MSSM Higgs particle can be discovered.
Because of the decoupling limit for large $M_A$, the MSSM cannot be distinguished
from the SM in this mass range.\\

\noindent
{\bf (ii)} For small values of $\tgb \lsim 3$ and pseudoscalar Higgs masses
between about 200 and 350 GeV, the heavy scalar Higgs boson can be searched
for in the `gold-plated' channel $H \to ZZ \to 4l^\pm$. Otherwise this
`gold-plated' signal does not play any role in the MSSM.
However, the  MSSM parameter region covered in this scenario hardly 
exceeds the anti\-cipated
exclusion limits of the LEP2 experiments.\\
 
\noindent
{\bf (iii)} For large and moderate values of $\tgb$ ($\gsim 3$), the decays $H,A
\to \tau^+\tau^-$
become visible at the LHC. Thus this decay mode plays a significant role for
the MSSM in contrast to the SM. Moreover, this mode can also be detected
 for 
small values of  $\tgb$ ($\gsim 1$--$2$) and $M_A$ ($\lsim 200$~GeV).\\

\noindent
{\bf (iv)} For $\tgb \lsim 4$ and 150 GeV $\!\!\lsim M_A \lsim 400$ GeV, the
heavy scalar Higgs
particle can be detected in the decay mode $H\to hh\to b\bar b\gamma\gamma$.
However, the MSSM parameter range for this signature is very limited.\\

\noindent
{\bf (v)} For $\tgb\lsim 3$--$5$ and 50 GeV $\!\!\lsim M_A\lsim 350$ GeV, the
pseudoscalar decay mode $A\to Zh \to l^+ l^- b\bar b$ will be visible, but
hardly exceeds the exclusion limits from LEP2.\\

\noindent
{\bf (vi)} For pseudoscalar Higgs masses $M_A\lsim 100$ GeV, charged Higgs
bosons, produced from top quark decays $t\to H^+ b$, can be discovered in 
the  decay mode $H^+ \to \tau^+ \bar\nu_\tau$. 
The search for charged Higgs bosons is
quite difficult in general if the mass
exceeds the top-quark mass and $t\to b+H^+$ 
decays are forbidden kinematically. Since $H^\pm$ 
bosons cannot be radiated off $Z$ or $W$ 
bosons, they must be produced in pairs
in the Drell--Yan process \cite{qqtohphm} or
in $gg$ collisions \cite{ggtohphm}. In the second
process, and equivalently in $W^\pm H^\mp$
final states, the effective couplings
are built up by loops of heavy quarks.\\

\begin{figure}[hbtp]

\vspace*{-5.0cm}
\hspace*{-3.5cm}
\epsfxsize=20cm \epsfysize=28cm \epsfbox{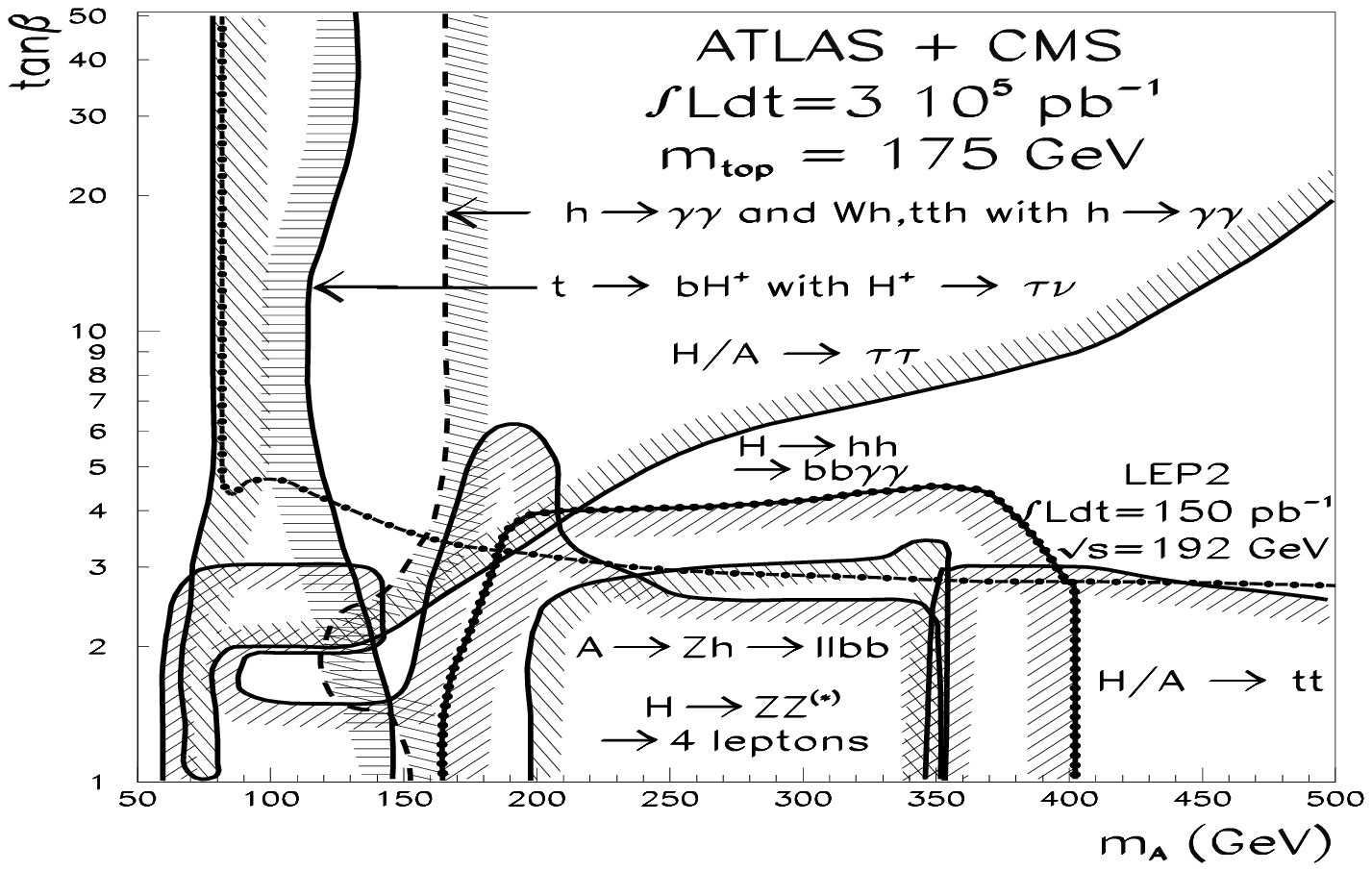}
\vspace*{-6.5cm}

\caption[]{\label{fg:atlascms} \it MSSM parameter space including the contours
of the various Higgs decay modes, which will be visible at the LHC after
reaching the anticipated integrated luminosity $\int {\cal L}  = 0.3~ab^{-1}$ 
and combining the experimental data of both LHC experiments,
ATLAS and CMS [taken from Ref. \cite{richter}].}
\end{figure}

\noindent
The final summary, Fig. \ref{fg:atlascms},  exhibits a difficult region for
the MSSM Higgs search at the
LHC. For $\tgb \sim 5$ and $M_A \sim 150$ GeV, the full
luminosity and the full data sample of both the ATLAS and CMS 
experiments at the
LHC are needed to cover the problematic parameter region \cite{richter}.
On the other hand, if no excess of Higgs events
above the SM background processes beyond 2 standard deviations will be found,
the MSSM Higgs bosons can be excluded at 95\% C.L.\\

The overall picture reveals several
difficulties, as evident from Fig. \ref{fg:atlascms}. Even though the
entire supersymmetric Higgs parameter
space may finally be covered by the
LHC experiments, the entire ensemble of individual Higgs
bosons is accessible only in part of
the parameter space. Moreover, the search
for heavy $H,A$ Higgs particles is very
difficult, because of the $t\bar t$
continuum background for masses  $\gsim 500$ GeV.

\subsection{Measuring the Parity of Higgs Bosons}
Once the  Higgs bosons  are
discovered, 
the properties of the particles must be established.
Besides the reconstruction of the supersymmetric Higgs potential \cite{66A},
which will be a very demanding effort, the external quantum
 numbers must be established, in particular the parity of the
heavy scalar and pseudoscalar Higgs particles $H$ and $A$ \cite{618}.\\[-0.1cm]

For large $H,A$ masses the decays $H,A\to t\bar t$ 
to top final states can be used to
discriminate between the different parity
assignments \cite{618}. For example, the $W^+$ and $W^-$
bosons in the $t$ and $\bar t$
decays tend to be emitted antiparallel
and parallel in the plane perpendicular
to the $t\bar t$ axis:
\begin{equation}
\frac{d\Gamma^\pm}{d\phi_*} \propto 1 \mp \left( \frac{\pi}{4} \right)^2
\cos \phi_*
\end{equation}
for $H$ and $A$ decays, respectively. \\[-0.1cm]

For light $H,A$ masses, $\gamma\gamma$
collisions appear to provide a viable
solution \cite{618}. The fusion of Higgs
particles in linearly polarized photon
beams depends on the angle between
the polarization vectors. For scalar $0^+$
particles the production amplitude
is non-zero for parallel polarization
vectors, while pseudoscalar $0^-$
particles require perpendicular
polarization vectors:
\begin{equation}
{\cal M}(H)^+  \sim  \vec{\epsilon}_1 \cdot \vec{\epsilon}_2  \hspace*{0.5cm}
\mbox{and} \hspace*{0.5cm}
{\cal M}(A)^-  \sim  \vec{\epsilon}_1 \times \vec{\epsilon}_2 ~.
\end{equation}
The experimental set-up for Compton
back-scattering of laser light can
be tuned in such a way that the
linear polarization of the hard-photon
beams approaches values close to 100\%.
 Depending on the $\pm$ parity 
of the resonance produced, the measured
asymmetry for photons of  parallel and perpendicular polarization, 
\begin{equation}
{\cal A} = \frac{\sigma_\parallel - \sigma_\perp}{\sigma_\parallel +
\sigma_\perp} ~, 
\end{equation}
is either positive or negative.

\subsection{Non-minimal Supersymmetric Extensions}
The minimal supersymmetric extension of the \SM may appear very
restrictive for supersymmetric theories in general, in particular in
the Higgs sector where the quartic couplings are identified with the
gauge couplings.  However, it turns out that the mass pattern of the
MSSM is quite typical if the theory is assumed to be valid up to the
GUT scale -- the motivation for supersymmetry {\it sui generis}.  This
general pattern has been studied thoroughly within the
next-to-minimal extension: the MSSM, incorporating two Higgs
isodoublets, is extended by introducing an additional isosinglet field $N$.
This extension leads to a model \citer{621,70A} that is generally
referred to as the (M+1)SSM.

\STS The additional Higgs singlet can solve the so-called
$\mu$-problem [i.e. $\mu \sim$ order $M_W$] by
eliminating the $\mu$ higgsino parameter from the potential and by 
replacing this parameter  by the vacuum expectation value of the $N$ field,
which can  naturally be related to the usual vacuum expectation values
of the Higgs isodoublet fields.  In this scenario the superpotential
involves the two trilinear couplings $H_1 H_2 N$ and $N^3$.  The
consequences of this extended Higgs sector will be outlined  in
the context of (s)grand unification, including the universal soft breaking
terms of the supersymmetry \cite{622}.\\

\GS The Higgs spectrum of the (M+1)SSM includes, besides the minimal
set of Higgs particles, one additional scalar and pseudoscalar Higgs
particle.  The neutral Higgs \ps are in general mixtures of 
iso scalar doublets, which couple to $W, Z$ bosons and fermions, and
the iso scalar singlet, decoupled from the non-Higgs sector.
The trilinear self-interactions contribute to the masses of the Higgs
particles; for the lightest Higgs boson of each species: 
\begin{eqnarray}
M^2 (h_1) & \le & M^2_Z \cos^2 2\beta + \lambda^2 v^2 \sin^2 2 \beta \\
M^2 (A_1) & \le & M^2 (A)   \nonumber \\
M^2 (H^{\pm}) & \le & M^2 (W) + M^2 (A) - \lambda^2 v^2 \nonumber
\end{eqnarray}
In contrast with the minimal model, the mass of the charged
Higgs \p could be smaller than the \W mass.  Since the trilinear \cps
increase with energy, upper bounds on the mass of the lightest neutral
Higgs boson $h_1^0$ can be derived, in analogy to the Standard Model,
from the assumption that the theory be valid up to the GUT scale:
$m(h_1^0) \lessim 140 $~GeV.  Thus, despite the additional
interactions, the distinct pattern of the minimal extension remains
valid also in more complex \ssy \sces.  In fact, the mass
bound of 140~GeV for the lightest Higgs particle is realized in almost
all \ssy theories \cite{623}.  If $h_1^0$ is (nearly) pure iso scalar, it
decouples from the gauge boson and fermion system and its role is
taken by the next Higgs \p with a large isodoublet component, implying
the validity of the mass bound again.\\

\begin{figure}[hbtp]
\begin{center}
\epsfig{file=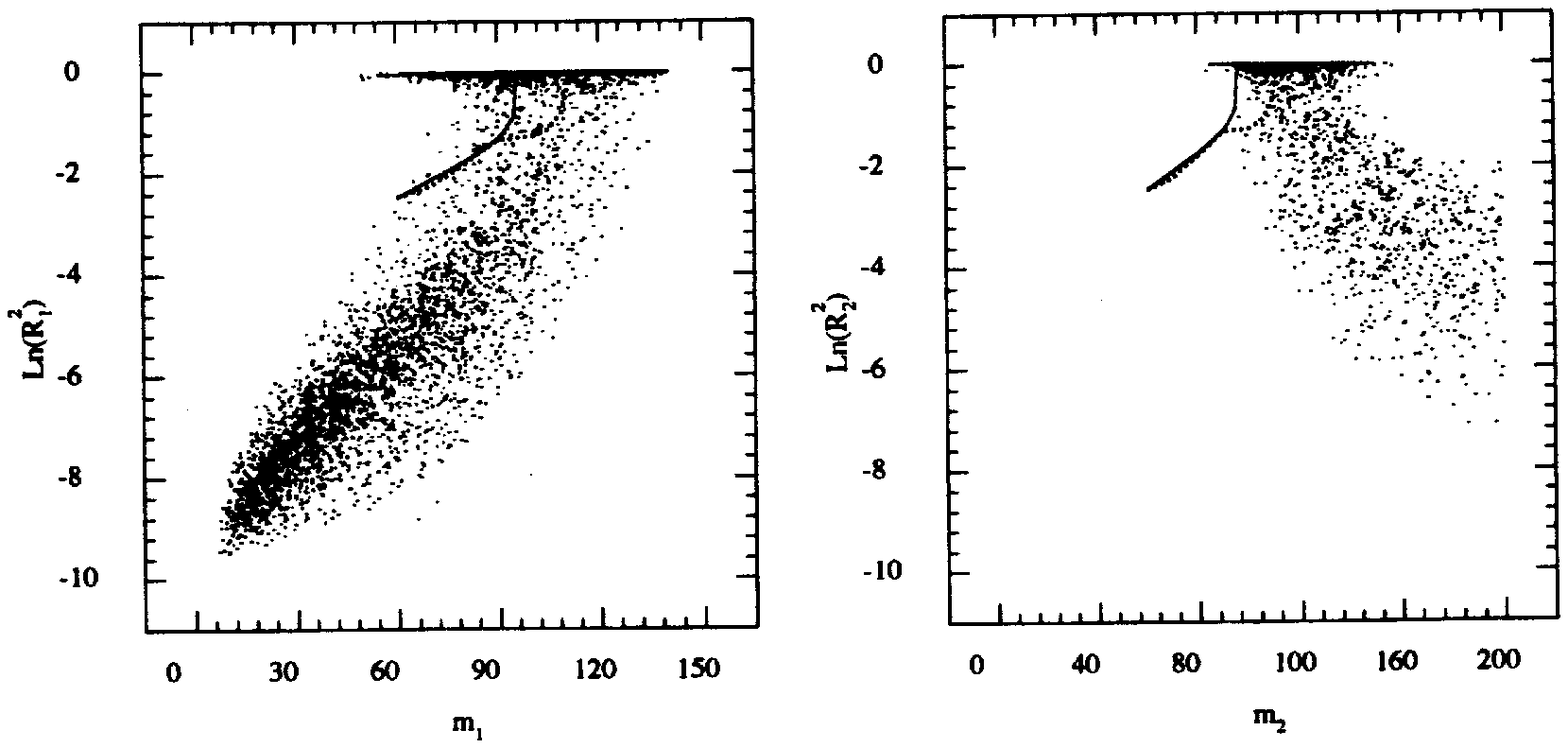,width=15.5cm}
\end{center}
\vspace{-.7cm}
\caption[]{\it 
  The couplings $ZZh_1$ and $ZZh_2$ of the two lightest ${\cal CP}$-even
  Higgs bosons in the next-to-minimal supersymmetric extension of
  the Standard Model, $(M + 1) SSM$. The solid lines indicate the
  accessible range at LEP2 for an energy of 192~GeV, the dotted lines 
for  205~GeV.
  The scatter plots are solutions for an ensemble of possible SUSY
  parameters defined at the scale of grand unification.
  Ref. \protect\cite{622}.  \protect\label{f621}\label{Ellw}}
\end{figure}

\STS The \cps $R_i$ of the ${\cal CP}$-even neutral Higgs \ps $h_i^0$
to the $Z$ boson, $ZZh_i^0$, are defined relative to the usual SM
coupling.  If the Higgs \p $h_1^0$ is primarily isosinglet, the \cp
$R_1$ is small and the \p cannot be produced by Higgs-strahlung.
However, in this case $h_2^0$ is generally light and couples with
sufficient strength to the $Z$ boson; if not, $h_3^0$ plays this role.
This \sce is quantified in Fig. \ref{f621}, where the \cps $R_1$ and
$R_2$ are shown for the ensemble of allowed Higgs masses $m(h_1^0)$
and $m(h_2^0)$ [adopted from Ref. \cite{10}; see also
Refs. \cite{622,L141A}].  Two different regions exist within the GUT
(M+1)SSM: a densely populated region with $R_1 \sim 1$ and $m_1 > 50
$~GeV, and a tail with $R_1 < 1$ to $\ll 1$ and small $m_1$.  Within
this tail, the lightest Higgs boson is essentially a gauge-singlet
state so that it can escape detection at LEP [full/solid lines].  If
the lightest Higgs boson is essentially a gauge singlet, the second
lightest Higgs \p cannot be heavy.  In the tail of diagram~\ref{f621}a
the mass of the second Higgs boson $h_2^0$ varies between 80~GeV and,
essentially, the general upper limit of $\sim 140 $~GeV;  $h_2^0$
couples with full strength to $Z$ bosons, $R_2 \sim 1$.  If this \cp becomes
 weak in the
tail of diagram~\ref{f621}b, the third Higgs
boson will finally take the role of the leading light particle.\\[0.5cm]

{\it In summa}. Experiments at $e^+e^-$ colliders are in a `no-lose' situation 
\cite{L141A} for detecting the Higgs particles in general supersymmetric
theories, even for c.m. energies as low as $\sqrt{s} \sim 300$ GeV.


\section{Strongly Interacting $W$ Bosons}
The Higgs mechanism is based on the
theoretical concept  of spontaneous symmetry
breaking \cite{1}. In the canonical 
formulation, adopted in the Standard
Model, a four-component {\it fundamental}
scalar field is introduced, which is
endowed with a self-interation such 
that the field acquires a non-zero
value in the ground state. The specific
direction in iso space, which is singled out by  the 
ground-state solution, breaks the
isospin invariance of the interaction
spontaneously. The interaction of the gauge fields with the 
scalar field in the ground state 
 generates the masses
of these fields. The longitudinal degrees
of freedom of the gauge fields are built
up by absorption of the Goldstone modes, 
which are associated with the spontaneous
breaking of the electroweak symmetries 
in the scalar field sector. Fermions
acquire masses through Yukawa interactions
with the ground-state field. While three
scalar components are absorbed by the 
gauge fields, one degree of freedom
manifests itself as a physical particle,
the Higgs boson. The exchange of this
particle in scattering amplitudes, including
longitudinal gauge fields and massive 
fermion fields, guarantees the unitarity
of the theory up to asymptotic energies.

In the alternative to this scenario 
based on a fundamental Higgs field, the
spontaneous symmetry breaking is generated
{\it dynamically} \cite{2}. A system of novel 
fermions is introduced, which interact
strongly at a scale of order 1 TeV. In
the ground state of such a system a scalar
condensate of fermion--antifermion pairs
may form. Such a process is  generally
expected to be realized in any non-Abelian gauge theory
of the novel strong interactions [and 
realized in QCD, for instance]. Since the
scalar condensate breaks the chiral
symmetry of the fermion system, Goldstone
fields will form, and these  can be absorbed
by the electroweak gauge fields to build
up the longitudinal components and the
masses of the gauge fields. Novel gauge
interactions must be introduced, which
couple the leptons and quarks of the
Standard Model to the new fermions in order
to generate lepton and quark masses
through interactions with the ground-state
fermion--antifermion condensate. In the
low-energy sector of the electroweak theory, 
the fundamental Higgs-field approach and
the dynamical alternative are equivalent.
However, the two theories are fundamentally
different at high energies. While the 
unitarity of the electroweak gauge theory
is guaranteed by the exchange of the scalar
Higgs particle in scattering processes, 
unitarity is restored in the dynamical
theory at high energies through the
non-perturbative strong interactions
between the particles. Since the longitudinal
gauge field components are equivalent to the
Goldstone fields associated with the microscopic
theory, their strong interactions at high
energies are transferred to the electroweak
gauge bosons. Since, by unitarity, the $S$-wave scattering
 amplitude of longitudinally polarized $W, Z$ bosons in the 
isoscalar channel $(2W^+W^- + ZZ) / \sqrt{3}$, 
$a^0_0 = \sqrt{2} G_F s/ 16 \pi$, is bounded by
1/2, the characteristic scale of the new strong interactions
must be close to 1.2 TeV. Thus near the critial energy of
1 TeV the $W, Z$ bosons interact strongly with each other.
Technicolour theories provide an elaborate form of such scenarios.

\subsection{Dynamical Symmetry Breaking}
Physical scenarios of dynamical
symmetry breaking may be based on  new
strong interaction theories, which extend
the  spectrum of matter particles and of the interactions 
beyond the degrees of freedom realized in the
Standard Model. If the new strong interactions are 
invariant under transformations of a
chiral $SU(2) \times SU(2)$
group, the chiral invariance is generally  broken
spontaneously down to the diagonal custodial isospin group
$SU(2)$. This process is associated with the
formation of a chiral condensate in the
ground state and the existence of three
massless Goldstone bosons.

\begin{figure}[hbt]
\begin{center}
\begin{picture}(60,10)(90,30)
\Photon(0,25)(50,25){3}{6}
\LongArrow(60,25)(75,25)
\put(-10,21){$V$}
\end{picture}
\begin{picture}(60,10)(70,30)
\Photon(0,25)(50,25){3}{6}
\put(55,21){$+$}
\end{picture}
\begin{picture}(60,10)(55,30)
\Photon(0,25)(25,25){3}{3}
\Photon(50,25)(75,25){3}{3}
\Line(25,24)(50,24)
\Line(25,26)(50,26)
\GCirc(25,25){5}{0.5}
\GCirc(50,25){5}{0.5}
\put(80,21){$+$}
\put(35,30){$G$}
\end{picture}
\begin{picture}(60,10)(20,30)
\Photon(0,25)(25,25){3}{3}
\Photon(50,25)(75,25){3}{3}
\Photon(100,25)(125,25){3}{3}
\Line(25,24)(50,24)
\Line(25,26)(50,26)
\Line(75,24)(100,24)
\Line(75,26)(100,26)
\GCirc(25,25){5}{0.5}
\GCirc(50,25){5}{0.5}
\GCirc(75,25){5}{0.5}
\GCirc(100,25){5}{0.5}
\put(130,21){$\cdots$}
\put(35,30){$G$}
\put(85,30){$G$}
\end{picture}
\end{center}
\caption[]{\label{fg:gaugemass} \it Generating  gauge-boson masses (V)
 through the interaction with the Goldstone bosons (G).}
\end{figure}
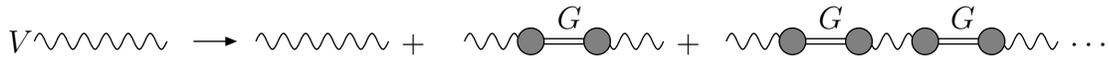
The Goldstone bosons can be absorbed by
the gauge fields, generating longitudinal
states and non-zero masses of the gauge bosons, as shown in
Fig. \ref{fg:gaugemass}. Summing up the geometric series
of vector-boson--Goldstone-boson transitions
in the propagator leads to in a shift of the  
mass pole:
\begin{eqnarray}
\frac{1}{q^2} & \to & \frac{1}{q^2} + \frac{1}{q^2} q_\mu \frac{g^2 F^2/2}{q^2}
q_\mu \frac{1}{q^2} + \frac{1}{q^2} \left[ \frac{g^2 F^2}{2} \frac{1}{q^2}
\right]^2 + \cdots \nonumber \\
& \to & \frac{1}{q^2-M^2}
\end{eqnarray}
The coupling between gauge fields and
Goldstone bosons has been defined as $ig F/\sqrt{2} q_\mu$.
The mass generated for the gauge field is related
to this coupling by
\begin{equation}
M^2 = \frac{1}{2} g^2 F^2 ~.
\end{equation}
The numerical value of the coupling $F$ must coincide
with $v=246$ GeV.\\

The remaining custodial $SU(2)$ symmetry
guarantees that the $\rho$
parameter, the relative strength between
$NC$ and $CC$ couplings, is one. Denoting the $W/B$ mass
matrix elements by
\begin{equation}
\begin{array}{rclcrcl}
\langle W^i | {\cal M}^2 | W^j \rangle & = & \displaystyle \frac{1}{2} g^2
F^2 \delta_{ij}
& \hspace*{1cm} & \langle W^3 | {\cal M}^2 | B \rangle & = & \langle B |
{\cal M}^2 | W^3 \rangle \\ \\
\langle B | {\cal M}^2 | B \rangle & = & \displaystyle \frac{1}{2} g'^2 F^2 &
& & = & \displaystyle \frac{1}{2} gg' F^2
\end{array}
\end{equation}
the universality of the coupling $F$ leads
to the ratio  $M_W^2/M_Z^2 = g^2/(g^2+g'^2) =
\cos^2\theta_W$ of the mass eigenvalues, equivalent to $\rho=1$.\\

Since the wave functions of longitudinally
polarized vector bosons grow with the
energy, the longitudinal field components are
the dominant degrees of freedom at high
energies. These states can, however, 
for asymptotic energies be identified with the
absorbed Goldstone bosons. This equivalence \cite{75} is apparent
in the 't Hooft--Feynman gauge where, for
asymptotic energies, 
\begin{equation}
\epsilon_\mu^L W_\mu \to k_\mu W_\mu \sim M^2 \Phi ~. 
\end{equation}
The dynamics of gauge bosons can therefore be
identified at high energies with the
dynamics of scalar Goldstone fields. An
elegant representation of the Goldstone
fields $\vec{G}$ in this context is provided by
the exponentiated form
\begin{equation}
U = \exp [-i \vec{G} \vec{\tau}/v ] ~, 
\end{equation}
which corresponds to an $SU(2)$ matrix field.\\

The Lagrangian of a system of strongly interacting
 bosons  consists in such a scenario 
of the Yang--Mills part ${\cal L}_{YM}$
and the interactions ${\cal L}_G$
of the Goldstone fields, 
\begin{equation}
{\cal L}={\cal L}_{YM}+{\cal L}_G ~. 
\end{equation}
The Yang--Mills part is written in the
usual form ${\cal L}_{YM} = -\frac{1}{4} {\rm Tr} [W_{\mu\nu} W_{\mu\nu} +
B_{\mu\nu} B_{\mu\nu} ]$.  
The interaction of the Goldstone fields can be systematically expanded
in chiral theories 
in the derivatives of the
fields, corresponding to expansions in
powers of the energy for scattering
amplitudes \cite{76}:
\begin{equation}
{\cal L}_G = {\cal L}_0 + \sum_{dim=4} {\cal L}_i + \cdots
\end{equation}
Denoting the SM covariant derivative of
the Goldstone fields by
\begin{equation}
D_\mu U = \partial_\mu U - i g W_\mu U + i g' B_\mu U ~, 
\end{equation}
the leading term ${\cal L}_0$, which is  
of dimension = 2, is given by
\begin{equation}
{\cal L}_0 = \frac{v^2}{4} {\rm Tr} [ D_\mu U^+ D_\mu U ] ~.
\end{equation}
This term generates the masses of the $W,Z$
gauge bosons: $M_W^2 = \frac{1}{4} g^2 v^2$ and
$M_Z^2 = \frac{1}{4} (g^2+g'^2) v^2$.
The only parameter in this part of the
interaction is $v$, which however is fixed
uniquely by the experimental value of the
$W$ mass; thus the amplitudes predicted by
the leading term in the chiral expansion
can effectively be considered as parameter-free.\\

The next-to-leading component in the expansion with
dimension = 4 consists of ten individual terms. If the
custodial $SU(2)$ symmetry is imposed, only two
terms are left, which do not affect propagators
and 3-boson vertices but only 4-boson vertices.
Introducing the vector field $V_\mu$ by 
\begin{equation}
V_\mu = U^+ D_\mu U
\end{equation}
these two terms are given by the interaction
densities
\begin{equation}
{\cal L}_4  =  \alpha_4 \left[Tr V_\mu V_\nu \right]^2 \hspace*{0.5cm}
\mbox{and} \hspace*{0.5cm}
{\cal L}_5  =  \alpha_5 \left[Tr V_\mu V_\mu \right]^2
\end{equation}

The two coefficients $\alpha_4,\alpha_5$
are free parameters that must be adjusted
experimentally from $WW$ scattering data.

Higher orders in the chiral expansion give
rise to an energy expansion of the scattering
amplitudes of the form ${\cal A} = \sum c_n (s/v^2)^n$.
This series  will diverge at energies for which
the resonances of the new strong interaction
theory can be formed in $WW$ collisions: $0^+$ `Higgs-like', 
$1^-$ `$\rho$-like' resonances, etc. The masses of these resonance
states are expected in the range $M_R \sim 4\pi v$ 
where chiral loop expansions diverge,
i.e. between about 1 and 3 TeV.

\subsection{$WW$ Scattering at High-Energy Colliders}
The (quasi-)elastic 2--2 $WW$ scattering
amplitudes can be expressed at high
energies by a master amplitude
$A(s,t,u)$, which depends on the three
Mandelstam variables of the scattering processes:
\begin{eqnarray}
A(W^+ W^- \to ZZ) & = & A(s,t,u) \\
A(W^+ W^- \to W^+ W^-) & = & A(s,t,u) + A(t,s,u) \nonumber \\
A(ZZ \to ZZ) & = & A(s,t,u) + A(t,s,u) + A(u,s,t) \nonumber \\
A(W^- W^- \to W^- W^-) & = & A(t,s,u) + A(u,s,t) ~. \nonumber
\end{eqnarray} ~\\

To lowest order in the chiral expansion, ${\cal L} \to {\cal L}_{YM} +
{\cal L}_0$, the master amplitude is given, in a
parameter-free form, by the energy squared $s$:
\begin{equation}
A(s,t,u) \to \frac{s}{v^2} ~.
\end{equation}
This representation is valid for energies $s \gg M_W^2$
but below the new resonance region, i.e. in
practice at energies $\sqrt{s}={\cal O}(1~\mbox{TeV})$.
Denoting the scattering length for the
channel carrying isospin $I$ and angular
momentum $J$ by $a_{IJ}$,
the only non-zero scattering channels
predicted by the leading term of the
chiral expansion correspond to
\begin{eqnarray}
a_{00} & = & + \frac{s}{16\pi v^2} \\
a_{11}   & = & + \frac{s}{96\pi v^2} \nonumber \\
a_{20}   & = & - \frac{s}{32\pi v^2} ~.
\end{eqnarray}
While the exotic $I=2$ channel is repulsive,
the $I=J=0$ and $I=J=1$ channels are attractive,
indicating the formation of non-fundamental
Higgs-type and $\rho$-type resonances.\\

Taking into account the next-to-leading terms
in the chiral expansion, the master amplitude
turns out to be \cite{24}
\begin{equation}
A(s,t,u) = \frac{s}{v^2} + \alpha_4 \frac{4(t^2+u^2)}{v^4}
+ \alpha_5 \frac{8s^2}{v^4} + \cdots ~,
\end{equation}
including the two parameters $\alpha_4$ and $\alpha_5$.\\

Increasing the energy, the
amplitudes will approach the resonance area.
There, the chiral character of the
theory does not provide any more guiding principle
for constructing  the scattering
amplitudes. Instead, {\it ad-hoc} hypotheses must
be introduced to define the nature of the
resonances; see e.g. Ref. \cite{24a}. A sample
of resonances is provided by the following
models:
\begin{description}
\item[(a)] \underline{SM heavy Higgs boson:}
\begin{eqnarray}
A & = & - \frac{M_H^2}{v^2}\left[1 + \frac{M_H^2}{s-M_H^2 + iM_H \Gamma_H}
\right] \\
& & \mbox{with}~~~\Gamma_H = \frac{3 M_H^3}{32 \pi v^2} \nonumber
\end{eqnarray}

\item[(b)] \underline{Chirally coupled scalar resonance:}
\begin{eqnarray}
A & = & \frac{s}{v^2} - \frac{g_s^2 s^2}{v^2} \frac{1}{s-M_S^2 - iM_S \Gamma_S}
\\
& & \mbox{with}~~~\Gamma_S = \frac{3g_s^2 M_S^3}{32 \pi v^2} \nonumber
\end{eqnarray}

\item[(c)] \underline{Chirally coupled vector resonance:}
\begin{eqnarray}
A & = & \frac{s}{v^2}\left[1-\frac{3a}{4}\right] + \frac{aM_V^2}{4v^2}
\left[\frac{u-s}{t-M_V^2 + iM_V \Gamma_V} + (u \leftrightarrow t) \right] \\
& & \mbox{with}~~~\Gamma_V = \frac{aM_V^3}{192 \pi v^2} \nonumber
\end{eqnarray}

\end{description}
For small energies, the scattering
amplitudes reduce to the leading chiral
form $s/v^2$.  In the resonance region they are 
described by two parameters, the mass
and the width of the resonance. The
amplitudes interpolate between the
two regions in a simplified smooth way.\\
 
\begin{figure}[hbt]
\begin{center}
\begin{picture}(60,50)(140,0)
\ArrowLine(0,50)(25,50)
\ArrowLine(25,50)(75,50)
\ArrowLine(0,0)(25,0)
\ArrowLine(25,0)(75,0)
\Photon(25,50)(45,30){-3}{3}
\Photon(25,0)(45,20){3}{3}
\Photon(55,20)(75,10){-3}{3}
\Photon(55,30)(75,40){3}{3}
\GBox(45,20)(55,30){0.5}
\put(-20,48){$q/e$}
\put(-20,-2){$q/e$}
\put(20,8){$W$}
\put(20,30){$W$}
\put(80,35){$W$}
\put(80,5){$W$}
\put(110,35){$I=0,2~~J~\mbox{even}$}
\put(110,15){$I=1~~~~~J~\mbox{odd}$}
\end{picture}
\begin{picture}(60,50)(-20,0)
\ArrowLine(25,25)(0,50)
\ArrowLine(0,0)(25,25)
\Photon(25,25)(75,25){3}{5}
\Photon(75,25)(100,50){3}{4}
\Photon(75,25)(100,0){3}{4}
\Photon(125,0)(175,0){3}{5}
\Photon(125,50)(175,50){3}{5}
\GBox(100,-5)(125,55){0.5}
\put(-15,48){$e^+$}
\put(-15,-2){$e^-$}
\put(70,2){$W^-$}
\put(70,40){$W^+$}
\put(180,48){$W^+$}
\put(180,-2){$W^-$}
\put(145,22){$I=J=1$}
\end{picture}
\end{center}
\caption[]{\label{fg:qqtoqqww} \it $WW$ scattering and rescattering 
at high energies at the LHC
and TeV $e^+e^-$ linear colliders.}
\end{figure}
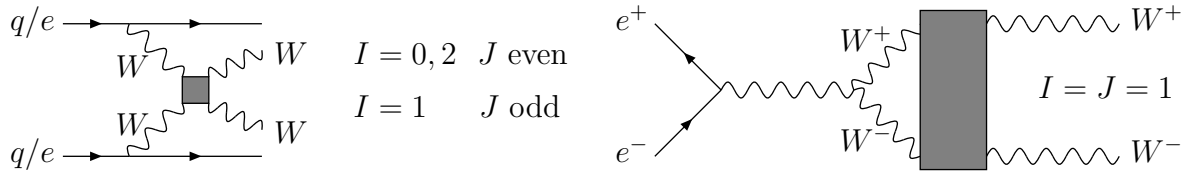

\begin{figure}[hbtp]
\begin{center}
\vspace*{-5.5cm}

\hspace*{-3.5cm}
\epsfig{file=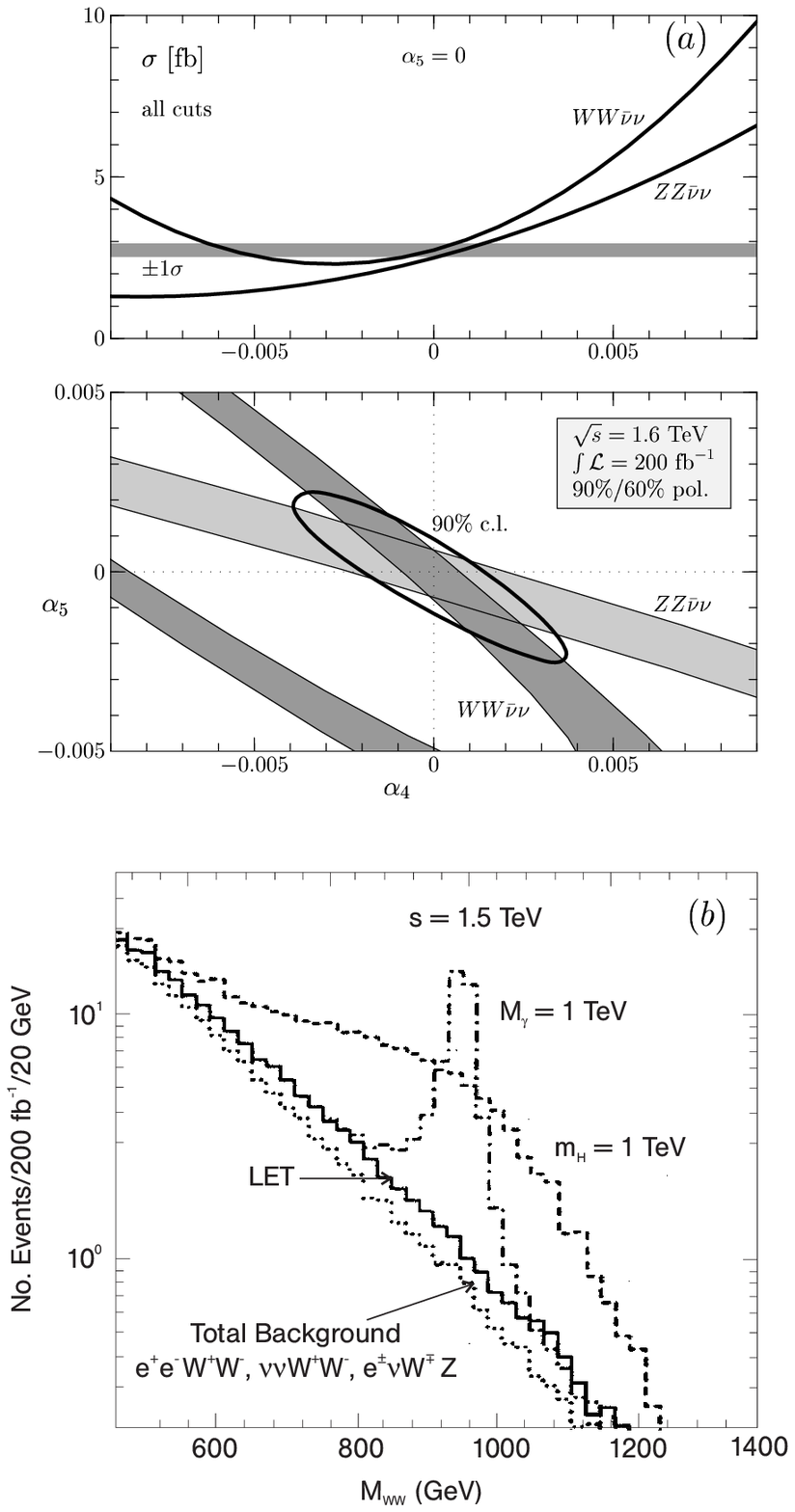,width=22.0cm,angle=0}
\vspace*{-8.5cm}

\end{center}
\caption[]{\it
  Upper part: Sensitivity to the expansion
  parameters in chiral electroweak models of $WW \to WW$ and $WW \to
  ZZ$ scattering at the strong-interaction threshold;
  Ref. \protect\cite{24}.
Lower part: The distribution of the $WW$ invariant energy in $e^+e^-
  \rightarrow \overline{\nu} \nu WW$ for scalar and vector resonance
  models [$M_H, M_V$ = 1 TeV];
  Ref. \protect\cite{24a}. 
\protect\label{17tt}\label{PKB}
}
\end{figure}
\begin{figure}[hbtp]

\vspace*{-2.0cm}
\hspace*{-2.0cm}
\epsfxsize=20cm \epsfbox{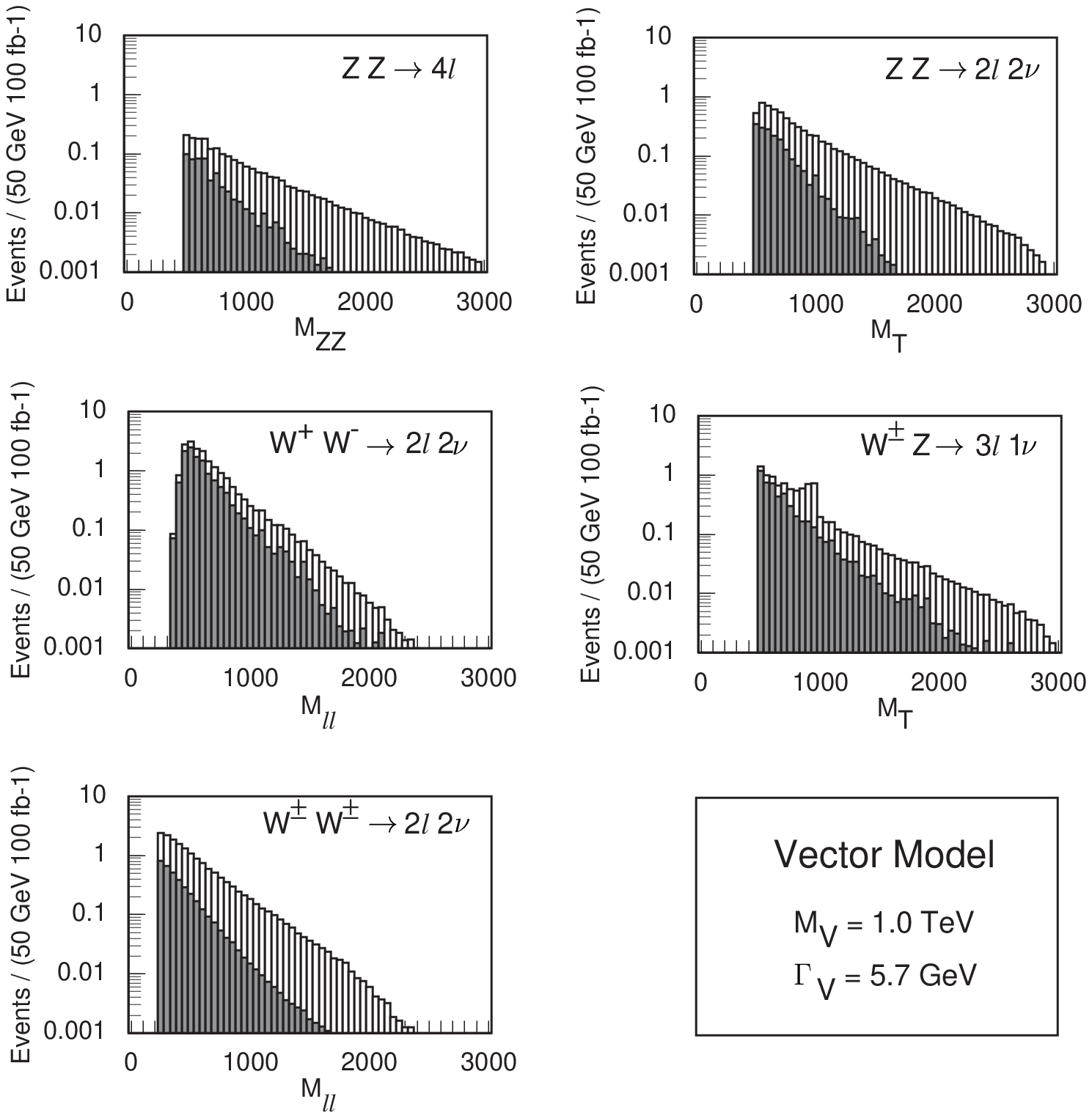}
\vspace*{-11cm}

\caption[]{\label{fg:vvto4l} \it Invariant mass distributions for the
gold-plated purely leptonic final states that arise from the processes
$pp\to ZZX \to 4\ell X, pp\to ZZX\to 2\ell 2\nu X, pp\to W^+W^-X, pp\to
W^\pm ZX$ and $pp\to W^\pm W^\pm X$, for the LHC (mass in  GeV).
The signal is plotted above the summed background. Distributions are shown
for a chirally coupled vector with $M_V=1$ TeV, $\Gamma_V=5.7$ GeV;
Ref. \protect\cite{23B}.}
\end{figure}
$WW$ scattering can be studied at the LHC
and at TeV $e^+e^-$ linear colliders. At high energies, 
equivalent $W$ beams accompany the quark
and electron/positron beams (Fig. \ref{fg:qqtoqqww})
in the fragmentation processes $pp\to qq \to qqWW$ and
$ee\to \nu\nu WW$; the spectra of the longitudinally
polarized $W$ bosons have been given in Eq. (\ref{eq:xyz}). 
In the hadronic LHC  environment the final-state 
$W$ bosons can only be observed in
leptonic decays. Resonance reconstruction
is thus not possible for charged $W$ final
states. However, the clean environment of
$e^+e^-$ colliders will allow the reconstruction 
of resonances from $W$ decays to jet pairs.
The results of three experimental 
simulations are displayed in Fig. \ref{PKB}.
In Fig. \ref{PKB}a the sensitivity to the parameters
$\alpha_4,\alpha_5$ of the chiral expansion is shown for $WW$
scattering in $e^+e^-$ colliders \cite{24}. The results of this analysis
can be reinterpreted as sensitivity to the
parameter-free prediction of the chiral
expansion, corresponding to an error of
about 10\% in the first term of the master
amplitude $s/v^2$.  These experiments test the basic concept
of dynamical symmetry breaking through
spontaneous symmetry breaking. The production
of a vector-boson resonance of mass $M_V=1$ TeV
is exemplified in Fig. \ref{PKB}b \cite{24a}. Expectations
for leptonic invariant energies of  $WW$ scattering final states
 at the LHC are compared in the vector model 
with the background in Fig. \ref{fg:vvto4l} \cite{23B}.\\

A second powerful method measures the elastic 
$W^+W^- \to W^+W^-$ scattering in the $I=1, J=1$ channel. The
rescattering of $W^+W^-$ bosons produced in $e^+e^-$
annihilation, cf. Fig. \ref{fg:qqtoqqww}, depends at high
energies on the $WW$ scattering phase $\delta_{11}$ 
\cite{77}. The production amplitude $F = F_{LO} \times R$
is the product of the lowest-order
perturbative diagram with the Mushkelishvili--Omn\`es rescattering amplitude
${\cal R}_{11}$,
\begin{equation}
{\cal R}_{11} = \exp \frac{s}{\pi} \int \frac{ds'}{s'}
\frac{\delta_{11}(s')}{s'-s-i\epsilon} ~,
\end{equation}
which is determined by the $I = J = 1$ $WW$ phase shift $\delta_{11}$.
The power of this method derives from the 
fact that 
the entire $e^+e^-$
collider energy is transferred to the $WW$ system
[while a major fraction of the energy
is lost in the fragmentation of $e \to \nu W$
if the $WW$ scattering is studied in the
process $ee\to \nu\nu WW$]. Detailed simulations \cite{78}
have shown that this process is sensitive
to vector-boson masses up to about $M_V \lsim 6$ TeV in technicolor-type 
theories. More elaborate scenarios \cite{78A} have been analysed in 
Ref. \cite{79}.

\newpage

\section{Summary}
The mechanism of electroweak symmetry
breaking can be established in the present
or the next generation of $e^+e^-$ and $p\bar p/pp$ 
colliders:
\begin{itemize}
\item[$\star$] Whether there exists a light fundamental Higgs boson;
\item[$\star$] The profile of the Higgs particle can be
  reconstructed, which reveals the physical
  nature of the underlying mechanism of
  electroweak symmetry breaking;
\item[$\star$] Analyses of strong WW scattering can be
  performed if the symmetry breaking is of a
  dynamical nature and generated in a novel
  strong interaction theory.
\end{itemize}
Moreover, depending on the experimental answer to these questions,
the electroweak sector will provide the
platform for extrapolations into physical
areas beyond the Standard Model:  either
to the low-energy  supersymmetry sector or, alternatively,
to a new strong interaction theory at a characteristic 
scale of order 1 TeV.


\end{document}